\documentclass[twocolumn,superscriptaddress,nofootinbib,preprintnumbers,floatfix]{revtex4}

\usepackage{amsmath}
\usepackage{graphicx}
\usepackage[hang,tight,raggedright]{subfigure}
\usepackage[hyperfootnotes=false,bookmarks=false]{hyperref}
\usepackage{xspace}
\usepackage{xcolor}

\newcommand{\eq}[1]{Eq.~\eqref{eq:#1}}
\newcommand{\eqs}[2]{Eqs.~\eqref{eq:#1} and \eqref{eq:#2}}
\renewcommand{\sec}[1]{Sec.~\ref{sec:#1}}

\newcommand{\subsec}[1]{Sec.~\ref{subsec:#1}}

\newcommand{\app}[1]{App.~\ref{app:#1}}
\newcommand{\fig}[1]{Fig.~\ref{fig:#1}}
\newcommand{\figs}[2]{Figs.~\ref{fig:#1} and \ref{fig:#2}}

\newcommand{\ord}[1]{\mathcal{O}(#1)}
\newcommand{\ORd}[1]{{\mathcal O}\Bigl(#1\Bigr)}

\newcommand{\al}{\alpha}
\newcommand{\bt}{\beta}
\newcommand{\de}{\delta}
\newcommand{\ga}{\gamma}
\newcommand{\Ga}{\Gamma}

\newcommand{\W}{\Omega}

\newcommand{\df}{\mathrm{d}}
\newcommand{\img}{\mathrm{i}}

\newcommand{\Tau}{\mathcal{T}}

\newcommand{\cI}{{\mathcal I}}
\newcommand{\cL}{{\mathcal L}}

\newcommand{\id}{\mathbf{1}}
\newcommand{\bT}{\mathbf{T}}

\newcommand{\hs}{\hat{s}}

\newcommand{\fr}{\frac}
\newcommand{\nn}{\nonumber}

\newcommand{\Li}{\mathrm{Li}}

\newcommand{\lqcd}{\Lambda_\mathrm{QCD}}

\newcommand{\cusp}{\mathrm{cusp}}
\newcommand{\cut}{\mathrm{cut}}
\newcommand{\Ecm}{E_\mathrm{cm}}

\newcommand{\kT}{k${}_T$\xspace}

\newcommand{\bq}{{\bar{q}}}

\newcommand{\GeV}{\mathrm{GeV}}
\newcommand{\Pythia}{\textsc{Pythia}\xspace}

\newcommand{\pTJ}{{p_T^J}}
\newcommand{\pTJs}{{p_T^{J\,2}}}

\newcommand{\geopT}{\mbox{$p_T$}\xspace}

\allowdisplaybreaks[4]

\begin{document}

%%%%%%%%%%%%%%%%%%%%%%%%%%%%%%%%%%%%%%%%%%%%%%%%%%%%%%%%%%%%%%%%%%%%%%%%%%%%%%%%
% Title page
%%%%%%%%%%%%%%%%%%%%%%%%%%%%%%%%%%%%%%%%%%%%%%%%%%%%%%%%%%%%%%%%%%%%%%%%%%%%%%%%

\title{Jet Mass Spectra in Higgs + One Jet at NNLL}

\preprint{\vbox{
\hbox{MIT--CTP 4427}
\hbox{DESY 12-222}
}}

\author{Teppo T.~Jouttenus}
\affiliation{Center for Theoretical Physics, Massachusetts Institute of 
Technology, Cambridge, MA 02139, USA\vspace{0.5ex}}

\author{Iain W.~Stewart}
\affiliation{Center for Theoretical Physics, Massachusetts Institute of 
Technology, Cambridge, MA 02139, USA\vspace{0.5ex}}

\author{Frank J.~Tackmann}
\affiliation{Theory Group, Deutsches Elektronen-Synchrotron (DESY), D-22607 Hamburg, Germany\vspace{0.5ex}}

\author{Wouter J.~Waalewijn\vspace{1.ex}}
\affiliation{Department of Physics, University of California at San Diego, 
La Jolla, CA 92093, USA\vspace{0.5ex}}

\date{February 4, 2013}

%%%%%%%%%%%%%%%%%%%%%%%%%%%%%%%%%%%%%%%%%%%%%%%%%%%%%%%%%%%%%%%%%%%%%%%%%%%%%%%%
\begin{abstract}
The invariant mass of a jet is a benchmark variable describing
the structure of jets at the LHC.
We calculate the jet mass spectrum for Higgs plus one jet at the LHC at
next-to-next-to-leading logarithmic (NNLL) order using a factorization
formula. At this order, the cross section becomes sensitive to perturbation
theory at the soft $m_{\rm jet}^2/p_T^{\rm jet}$ scale.  Our calculation is
exclusive and uses the 1-jettiness global event shape to implement a veto on
additional jets. The dominant dependence on the jet veto is removed by
normalizing the spectrum, leaving residual dependence from non-global
logarithms depending on the ratio of the jet mass and jet veto variables.  For
our exclusive jet cross section these non-global logarithms are parametrically
smaller than in the inclusive case, allowing us to obtain a complete NNLL
result.  Results for the dependence of the jet mass spectrum on the
kinematics, jet algorithm, and jet size $R$ are given.  Using individual
partonic channels we illustrate the difference between the jet mass spectra
for quark and gluon jets.  We also study the effect of hadronization and
underlying event on the jet mass in \Pythia. To highlight the similarity of
inclusive and exclusive jet mass spectra, a comparison to LHC data is
presented.
\end{abstract}
%%%%%%%%%%%%%%%%%%%%%%%%%%%%%%%%%%%%%%%%%%%%%%%%%%%%%%%%%%%%%%%%%%%%%%%%%%%%%%%%

\maketitle

%%%%%%%%%%%%%%%%%%%%%%%%%%%%%%%%%%%%%%%%%%%%%%%%%%%%%%%%%%%%%%%%%%%%%%%%%%%%%%%%
\section{Introduction}
%%%%%%%%%%%%%%%%%%%%%%%%%%%%%%%%%%%%%%%%%%%%%%%%%%%%%%%%%%%%%%%%%%%%%%%%%%%%%%%%

There has been a rapidly expanding theoretical and experimental effort on
techniques that exploit the substructure of jets (for a recent review see
Ref.~\cite{Altheimer:2012mn}). Jet substructure is of interest both for testing
QCD and for identifying new physics. Much of the excitement in this field has
been driven by the excellent performance of the ATLAS and CMS detectors, and the
sophisticated jet measurements this has made possible at the LHC. Jet
substructure measurements can for example be used to tag boosted heavy
particles, whose decay products get collimated into a fat jet, or to test and
tune Monte Carlo programs. Most theoretical work has focused on designing these
techniques and observables with the help of Monte Carlo programs. At the same
time, one would also like to know that these methods are under theoretical
control and build confidence that higher-order effects are not significant. (For
some recent progress in this direction see e.g.
Refs.~\cite{Ellis:2009wj,Bauer:2011uc,Feige:2012vc,Larkoski:2012eh,Krohn:2012fg,Waalewijn:2012sv}.)

As our underlying hard process we consider $pp\to H+1$ jet with gluon fusion
$gg\to H$ as the underlying Higgs production mechanism. This process is
convenient as it provides a clean setup with a single quark or gluon jet in the
final state via the three basic partonic channels $gg\to Hg$, $gq\to Hq$, and $q\bar q \to Hg$.
Of course, it is also important in its own right for Higgs measurements at the LHC,
which rely on exclusive jet channels.

Here we focus on one of the simplest jet substructures: the invariant mass of a
jet.  A successful calculation of this benchmark observable will instill
confidence in our ability to carry out analogous calculations for other more
complicated jet substructure observables. Such analyses require incorporating
both a resummation of large logarithms $\alpha_s^i\ln^j(m_J^2/\pTJs)$ where
$m_J$ is the jet mass and $\pTJ$ is the transverse momentum of the jet, as well
as fixed-order perturbative corrections.  This is made intricate by the
dependence on multiple variables. There has been a lot of recent work on the
calculation (resummation) of the jet invariant mass spectrum for jets with a
realistic angular size~\cite{Ellis:2009wj, Ellis:2010rwa, Jouttenus:2009ns,
  Banfi:2010pa, Kelley:2011tj, Kelley:2011aa, Li:2012bw, Dasgupta:2012hg,
  Chien:2012ur} which we will review in more detail below.  Some of the key
theoretical issues that must be addressed for the LHC case include:
\begin{itemize} \setlength{\itemsep}{0.3ex}
    \setlength{\parskip}{0.5ex}
\item 
  Impact of summing large logarithms, $\ln(m_J^2/\pTJs)$
\item
  Soft radiation effects at the scale $m_J^2/\pTJ$
\item
  Impact of initial-state radiation
\item
  Color flow and hard process dependence
\item 
  Dependence on kinematics including rapidity cuts
\item
  Jet algorithm and dependence on jet size $R$
\item
  Inclusive ($\geq N$ jets) vs.~exclusive ($=N$ jets)
\item
  Impact of non-global logarithms (NGLs)
\item
  Effect of hadronization on the spectrum
\item
  Effect of underlying event on the spectrum
\item
  Effect of pile-up on the spectrum
\item 
  Utility of using groomed jets with trimming~\cite{Krohn:2009th},
  filtering~\cite{Butterworth:2008iy}, or pruning~\cite{Ellis:2009su}
\end{itemize}
We now elaborate on several of these items. For a jet with $\pTJ\sim
300\,{\rm GeV}$, the jet mass peaks at $m_J\sim 50\,{\rm GeV}$, leading to large
logarithms of $\pTJs/m_J^2\sim 36$. Therefore, a description of the peak region
of the jet mass spectrum requires the all-order resummation of these logarithms.
Soft radiation with momentum $k^\mu \sim m_J^2/\pTJ$ is generated by both
initial and final-state particles and contributes at leading order in the power
expansion to the jet mass. Since fixed-order corrections start to become
relevant for resummation at next-to-next-to-leading logarithmic (NNLL) order, a
proper treatment of the soft scale $\sim m_J^2/\pTJ$ is crucial at this
order~\cite{Korchemsky:1999kt, Bauer:2001yt, Fleming:2007qr, Ellis:2009wj}.
Numerically, the importance of these fixed-order soft corrections is also well
known from recent work up to N$^3$LL for event shapes in $e^+e^-\!\to$
jets~\cite{Chien:2010kc, Abbate:2010xh, Chien:2012ur, Gehrmann:2012sc}.  For
processes with $\geq 2$ jets at hadron colliders there are multiple color
structures, and the corresponding color flow must be taken into account starting
at next-to-leading-logarithmic (NLL) order~\cite{Dixon:2008gr}.

The available freedom in defining a jet introduces a dependence of the jet mass
spectrum on the choice of algorithm/clustering method and the jet size parameter
$R$.  There is also a choice of whether to use an inclusive or exclusive jet
cross section, where the latter involves a veto on additional jets.  The
inclusive case has been studied at the LHC~\cite{ATLAS:2012am}, and inclusive
calculations tend to focus on the anti-\kT algorithm~\cite{Cacciari:2008gp}.
(Use of the anti-\kT jet algorithm avoids issues associated to clustering
effects~\cite{Banfi:2005gj,Delenda:2006nf,KhelifaKerfa:2011zu,Kelley:2012kj}.)
As we will emphasize further below, a key difference between the inclusive and
exclusive cases are the form of the non-global
logarithms~\cite{Dasgupta:2001sh,Dasgupta:2002dc} that arise at ${\cal
  O}(\alpha_s^2)$ beyond the Born cross section due to multiple restrictions on
phase space.

Let us summarize how the above issues have been studied so far in the literature
on jet mass calculations. The first calculations were carried out for event
shapes in $e^+e^-\!\to$ jets using hemisphere jet masses. Here factorization
theorems are well established and calculations exist up to
N$^3$LL~\cite{Clavelli:1979md, Chandramohan:1980ry, Clavelli:1981yh, Korchemsky:1999kt,
Berger:2003iw, Fleming:2007qr, Fleming:2007xt, Chien:2012ur}.
In Refs.~\cite{Ellis:2009wj,Ellis:2010rwa} a factorization formula for exclusive
$N$-jet cross sections at $e^+e^-$ colliders was derived, where the angularity
of a jet (which includes the jet mass as a special case) is measured. This
result only depends on the class of the jet algorithm (such as cone or
\kT-type), but suffers from non-global logarithms involving the jet veto and jet
size $R$.  The resummation of the jet mass in $e^+e^-\!\to 2$ jets with a jet
veto was carried out at NLL in Ref.~\cite{Banfi:2010pa}, which includes a
resummation of NGLs in the large-$N_c$ approximation. This same process was
considered in Ref.~\cite{Kelley:2011tj}, where the dominant $R$ dependence of
asymmetric thrust (which is related to jet mass) was obtained using a
refactorization of the soft function. In Ref.~\cite{Kelley:2011aa}, this
refactorization was verified at ${\cal O}(\alpha_s^2)$ and the leading NGLs were
obtained at this order.

For jet mass calculations in $pp$ collisions one considers jets with large
transverse momentum, $p_T^J$, and with rapidities $\eta^J$ away from the beam
axis.  Recently, several inclusive jet mass calculations have been carried
out~\cite{Li:2012bw,Dasgupta:2012hg,Chien:2012ur}. In Ref.~\cite{Li:2012bw}, the
jet mass was calculated using only a jet function. This ignores important
contributions from wide-angle soft radiation, which couples together multiple
hard partons, depends on the choice of jet algorithm, and contains NGLs.  In
Ref.~\cite{Dasgupta:2012hg}, the jet mass in $pp \to 2$ jets and $Z+1$ jet were
calculated at NLL, including a resummation of NGLs in the large-$N_c$
approximation. Although this is an inclusive calculation (no jet veto), one
should also note that hard emissions giving rise to additional jets are beyond
the NLL order considered. In this case the dominant effect of the NGLs is on the peak of the
jet mass distribution. Another inclusive calculation of the jet mass was carried
out to obtain partial NNLL results in Ref.~\cite{Chien:2012ur}, by expanding
around the threshold limit.  Here dynamical threshold
enhancement~\cite{Appell:1988ie, Catani:1998tm, Becher:2007ty} was used to argue
that additional hard emissions are suppressed.
Although NGLs were not resummed, their size was estimated, and found to mainly
affect the peak region of the jet mass, as in Ref.~\cite{Dasgupta:2012hg}.

Our calculation at NNLL is for the exclusive jet mass spectrum, so it is useful
to highlight differences with the inclusive case.  At NLL, for a given partonic
channel and fixed momenta of the hard partons, the two cases simply differ by a
multiplicative factor, except for their respective NGLs. In both cases the
lowest order NGLs involve terms of the form
%%%
\begin{align} \label{eq:roughNGL}
  \al_s^2 \ln^2\Big( \frac{m_J^{\cut\,2}}{p_{\rm cut}^2}\Big) \,.
\end{align}
%%%
for the cumulant jet mass spectrum integrated up to $m_J^\cut$. For
the inclusive jet mass spectrum, $p_{\rm cut}$ is a hard scale $\simeq p_T^J$
and the NGLs are therefore large logarithms that are parametrically of the same
size as other $\alpha_s^i \ln^j(m_J^2/\pTJs)$ terms, and are thus part of the
NLL result.  Hence, in this case a complete resummation at NLL (or beyond)
requires the NGLs to be resummed to all orders, which practically is currently
only possible in the large-$N_c$ approximation.  In contrast, in the exclusive
case $p_{\rm cut}$ is an adjustable parameter and is related to the jet veto 
(in our analysis below we will have $p_\cut^2\simeq \pTJ \Tau^\cut$ where $\Tau^\cut$
implements the jet veto). In this case we have both $m_J^2 \ll \pTJs$ and $p_{\rm cut}^2\ll \pTJs$, so the
logarithms in \eq{roughNGL} are smaller than in the inclusive case. In
particular, for fixed $p_{\rm cut}$ there is a point in the $m_J$ spectrum
where the NGLs vanish, and there is a region about this point where the NGLs are
not large logarithms. An estimate
for the size of this region can be obtained from the series of three NGL terms
(log-squared, log, and non-log) that are known for the hemisphere jet
masses~\cite{Kelley:2011ng, Hornig:2011iu}. When all the terms in this series
are of similar magnitude the logarithmic enhancement is not dominant, and the
NGLs do not need to be resummed. This occurs for $1/8 \le m_J^{\cut\,2}/p_{\rm
  cut}^2 \le 8$. We will numerically explore the size of this region in our
exclusive jet mass calculation, and demonstrate that the region is large enough
that we may consider the non-global logarithms to not be large.  This can be
contrasted with Fig.~3 of Ref.~\cite{Banfi:2010pa}, which shows that the
presence of an unmeasured region of phase space makes large NGLs unavoidable in
the inclusive case~\cite{Dasgupta:2012hg, Chien:2012ur}.

It should also be noted that although exclusive jet cross sections are not
necessary for jet mass spectra, they are important in their own right because
many Higgs and new physics searches categorize the data by the number of jets to
improve their sensitivity. For example, the importance of the Higgs + 1 jet
channel in $H \to \tau \tau$ and $H \to WW^*$ was pointed out in
Refs.~\cite{Mellado:2004tj, Mellado:2007fb}. Recently a NLL resummation of jet
veto logarithms was carried out in the context of Higgs plus jets in
Ref.~\cite{Liu:2012sz}.

Our calculation of the jet mass is centered on using the $N$-jettiness global
event shape~\cite{Stewart:2010tn} to define jets, instead of a more traditional jet algorithm.
For an event with $N$ jets, $N$-jettiness assigns all particles to $N+2$
regions, corresponding to the $N$ jets and two beams. We calculate the cross
section for $pp \to H+1$ jet at NNLL, fully differential in the contributions of
each region to $1$-jettiness.  For the jet region, this contribution yields the
jet invariant mass.  The contribution from the remaining two beam regions are used to
implement the jet veto.  In each of these variables there is a series of large
double logarithms that must be summed.

An advantage of using $N$-jettiness is that the jet veto is made through a jet
mass-type variable, rather than a $p_T$ variable. Therefore, the structure of the
perturbation theory, which is simultaneously differential in these two kinematic
variables, is simpler. In particular, there is a QCD factorization formula for
this cross section~\cite{Stewart:2010tn, Jouttenus:2011wh}, obtained by making
use of Soft-Collinear Effective Theory (SCET)~\cite{Bauer:2000ew, Bauer:2000yr,
  Bauer:2001ct, Bauer:2001yt}. For the experimentally more realistic case of
measuring $m_J$ with a $p_T$ veto variable one must simultaneously deal with a
thrust-like invariant mass resummation and a $p_T$-type resummation.

Returning to our list of theoretical issues from the beginning, the use of
$N$-jettiness allows us to carry out the summation of large logarithms at NNLL
while properly accounting for soft radiation effects and initial-state
radiation.  We also use it to calculate the dependence of the jet mass spectrum
on the jet kinematics, the jet size, and the definition of the jet region.
Results are shown for individual partonic channels, $gg\to Hg$ and $gq\to Hq$,
illustrating the differences between quark and gluon jets, as well as the full
$pp\to H + 1$ jet process from the Higgs coupling through a top quark loop.  To
investigate the differences between exclusive and inclusive jet mass measurements we
compare our results with \Pythia and also to ATLAS jet mass
data~\cite{ATLAS:2012am}. We also analytically explore the effect
of NGLs on the jet mass spectrum, and the effect of hadronization and
underlying event with \Pythia~\cite{Sjostrand:2006za, Sjostrand:2007gs}.

Thus, we address all items in the bullet list of issues except for the
last two, for which some brief comments are in order.
Methods for removing pile-up contributions to jet observables have been
discussed in e.g.~Refs.~\cite{Cacciari:2007fd,Soyez:2012hv}, and direct pile-up
calculations are beyond the scope of our work. Finally, it is known that
grooming jets has a large impact on their soft radiation and causes significant
changes to the jet mass spectrum. We do not attempt to analytically control the
effects of jet-grooming methods here.

In calculating the jet mass we consider both absolute and normalized spectra.
Normalizing the jet mass spectrum reduces the perturbative uncertainty, and
turns out to remove the dominant dependence on the jet veto variable. In
particular, the jet veto dependence cancels up to NLL if we consider a
particular partonic channel and fixed jet kinematics.  We will show that this
cancellation remains effective when summing over partonic channels and
integrating over a range of kinematic variables.

In \sec{kin}, we discuss the kinematics and several jet definitions based on
$N$-jettiness, exploring their features. The technical details of our
calculation are presented in \sec{calc}. Here we discuss the factorization
formula for the cross section, the refactorization of the soft function,
non-global logarithms, and the choice of running scales.  \sec{results_part} and
\sec{results_Hj} contain our numerical results for the individual partonic
channels and for $pp\to H+1$ jet, showing the dependence of the jet mass
spectrum on the jet veto cut, the order in perturbation theory, the jet
kinematics, the jet definition, the jet area, on gluon versus quark jets, and on
NGLs. Using \textsc{Pythia}8, in \sec{compare} we analyze the hard
process dependence for gluon jets, compare inclusive versus exclusive jet mass
spectra, study the dependence on classic jet algorithms, and look at the
impact of hadronization and underlying event. We also compare our NNLL
exclusive jet results with \Pythia for the same jet definition and kinematics,
and compare them with inclusive jets from the LHC data. We conclude in
\sec{conc}.  Detailed ingredients for the NNLL cross section are summarized in
appendices.

%%%%%%%%%%%%%%%%%%%%%%%%%%%%%%%%%%%%%%%%%%%%%%%%%%%%%%%%%%%%%%%%%%%%%%%%%%%%%%%%
\section{Kinematics and Jet Definitions}
\label{sec:kin}
%%%%%%%%%%%%%%%%%%%%%%%%%%%%%%%%%%%%%%%%%%%%%%%%%%%%%%%%%%%%%%%%%%%%%%%%%%%%%%%%

\begin{figure*}[t]
\hfill
\subfigure[Geometric $E$ and $p_T$ measures with $\rho=0.5$.]
{\label{fig:geopte}\includegraphics[width=0.8\columnwidth]{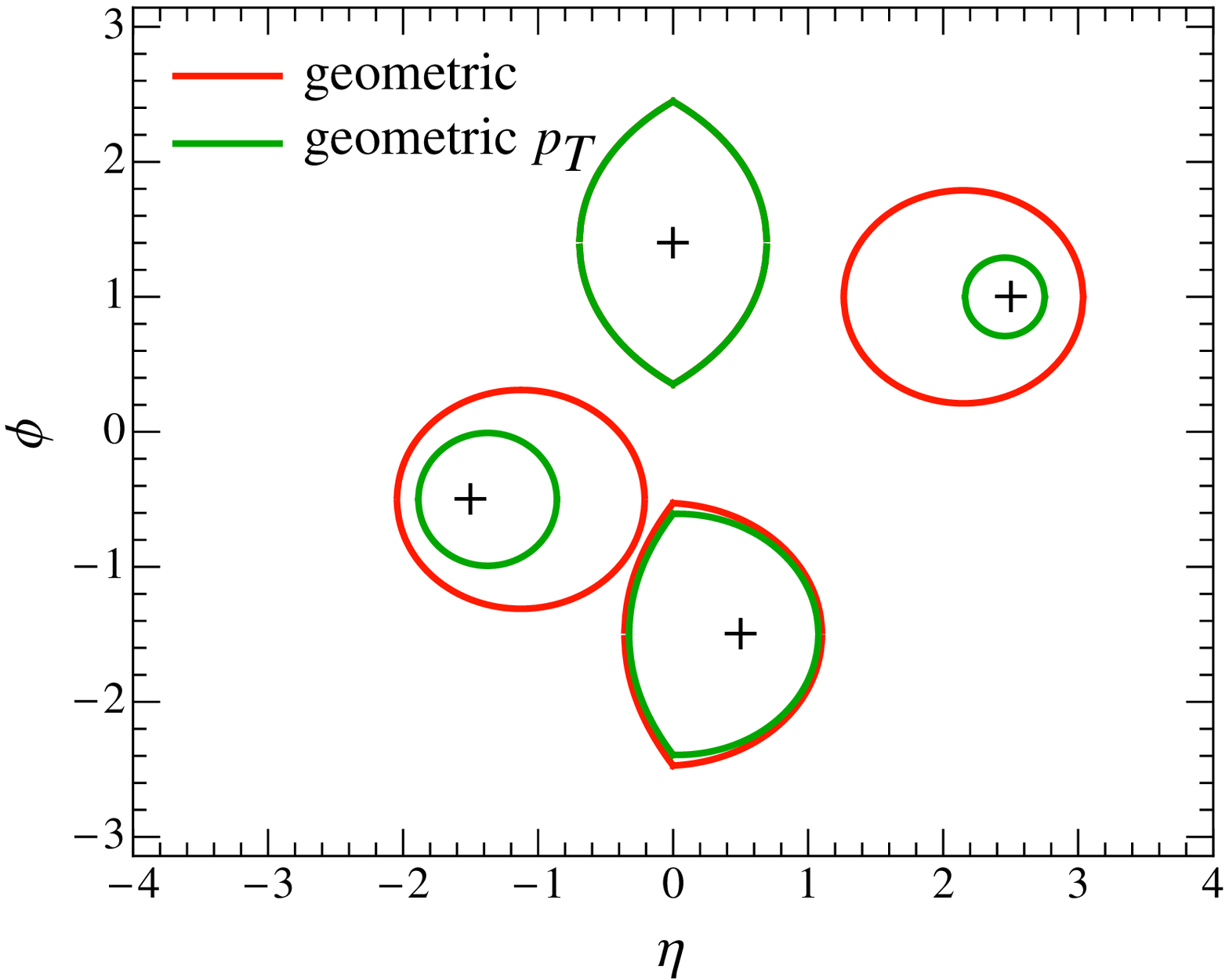}}%
\hfill
\subfigure[Anti-\kT and geometric $R$ for $R=1$.]
{\label{fig:geor}\includegraphics[width=0.8\columnwidth]{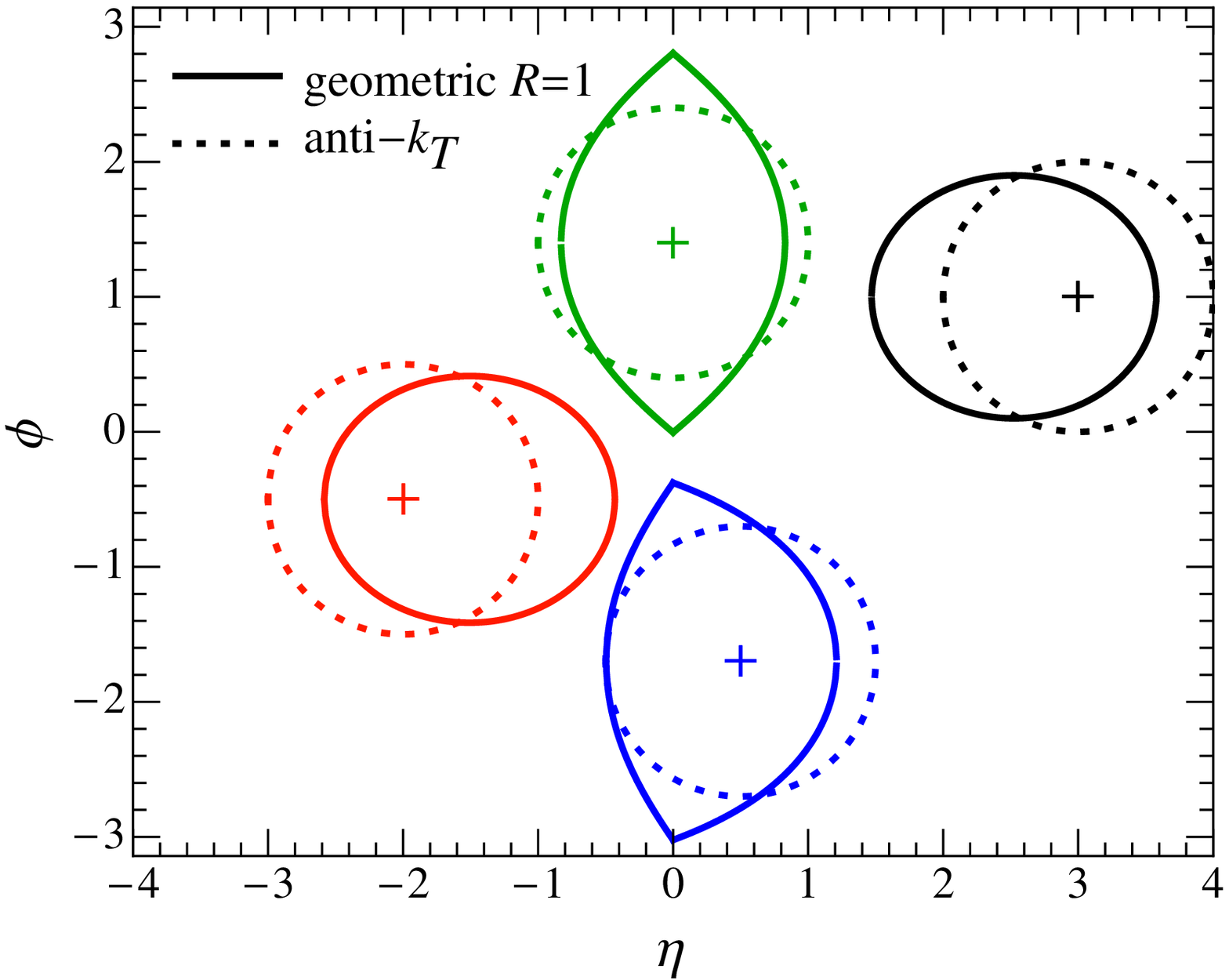}}%
\hspace{\fill}
\vspace{-1ex}
\caption{Comparison of the jet regions for different jet measures at different $\eta$ and $\phi$. The ``$+$'' marks the jet direction $\vec{n}_J$.}
\vspace{-1ex}
\end{figure*}

We describe the process $pp\to H+1$ jet using the transverse momentum $p_T^J$ of the jet, the pseudorapidity $\eta_J$ of the jet, and the rapidity $Y$ of the hard collision relative to the CM frame of the colliding protons.
The $1$-jettiness event shape is defined as~\cite{Stewart:2010tn}
%%%
\begin{equation}\label{eq:TauN_def}
\Tau_1 = \sum_k
\min \Bigl\{ 
     \frac{2 q_J\cdot p_k}{Q_J} ,
     \frac{2 q_a\cdot p_k}{Q_a} ,
     \frac{2 q_b\cdot p_k}{Q_b}    \Bigr\}
\,,\end{equation}
%%%
where $a, b$ denote the two beams and $J$ the jet, the $q_i$ are massless
reference momenta and the $Q_i$ are normalization factors. For the reference
momenta we take
%%%
\begin{align} \label{eq:qis}
q_J^\mu = E_J\, (1,\vec n_J)
\,,\qquad
q_{a,b}^\mu = x_{a,b}\frac{\Ecm}{2} (1, \pm \hat z)
\,.\end{align}
%%%
The jet energy $E_J$ and jet
direction $\vec n_J$ can be predetermined with a suitable jet algorithm. The jet
algorithm dependence this induces on $\Tau_1$ is power
suppressed~\cite{Stewart:2010tn}, and we will use anti-\kT.\footnote{If $Q_J =2
  E_J$ then an equally good choice would be to minimize $\Tau_1$ with respect to
  the axis $\vec n_J$. A fast algorithm to carry out this minimization has been
  devised in Ref.~\cite{Thaler:2011gf}, using a slightly different $N$-jettiness
  measure than the ones we use here.}  The unit vector $\hat z$ points along the
beam axis, and the momentum fractions $x_a$ and $x_b$ are fixed in terms of the
total invariant mass $Q$ and rapidity $Y$,
%%%
\begin{align}
x_a x_b \Ecm^2 &= Q^2 = (q_J + q_H)^2
\,,\nn\\
\ln\frac{x_a}{x_b} &= 2 Y  = \ln\frac{(1, -\hat z) \cdot (q_J+q_H)}
  {(1, \hat z) \cdot (q_J+q_H)}
\,,\end{align}
%%%
where $q_H^\mu$ denotes the momentum of the Higgs. For later convenience we also
introduce the notation
\begin{align}
  s_{ij} = 2q_i \cdot q_j \,.
\end{align}

The minimum in \eq{TauN_def} divides the total phase space into $3$ regions, one
for each beam and one for the jet. We denote their contributions to $\Tau_1$ as
$\Tau_a$ and $\Tau_b$ for the two beam regions, and $\Tau_J$ for the jet region,
so
%%%
\begin{align}
  \Tau_1 =  \Tau_J + \Tau_a + \Tau_b
  \,.
\end{align}
%%%
The contribution of the jet, $\Tau_J$, is directly related to the jet's invariant mass $m_J$
%%%
\begin{align} \label{eq:M2}
m_J^2 &= p_J^2 = (\bar n_J \cdot p_J) (n_J\cdot p_J) - \vec p_{J\perp}^{\:2}
\nn\\
&= 2 q_J \cdot p_J\, [1 + \ord{\lambda^2}]
\nn\\
&= Q_J \Tau_J\, [1 + \ord{\lambda^2}]
\,,\end{align}
%%%
where $p_J^\mu$ is the full jet momentum defined by summing all particles in the
$\Tau_J$-region, $n_J^\mu=(1,\vec n_J)$ and $\bar n_J^\mu=(1,-\vec n_J)$ are defined
by the predetermined jet direction $\vec n_J$, and the power counting parameter $\lambda$
scales as $\lambda^2 \sim \Tau_J/E_J \sim m_J^2/E_J^2$. In the second line of
\eq{M2} we used the fact that $\vec n_J$ and the exact direction of the $N$-jettiness
jet, $\vec p_J$, differ by very little, such that $p_{J\perp}/(\bar n_J \cdot
p_J)\sim\lambda^2$. The difference between these two jet directions affects
the jet boundary, which changes the contribution of soft radiation to the jet $p_T$,
but only by a small amount $\sim \lambda^2$.  We also used that the large jet
momentum $\bar n_J\cdot p_J = \bar n_J\cdot q_J [1 + \ord{\lambda^2}]$. For a
jet with $\pTJ\sim 300\,{\rm GeV}$ these $\ord{\lambda^2}$ power corrections are
$1/36\sim 3\%$ in the peak region, and hence negligible relative to the
perturbative uncertainties at NNLL.  Investigating the jet
mass spectra for the exact $m_J^2 = p_J^2$ vs.~using $m_J^2 = Q_J \Tau_J$ in \Pythia,
we also find that they are indistinguishable.

\begin{figure*}[t]
\hfill
\includegraphics[width=0.75\columnwidth]{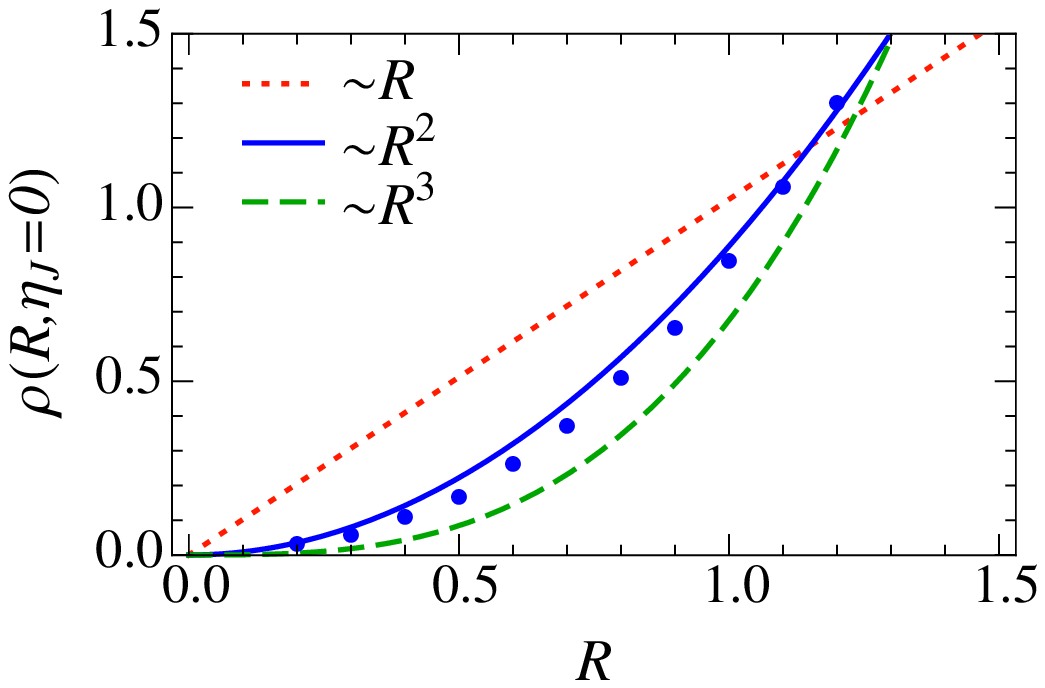}%
\hfill
\includegraphics[width=0.75\columnwidth]{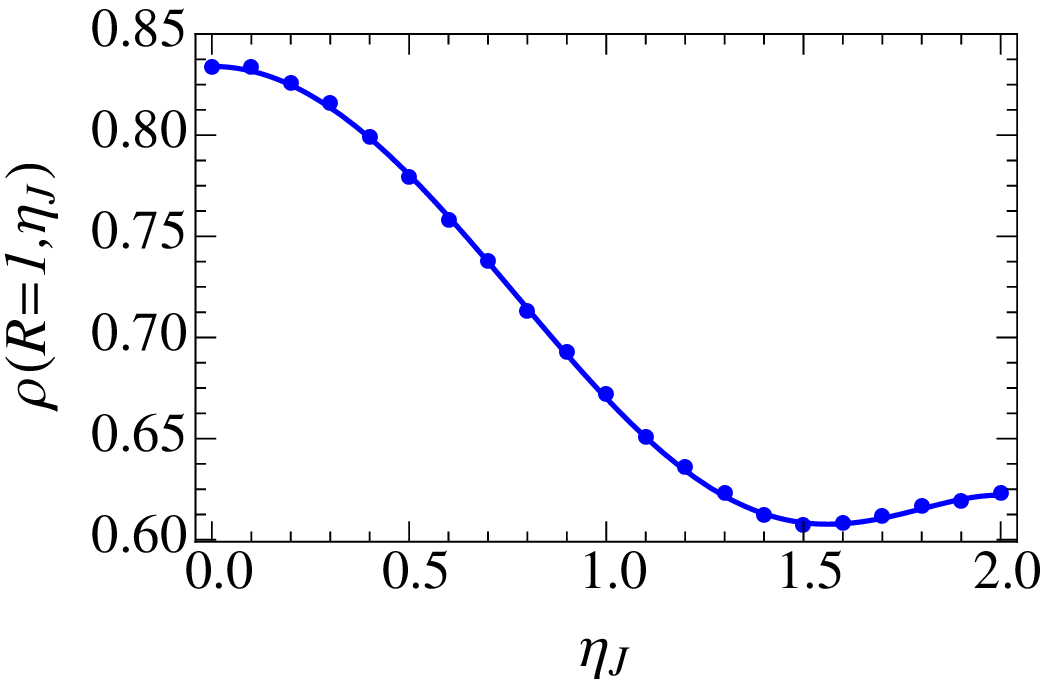}%
\hspace*{\fill}
\vspace{-2ex}
\caption{Numerical results for $\rho(R,\eta_J)$ in the geometric $R$ measure.
  Left: Dependence on $R$ for $\eta_J=0$, which is $\sim R^2$ as expected.
  Right: Dependence on $\eta_J$ for $R=1$. To solve for $\rho$ we use a fit
  (solid line) to the true $\eta_J$ dependence (dots).}
\label{fig:rho}%
\vspace{-2ex}
\end{figure*}

The details of the beam and jet regions selected by the minimum condition in
\eq{TauN_def} depend on the normalization factors $Q_i$. Since their values
affect which particles are grouped into the beam and jet regions, they
constitute a jet measure. They also impact the geometric shape of
  the jet area. Differences between measures are therefore similar to the
different choices for jet-algorithms (anti-\kT, Cambridge-Aachen, cone, etc.).
We will consider a variety of choices:
\begin{itemize}
\item %\vspace{-1ex}
invariant-mass measure: 
 \begin{align}
   Q_J = Q_a= Q_b =Q
 \end{align}
\item \vspace{-1ex}
geometric \geopT measure: 
\begin{align}
 Q_J &= 2 \rho \, |\vec{q}_{iT}| = 2 \rho\, E_J/\cosh \eta_J \\
 Q_{a,b} &= x_{a,b} \Ecm = e^{\pm Y} Q \nn
\end{align}
\item \vspace{-1ex}
geometric measure:
\begin{align}
 Q_J &= 2 \rho\, E_J \\
 Q_{a,b} &= x_{a,b} \Ecm = e^{\pm Y} Q \nn
\end{align} 
\item \vspace{-1ex}
geometric $R$ measure:
\begin{align}
  Q_J &= 2\rho(R,\eta_J)\, E_J\\
 Q_{a,b} &= x_{a,b} \Ecm = e^{\pm Y} Q \nn
\end{align}
where $\rho(R,\eta_J)$ fixes the area of the jet in $(\eta,\phi)$-space to be
$\pi R^2$.
\end{itemize}
In all cases $\rho$ is a dimensionless parameter that allows one to change the size
of the jet region. In the geometric $R$ case $\rho$ is fixed in terms of the
jet radius parameter $R$.\footnote{For the multijet case we would 
use the same $\rho(R,\eta_J)$ for each jet that is determined when they do not overlap.}
The choice of $Q_{a,b}$ in the
geometric measures removes the dependence in $q_a^\mu/Q_a$ and $q_b^\mu/Q_b$ on
the total rapidity $Y$. This is useful in the presence of missing energy, which
prohibits the measurement of the boost $Y$ of the partonic center-of-mass frame. Note that the 
definitions of the measures through the $Q_i$ is influenced by the convention to use energies inside 
the $q_i^\mu$s in \eq{qis}, since only the ratio $q_i^\mu/Q_i$ appears.
Since for the geometric measures $Q_J \sim E_J$, they are all insensitive to the total jet energy. For the
geometric \geopT case the jet is weighted by $E/p_T$ and we have explicitly
%%%
\begin{align} \label{eq:geometric}
\frac{2 q_i\cdot p_k}{{q}_{iT}}
&= p_{kT}\, \Bigl(2 \frac{m_{kT}}{p_{k T}}
\cosh \Delta y_{ik} \!-\! 2\cos \Delta \phi_{ik}\Bigr)
\end{align}
where $\Delta y_{ik} = y_i - y_k$, $\Delta\phi_{ik} = \phi_i - \phi_k$ are the
differences in rapidity and azimuthal angle between the direction of jet $i$ and
particle $k$, and $m_{kT}^2=p_{kT}^2 + m^2$ for a particle of mass $m$. For
massless particles we thus get
\begin{align}
\frac{2 q_i\cdot p_k}{{q}_{iT}}
&=  p_{kT}\, (2\cosh \Delta y_{ik} - 2\cos \Delta \phi_{ik})
\nn\\
&\approx p_{kT}\,\bigl[ (\Delta y)^2 + (\Delta \phi_{ik})^2 \bigr]
\,.\end{align}
%%%
The jet regions for geometric \geopT and geometric are roughly circular, as shown
in \fig{geopte}. They become smaller at large rapidities for geometric \geopT,
while they stay of comparable size for the geometric case.

For geometric $R$, numerical results for the parameter $\rho(R, \eta_J)$ as function of $R$ and $\eta_J$ are shown
in \fig{rho}. The left panel shows that the dependence on the jet radius $R$ is
approximately $\rho \propto R^2$, as expected. The right panel illustrates the dependence on $\eta_J$ for
fixed $R = 1$, showing that $\rho$ approaches a constant for large $\eta_J$, i.e.~when the jet becomes close to the beam. When using geometric $R$ in our results below,
we use for convenience a fit of the $\eta_J$ dependence for fixed value of $R$. For example,
for $R=0.5, 0.7, 1, 1.2$ we have for $|\eta_J|\leq2$
%%%
\begin{align}
  \rho(R=0.5,\eta_J)
  &= 0.164 + 0.037 \eta_J^2 - 0.009 \eta_J^4 + 0.0008 \eta_J^6
  \,, \nn \\
  \rho(R=0.7,\eta_J)  
 &=0.357 - 0.040 \eta_J^2 + 0.031 \eta_J^4 - 0.005 \eta_J^6
  \,, \nn \\
  \rho(R=1,\eta_J)
 &= 0.834 - 0.233 \eta_J^2 + 0.077 \eta_J^4 - 0.008 \eta_J^6
  \,, \nn \\
  \rho(R=1.2,\eta_J) 
  & = 1.272 - 0.377 \eta_J^2 + 0.101 \eta_J^4 - 0.010 \eta_J^6 
\,.\end{align}
%%%
Note that for $R=0.5$ the parameter $\rho$ increases rather than 
decreases with $\eta_J$. A comparison of the jet regions for
geometric $R$ with anti-\kT jets is shown in \fig{geor}. Although their
areas are chosen to be the same, the geometric $R$ jets are not perfectly
circular and have an ``offset'' between the jet direction and the center of the
jet region. The former (latter) effect decreases (increases) with $|\eta_J|$.
For a smaller jet radius of $R = 0.5$ the geometric $R$ jets become more circular
also at central rapidities and are very close to anti-\kT jets.
In Ref.~\cite{Thaler:2011gf} a modification of $N$-jettiness was introduced that
matches anti-\kT closely for any $R$. However, this definition reintroduces a region of
phase space that belongs neither to the jet nor the beams, making it more
complicated for calculations.

%%%%%%%%%%%%%%%%%%%%%%%%%%%%%%%%%%%%%%%%%%%%%%%%%%%%%%%%%%%%%%%%%%%%%%%%%%%%%%%%
\vspace{2ex}
\section{Calculation}
\vspace{-1ex}
\label{sec:calc}
%%%%%%%%%%%%%%%%%%%%%%%%%%%%%%%%%%%%%%%%%%%%%%%%%%%%%%%%%%%%%%%%%%%%%%%%%%%%%%%%

%%%
\begin{table}[t!]
 \begin{tabular}{l|ccc}
 \hline\hline
 channel & $\kappa_a$ & $\kappa_b$ & $\kappa_J$ \\
 \hline
  $gg \to Hg$ & $g$ & $g$ & $g$ \\
  $gq \to Hq$ & $g$ & $q$ & $q$ \\
  $qg \to Hq$ & $q$ & $g$ & $q$ \\
  $g\bar q \to \bar Hq$ & $g$ & $\bar q$ & $\bar q$ \\
  $\bar q g \to \bar Hq$ & $\bar q$ & $g$ & $\bar q$ \\
  $q \bar q \to Hg$ & $q$ & $\bar q$ & $g$ \\
  $\bar q q \to Hg$ & $\bar q$ & $q$ & $g$ \\
  \hline\hline
 \end{tabular}
 \caption{Values of $\kappa$ for the different partonic channels.}
\label{tab:kappa}
\end{table}
%%%

%===============================================================================
\subsection{Factorization Formula}
\vspace{-2ex}
%===============================================================================

We start by rewriting the phase space integrals for the hard kinematics in terms of the rapidity $\eta_J$ and transverse momentum $p_T^J$ of the jet and the total rapidity $Y$,
%%%
\begin{align} \label{eq:phsp}
 & \int\!\frac{\df x_a}{x_a}\! \int\! \frac{\df x_b}{x_b}\! \int \fr{\df^3 \vec q_H}{(2\pi)^3} \frac{1}{2E_H}\! \int\! \frac{\df^3 \vec q_J}{(2\pi)^3} \frac{1}{2E_J}\,
\nn \\ 
& \quad \times  (2\pi)^4 \delta^4\bigl(q_a + q_b - q_J  - q_H\bigr)
\nn \\
 & \qquad
= \int\! \df \eta_J\, \df p_T^J\, \df Y\, \frac{1}{2 \pi} \frac{p_T^J}{Q^2+m_H^2}
\,.\end{align}
%%%
The variables were defined in \sec{kin}, and we used azimuthal symmetry and the relations
%%%
\begin{align} \label{eq:pTQ}
  p_T^J &= \frac{Q^2 - m_H^2}{2Q \cosh(\eta_J-Y)}
  \,, \\
  Q &= p_T^J\cosh (\eta_J-Y) + \sqrt{p_T^{J\,2} \cosh^2 (\eta_J-Y) + m_H^2}
\,. \nn \end{align}
%%%
Many of our plots will be normalized and for fixed values of $\eta_J$, $p_T^J$, and $Y$,
in which case the phase space factor in \eq{phsp} drops out.

%%%
\begin{widetext}
Our calculation relies on the $N$-jettiness factorization formula in
Ref.~\cite{Jouttenus:2011wh}, which we here specialize to the case of
$1$-jettiness:
\begin{align} \label{eq:sigmaTau1}
\frac{\df^3\sigma^{H+1j}}
  {\df \eta_J\, \df p_T^J\, \df Y\, \df \Tau_a\, \df\Tau_b \, \df\Tau_J}
&= \frac{p_T^J}{4\pi \Ecm^2 (Q^2+m_H^2)}
\sum_\kappa H_\kappa(\{q_i^\mu\}, \mu) 
\int\!\df t_a\, B_{\kappa_a}(t_a, x_a, \mu)
\int\!\df t_b\, B_{\kappa_b}(t_b, x_b, \mu)
\nn \\* &\quad\times
\int\!\df s_J\, J_{\kappa_J}(s_J, \mu)\,
S_\kappa \biggl(\Tau_a - \frac{t_a}{Q_a}, \Tau_b - \frac{t_b}{Q_b},  \Tau_J - \frac{s_J}{Q_J} , \Big\{\frac{q_i^\mu}{Q_i}\Big\}, \mu\biggr)\,
\,. \end{align}
\end{widetext}
%%%
The $N$-jettiness variables $\Tau_a$, $\Tau_b$, and $\Tau_J$ were defined in
\sec{kin}. The hard function $H_{\kappa}$ contains the short-distance
matrix element for producing a Higgs plus a jet, the beam functions
$B_{\kappa_a}$ and $B_{\kappa_b}$ describe the collinear initial-state radiation
and contain the PDFs, the jet function $J_{\kappa_J}$ characterizes the
collinear final-state radiation, and the soft function $S_{\kappa}$ describes
soft radiation effects.\footnote{Note that we do not call \eq{sigmaTau1} a
  factorization theorem since the decoupling of Glauber gluons for hadron
  collider processes with a specific number of jets has not been proven.}  The
sum over $\kappa = \{\kappa_a,\kappa_b,\kappa_J\}$ runs over the possible
flavors $\kappa_i\in \{g,u,\bar u, d,\dots\}$ of the two incoming and one outgoing
parton. The possible combinations, corresponding to the various partonic channels,
are listed in Table~\ref{tab:kappa}.

The power of factorization is that it allows one to evaluate the various
fixed-order pieces at their natural scales, where they contain no large
logarithms.  We then use the RG evolution of each of these functions to evolve
them to a common scale $\mu$, resumming the logarithms of $m_J^2/\pTJs$ and
$Q_i\Tau_i/\pTJs$.  This evolution is implicit in \eq{sigmaTau1}, by writing all
functions as evaluated at the common scale $\mu$. The factorization formula with
all evolution factors written out explicitly is given in \eq{sigmaTau1evo} below. Our
choice of scales is discussed in \subsec{scales}.  Power corrections to
\eq{sigmaTau1} arise from so-called nonsingular corrections, which are
suppressed by a relative $\ord{m_J^2/Q^2}$ in this differential cross section, and are
not considered here.

The cross section in \eq{sigmaTau1} is differential in the $1$-jettiness
contributions from the jet and the beams $\Tau_J$, $\Tau_a$, and $\Tau_b$. As we
will see, the shape of the jet mass spectrum is independent of the jet veto for
a reasonable range of $\Tau_{a,b}$ values. For simplicity we impose a common
cut $\Tau_{a,b} \leq \Tau^{\cut}$. We also convert $\Tau_J$ to the jet mass
$m_J$ using \eq{M2}, and so consider
%%%
\begin{align} \label{eq:tauchange}
\sigma(m_J^\cut,\Tau^\cut) &= 
  \int_0^{\Tau^{\cut}}\!\! \!\!\! \!\!\!\!\! \df  \Tau_a
  \!\int_0^{\Tau^{\cut}}\!\! \!\!\! \!\!\!\!\! \df \Tau_b 
  \!\int_0^{m_J^{\cut\,2}/Q_J}\!\! \!\!\! \!\!\!\!\!\!\!\!\! \df \Tau_J\:\,
%\nn \\ & \quad \times
\frac{\df^3\sigma}{\df \Tau_a\, \df\Tau_b \, \df\Tau_J}
\,.\end{align}
%%%
The differential jet mass cross section, $\df\sigma/\df m_J$, is obtained by taking
the numerical derivative of this cumulant cross section.  We define the
normalized jet mass spectrum over the range $[0,m_J^\cut]$ as $\df\hat\sigma/\df
m_J$, so
%%%
\begin{align} \label{eq:norm}
& \frac{\df \hat{\sigma}}{\df m_J}(m_J^\cut, \Tau^\cut) 
\equiv \frac{1}{ \sigma(m_J^\cut, \Tau^\cut) }
  \frac{\df \sigma(\Tau^\cut)}{\df m_J} 
\,.\end{align}
%%%

The ingredients in the resummed cross section are needed at different orders in
perturbation theory, as summarized in Table~\ref{tab:counting}, where the
columns correspond to the fixed-order matching, non-cusp anomalous dimension
$\ga_x$, cusp anomalous dimension $\Ga_\cusp$, the $\beta$ function, and the
PDFs.  All ingredients necessary for a NNLL resummation of the global logarithms
are known and are collected in \app{inputs}: The one-loop hard function for the
three basic processes $gg\to Hg$, $gq\to Hq$, and $q\bar q\to
  Hg$ via gluon fusion (in the large $m_t$ limit) are obtained from the
one-loop helicity amplitudes calculated in Ref.~\cite{Schmidt:1997wr} following
the procedure in Ref.~\cite{Stewart:2012yh}.  The one-loop quark and gluon jet
function were calculated in Refs.~\cite{Bauer:2003pi, Fleming:2003gt,
  Becher:2009th}, the one-loop quark and gluon beam functions in
Refs.~\cite{Stewart:2009yx, Stewart:2010qs, Mantry:2009qz, Berger:2010xi}, and
the one-loop soft function in Ref.~\cite{Jouttenus:2011wh}. We also require the
cusp anomalous dimension to three loops~\cite{Korchemsky:1987wg, Moch:2004pa},
and the non-cusp anomalous dimensions to two loops, which are known from
Refs.~\cite{Kramer:1986sg, Harlander:2000mg, MertAybat:2006mz, Becher:2006qw,
  Becher:2009th, Stewart:2010qs}.

%%%
\begin{table}[t!]
  \centering
  \begin{tabular}{l | c c c c c c}
  \hline \hline
  & matching & $\gamma_x$ & $\Gamma_\cusp$ & $\beta$ & PDF \\ \hline
  LL & $0$-loop & - & $1$-loop & $1$-loop & NLO \\
  NLL & $0$-loop & $1$-loop & $2$-loop & $2$-loop & NLO \\
  NNLL & $1$-loop & $2$-loop & $3$-loop & $3$-loop & NLO \\
  \hline\hline
  \end{tabular}
  \caption{Perturbative ingredients at different orders in resummed perturbation theory.}
\label{tab:counting}
\end{table}
%%%

\begin{figure}[t!]
  \includegraphics[width=0.95\columnwidth]{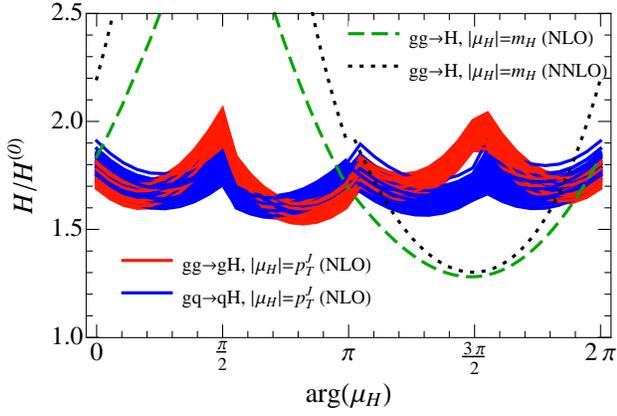}%
\caption{Hard functions for $gg\to H$ at NLO and NNLO, and for $gg\to Hg$ and $gq\to Hq$ at NLO as a 
function of the phase used in their scale $\mu_H$.  For the $gg\to Hg$ and $gq\to Hq$ the results are bands
because we scan over kinematics in the range $200<\pTJ<600\,{\rm GeV}$, $0<\eta_J<1$, and $0<Y<1$.}
\label{fig:Hphase}
\end{figure}

There is some freedom in how to treat products of the fixed-order corrections in
\eq{sigmaTau1}, specifically the higher-order cross terms that are generated,
such as the one-loop correction to $H$ times the one-loop correction to $J$,
which we denote $H^{(1)} J^{(1)}$.  The series for the individual objects are
fairly convergent, except for the hard function whose one-loop correction is
known to be rather large.  For the hard function in $pp \to H+0$ jets the 
use of a complex scale, $\mu_H= -{\img}\, m_H$ improves the perturbative convergence~\cite{Ahrens:2008qu}, since this $H$ 
is related to the time-like scalar form factor.  
For $pp\to H+1$ jet the hard functions contain logarithms with both positive and negative 
arguments, so some logarithms are minimized by an imaginary $\mu_H$ and others by a 
real $\mu_H$.  The convergence for the hard functions for both $pp\to H+0$ jets and 
$pp\to H+1$ jet are shown in \fig{Hphase} as a function of the complex phase chosen for $\mu_H$.  For $pp\to H+0$ jets the improvement in the convergence for $\arg(\mu_H)=3\pi/2$ is clearly visible, while for $pp\to H+1$ jets the 
convergence is only marginally affected by the choice of $\arg(\mu_H)$.   Therefore we alway use $\arg(\mu_H)=0$ for our analysis here.
When combining the perturbative series from different functions in the factorization theorem, we always expand the convolutions of the fixed-order
$B$, $J$, and $S$ functions order by order in $\alpha_s$ to the order needed,
but consider two possibilities for the hard function $H^{(0)}+H^{(1)}$, either expanded along with the other functions or kept as an overall multiplicative factor.
The difference between expanding the hard function or
treating it as multiplicative is within our perturbative uncertainty, being a
$\lesssim 20\%$ effect for the unnormalized $m_J$ spectrum, and only a
$\lesssim 2\%$ effect for the normalized $m_J$ spectrum.  When $H$ is expanded out
there is also $\sim 2\%$ increase in the perturbative uncertainties for the normalized 
$m_J$ spectrum for $gg\to Hg$, so we pick this convention as our default in order 
to be conservative. Schematically, this means that the fixed-order
components of our cross section take the form
%%%
\begin{align} \label{eq:Hmult}
&  H^{(0)}  \Bigl[(B^{(0)}B^{(0)}J^{(0)})\otimes S^{(0)}
\\ &\quad
  + (B^{(1)} B^{(0)}J^{(0)}) \otimes S^{(0)}
  +  (B^{(0)} B^{(1)}J^{(0)})  \otimes S^{(0)}
  \nn \\ &\quad
  +(B^{(0)} B^{(0)} J^{(1)}) \otimes S^{(0)}
  + (B^{(0)} B^{(0)} J^{(0)}) \otimes S^{(1)} \Bigr]
  \nn\\
 &\quad + H^{(1)} (B^{(0)}B^{(0)}J^{(0)})\otimes S^{(0)}
\nn\,.\end{align}
%%%

%===============================================================================
\subsection{Refactorization of the Soft Function}
\label{subsec:sfact}
%===============================================================================

For a process with one or more jets there are multiple directions for collinear
radiation and various kinematic variables so a few additional hierarchies become
possible. The factorization formula assumes that there are no additional strong
hierarchies beyond the collinearity of the jet $m_J^2\ll \pTJs$, and the
absence of additional central jets away from the beam directions, $Q_a
\Tau_a \ll \pTJs$ and $Q_b \Tau_b \ll \pTJs$.  Physically, the absence of no additional 
strong hierarchies corresponds to the following four assumptions
\begin{enumerate}
 \item [1)]\  $Q_i \Tau_i\sim Q_j \Tau_j$\qquad \text{commensurate $m_J$ and jet veto}
 \item [2)]\ $\dfrac{q_i\cdot q_j}{E_i E_j} \sim \dfrac{q_i\cdot q_k}{E_i E_k}$
   \quad  \text{well separated jet and beams}
 \item [3)]\ $E_i \sim E_j$  \qquad \text{jet and beam-jets of similar energy}
 \item [4)]\ $Q_i \sim Q_j$  \qquad \text{jet and beam regions of similar size}
\end{enumerate}
Assumption 1) ensures that we are in the region where NGLs are not large logarithms.
Assumption 2) implies that the jet is not too close to the beam direction, and
avoids having large angular logarithms, which would require an additional
``ninja summation''~\cite{Bauer:2011uc}.

Three combinations of these four assumptions are necessary to avoid introducing
additional large logarithms that are not summed by the renormalization group
evolution of terms in the factorization formula, namely
\begin{align} \label{eq:logass}
 &  \frac{s_{ij}}{s_{ik}} \sim 1 \,,
 &  \frac{\Tau_i}{\Tau_j}&  \sim 1 \,, 
  &  \frac{Q_i}{Q_j} & \sim 1 \,.
\end{align}
The first implies that the logarithms in the hard function can be minimized with
a common scale $\mu$, and all three combine to imply that a common scale also
minimizes all logarithms in the soft function.  One combination of assumptions,
$E_i/Q_i\sim E_j/Q_j$, does not appear explicitly in arguments of functions in
the factorized cross section, and hence does not show up in logarithms for the
leading power result. However, it is in general necessary as part of the
derivation of \eq{sigmaTau1} to ensure that certain neglected terms are power
suppressed.

An important consideration in carrying out the summation of large logarithms is
the order in $\alpha_s$ and logarithms at which violations of \eq{logass} first
become apparent. For the soft function the first terms that appear for the
various logarithms are
\begin{align} \label{eq:softlogs}
 & \alpha_s \ln^2\Bigl(\frac{Q_i\Tau_i^c Q_j\Tau_j^c}{\mu^2 s_{ij}}\Bigr) \,,
 & \alpha_s \ln\Bigl( \frac{s_{ij}Q_k}{s_{ik}Q_j} \Bigr) & \,,
 & \alpha_s^2 \ln^2\Bigl(\frac{\Tau_i^c}{\Tau_j^c}\Bigr) \,,
\end{align}
where we integrate the soft function over $\Tau_i$ up to the cumulant variable $\Tau_i^c$.
The first of these is
part of the LL series.  The second is an angular logarithm. It is part of the
NLL series if it counts as a large logarithm.  Otherwise, it is part of the
$\sim\alpha_s$ fixed-order terms that start to contribute at NNLL.  The third is
a NGL. It is part of the NLL series if it is a large logarithm. Otherwise it is
part of the $\sim\alpha_s^2$ fixed-order terms that start to contribute at
N${}^3$LL.  Therefore, there is a nontrivial constraint on the choice of scales
$\mu$ in the soft function. The scales must be chosen to minimize the first type
of logarithm in \eq{softlogs} {\em without} inducing terms of the form of the
second and third types already at LL order.  In particular, this implies that a
poor scale choice could introduce unphysical angular logarithms or NGLs into the
LL series. For our choice of kinematics and $Q_i$ the second type of angular
logarithm in \eq{softlogs} is never large. However, since we are exploring a
spectrum in $m_J^2 = Q_J \Tau_J$ the third term in \eq{softlogs} will grow as
the parameters are varied.  To surmount this problem requires a refactorization
of the soft function which we will consider below.

For the hard function the series of leading double logarithms
involves terms of the form
\begin{align}
 & \alpha_s \ln^2\Bigl( \frac{\mu^2}{s_{ij}} \Bigr) \,,
 & \alpha_s \ln^2\Bigl( \frac{s_{ij}}{s_{ik}} \Bigr) \,.
\end{align}
For the choice of jet kinematics explored in this paper we will always satisfy
the assumption $s_{ij}\sim s_{ik}$, so there is no additional constraint on the
scale associated with the hard function.

The hierarchy between $\Tau_J$ and $\Tau^\cut$ leads to unphysical large
logarithms if a single scale $\mu_S$ is used for the initial conditions for the
soft function evolution. Here we address how these can be removed by a
refactorization of the soft function, with corrections from the true higher
order non-global logarithms
(see Refs.~\cite{Fleming:2007xt, Ellis:2010rwa, Kelley:2011aa, Chien:2012ur} for
earlier refactorization discussions).

In general, the all-order soft function has the form
%%%
\begin{align} \label{eq:nlofact}
&  S(\{k_i\},\{\hat q_i^\mu\},\mu)
\nn\\* & \quad
= \prod_i S_i(k_i,\{\hat q_i^\mu\},\mu)
   + S_{\rm NGL}(\{k_i\},\{\hat q_i^\mu\},\mu)
\,,\end{align}
%%%
where $\hat q_i^\mu = q_i^\mu/Q_i$. Here $S_{\rm NGL}$ contains all non-global
terms, and hence has an intrinsic dependence on the ratios $k_i/k_j$.  At NLO
there is only one soft gluon emitted, which can contribute to only one of the
$\Tau_i$ at a time. Thus the NLO soft function factorizes, and $S_{\rm NGL}\sim
\ord{\al_s^2}$.  Truncating to $\ord{\al_s}$
there is still some freedom in the definition of the $S_i$.  Whereas the
terms with explicit $k_i$ dependence in $S(\{k_i\},\mu)$ clearly belong to $S_i(k_i,\mu)$, the
pure delta function terms $\delta(k_J)\delta(k_a)\delta(k_b)$ can in principle
be split in multiple ways between the various $S_i(k_i,\mu)$. We choose to
split these terms evenly, as detailed in \app{soft}, and we introduce an
additional parameter $r$ in the scale variation to estimate uncertainty from
this freedom as discussed further below and in detail in \subsec{scales}.

Due to the consistency of the factorization formula, the evolution of the soft
function factorizes exactly to all orders in perturbation theory,
%%%
\begin{align} \label{eq:evofact}
U_S(\{k_i\},\mu,\mu_0)  &= 
U_H(\mu_0,\mu) \prod_i Q_i U_{J_i}(Q_i k_i,\mu_0,\mu)
\nn \\
&=\prod_i U_{S_i}(k_i,\mu,\mu_0)
\,.\end{align}
%%%
Note that this result does not rely on the refactorization of the soft function discussed above.
(Here we used the fact that the beam and jet functions have the same
evolution~\cite{Stewart:2010qs}.)  Equation~(\ref{eq:evofact}) involves the factorization of the
evolution of the hard function $H=CC^\dagger$, which follows from the form of
the anomalous dimension for $C$~\cite{Chiu:2008vv, Becher:2009qa},
%%%
\begin{align}
 \widehat\ga_C(\mu) &= -\Ga_\cusp[\al_s(\mu)]
 \bigg[ \sum_i \bT_i^2 \ln \frac{\mu}{\mu_0} 
 \nn \\ & \quad
 +  \sum_{i < j} \bT_i \!\cdot\! \bT_j\, 
  \ln\Bigl(-\frac{s_{ij}}{\mu_0^2}\!-\!\img 0\Bigr) \bigg]
 + \widehat\ga_C[\al_s(\mu)]
\,.\end{align}
%%%
The sum on $i$ and $j$ runs over the colored partons participating in the
short-distance interaction and $\bT_i$ denotes the corresponding color charge
matrix. (For $pp\to H+1$ jet the color space is still trivial, so these color
matrices are just numbers.) To associate
the $\ln \mu$ terms to individual partons we introduced a dummy variable $\mu_0$
and used color conservation. It is not a priori clear how to associate the
remaining terms within the $\sum_{i<j}$ to each $U_{S_i}$, and we choose to split
each term evenly between $i$ and $j$. The explicit expression for the factorized
hard function evolution that we employ is given in \app{evo}.  Other potential
choices of splitting up these terms are again probed by the scale parameter $r$,
which is discussed in more detail around \eq{softmix}, and the
corresponding uncertainty is found to be small except on the large $m_J$ tail of the 
distribution.  The two-loop non-cusp anomalous
dimension has the structure $\widehat\ga_C(\alpha_s) = n_q \ga_q + n_g \ga_g$,
where $n_g$ and $n_q$ are the number of gluon and (anti)quark legs, so it
naturally factors.

\begin{figure}[t!]
\includegraphics[width=0.5\columnwidth]{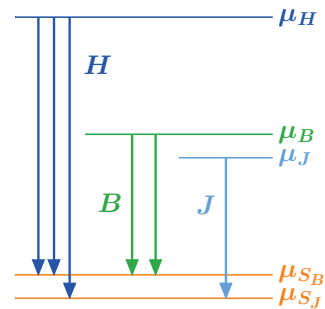}
\caption{Illustration of the different fixed-order scales appearing in the factorized cross section and our evolution strategy.  The figure has $Y=0$ where there is a common $\mu_B$ scale.}
\label{fig:running}
\end{figure}
The factorization of the evolution and fixed-order soft function in
\eqs{nlofact}{evofact} suggests that we can evaluate the piece of the soft
function corresponding to $\Tau_i$ at a scale $\mu_{S_i}$,
%%%
\begin{align}
S(\{k_i\},\mu) = \prod_i\! \int \! \df k_i'\, U_{S_i}(k_i - k_i', \mu, \mu_{S_i}) S_i(\{k_i'\},\mu_{S_i})
\,.\end{align}
%%%
This factorization does not hold for all the terms at order $\al_s^2$, since
there are diagrams that contribute to multiple $\Tau_i$, leading to non-global
logarithms of the form $\al_s^2 \ln^2(k_i^c/k_j^c)$ appearing in $S_{\rm NGL}$
in \eq{nlofact}. We discuss in \subsec{ngl} how we estimate the size of these NGL
contributions in the jet mass spectrum.

In our implementation we find it simplest to run the hard, jet, and beam
functions, rather than the soft function, as summarized in \fig{running}.
The final results are completely independent of this choice.  Since the cut on both
beams is the same, they have a common $\mu_{S_B}$, and a common $\mu_B$ for $Y=0$.  
We summarize the work in this section by presenting the factorization formula valid at NNLL which
includes the evolution factors and refactorization of $S$,
%%%
\begin{widetext}
\begin{align} \label{eq:sigmaTau1evo}
\frac{\df^3\sigma^{H+1j}(\Tau^\cut)}{\df \eta_J\, \df p_T^J\, \df Y\, \df m_J}
&= 
\frac{(2 p_T^J m_J/Q_J) }{4\pi \Ecm^2 (Q^2+m_H^2)}
\sum_\kappa H_\kappa(\{q_i^\mu\}, \mu_H) U_{H_{\kappa_a}}(\{q_i^\mu\},\mu_{S_B},\mu_H)
 U_{H_{\kappa_b}}(\{q_i^\mu\},\mu_{S_B},\mu_H)  U_{H_{\kappa_J}}(\{q_i^\mu\},\mu_{S_J},\mu_H) 
\nn \\ &\quad\times 
\int\!\df t_a\, \df t_a'\, U_{J_{\kappa_a}}(t_a - t_a',\mu_{S_B},\mu_B) B_{\kappa_a}(t_a', x_a, \mu_B)
\int\!\df t_b\, \df t_b'\, U_{J_{\kappa_b}}(t_b - t_b',\mu_{S_B},\mu_B) B_{\kappa_b}(t_b', x_b, \mu_B)
\nn \\ &\quad\times 
\int\!\df s_J\, \df s_J'\, U_{J_{\kappa_J}}(s_J - s_J',\mu_{S_J},\mu_J)J_{\kappa_J}(s_J', \mu_J)\,
\int_0^{\Tau^\cut}\!\! \df \Tau_a\, S_a \biggl(\Tau_a - \frac{t_a}{Q_a}, \Big\{\frac{q_i^\mu}{Q_i}\Big\}, \mu_{S_B} \biggr)
\nn \\ &\quad\times 
\int_0^{\Tau^\cut}\!\! \df \Tau_b\, S_b \biggl(\Tau_b - \frac{t_b}{Q_b}, \Big\{\frac{q_i^\mu}{Q_i}\Big\}, \mu_{S_B} \biggr)
S_J \biggl(\frac{m_J^2 - s_J}{Q_J}, \Big\{\frac{q_i^\mu}{Q_i}\Big\}, \mu_{S_J}\biggr)\,
\,. \end{align}
\end{widetext}
%%%
All necessary perturbative results for $H_\kappa$, $J_{\kappa_J}$, $S_i$, and the $U_i$ are collected in \app{inputs}.

%===============================================================================
\subsection{Choice of Running Scales}
\label{subsec:scales}
%===============================================================================

The factorization formula in \eq{sigmaTau1evo} sums the large logarithms of $Q_i
\Tau^\cut/\pTJs$ from the cuts on the beams and of $Q_J\Tau_J/\pTJs=m_J^2/\pTJs$
from the jet mass measurement. This is accomplished by carrying out perturbation theory for the hard,
beam, jet, and soft functions at their natural scales and then running them to an 
arbitrary common scale. Examining the fixed-order expressions from \app{inputs} we find
that the canonical scaling relations are
\begin{align}  \label{eq:canonicalscales}
 & \mu_H \simeq \pTJ\,, \qquad
  \mu_J  \simeq m_J\,, \qquad
  \mu_{S_J}  \simeq \frac{m_J^2}{\sqrt{\pTJ Q_J}} 
  \,, \nn\\
&  \mu_{B_{a,b}}  \simeq \sqrt{Q_{a,b} \Tau^\cut}  \,,\qquad 
 \mu_{S_{B}}  \simeq \Tau^\cut\,.
\end{align}
The situation for the beam and jet scales are fully analogous with  $\mu^2 \simeq Q_i \Tau_i$ for $i=a,b,J$. 
To ensure we have the correct leading logarithms we cannot use a common scale for $\{\mu_{B_i},\mu_J\}$ or for $\{\mu_{S_B},\mu_{S_J}\}$ (as discussed above in section~\ref{subsec:sfact}), and we see from \eq{canonicalscales} that they have different dependence on kinematic variables.  In deriving these scaling relations for the soft scales we have assumed certain $\eta_J$ dependence gives ${\cal O}(1)$ factors. This implies that we are not attempting to sum the additional rapidity logarithms that appear when the jet is in a forward region. In particular, for the global logarithms in the soft function that involve $m_J$ the full dependence that appears is
\begin{align}
  \ln\bigg( \frac{m_J^2}{\mu_{S_J} Q_J \sqrt{\hat s_{Ji}}}\bigg)  
     &= \ln \bigg( \frac{m_J^2 \: e^{\pm \eta_J/2} }{\mu_{S_J} \sqrt{\pTJ Q_J }}\bigg)  \,,
\end{align}
and to obtain the scaling in \eq{canonicalscales} we neglect the $\exp(\pm \eta_J/2)= {\cal O}(1)$ angular factor. Here $\hat s_{ij}=q_i\cdot q_j/(Q_i Q_j)$. Through $Q_J$ the $\mu_{S_J}$ soft scale still depends on the jet algorithm, jet size $R$, and mildly on $\eta_J$.  For the global logarithms in the soft function that involve $\Tau^\cut$, there are two forms that appear
\begin{align}
  & \ln\Big( \frac{\Tau^\cut}{\mu_{S_B}}\Big)  \,,
  &  \ln\Big( \frac{\Tau^\cut}{\mu_{S_B} \sqrt{\hat s_{iJ}} }\Big) \,.
\end{align}
Here to get the scale choice in \eq{canonicalscales} we neglect the $R$-dependent $\sqrt{\hat s_{iJ}}\sim 1$ factor.  This choice has very little impact on our main results for normalized cross sections (including a factor of $\sqrt{\hat s_{iJ}}$ into the canonical $\mu_{S_B}$ gives equivalent numerical results within our uncertainties).

The dependence of the cross section on the jet algorithm and jet radius through $Q_J$ and $Q_{a,b}$ occurs due
to their impact on the boundaries between the jet and beam regions. For $1$-jettiness these are all induced by the soft function. For example, for the geometric-$R$ algorithm we find that $\mu_{S_J}\propto m_J^2/(R\, \pTJ)$, so in this case the ratio of scales $\mu_{S_J}/\mu_J$ sums logarithms $\ln[m_J/(R \pTJ)]$, while the ratios of scales $\mu_J/\mu_H$ sums logarithms $\ln(m_J/\pTJ)$.   Beyond the dependence in the logarithmic resummation there is also jet algorithm dependence that is  encoded in the fixed-order terms in the soft function through dependence on $\hat s_{aJ}$ and $\hat s_{bJ}$.  The fixed-order terms in the factorized cross section reproduce the correct $\eta_J$ dependence for the singular ${\cal O}(\alpha_s)$ corrections.

If any $\Tau_i$ becomes very small, $\sim \lqcd$, the nonperturbative corrections
to the soft function become important.  Since anomalous dimensions are only valid in perturbative regions, the scales in SCET must be frozen before they enter the nonperturbative regime $\mu_{S_i} \gtrsim \lqcd$, $\mu_J^2\gtrsim \sqrt{\pTJ Q_J}\lqcd$, and $\mu_{B_i}^2\gtrsim Q_i\lqcd$. This is often referred to as the peak region since it occurs near the cross section peak for quark jets (for gluon jets it occurs to the left of the peak) . We will refer to it as the nonperturbative region here.

At the other end of spectra, for large $m_J^2\sim \pTJ Q_J$ and/or large $\Tau^\cut\sim \pTJ$, a part of the resummation is not important and must be turned off by having the SCET scales merge into a single fixed-order scale, $\mu_J = \mu_{S_J} =\mu_H$ and/or $\mu_{B_i}=\mu_{S_B}=\mu_H$. We will refer to this as the fixed-order scaling region.   To determine the location of this region for the scales depending on $m_J$ we note that the size of the jet puts an effective upper boundary on its mass $m_J\lesssim p_T^J R/\sqrt{2}$.  For a jet with two particles of separation $R$ the bound is $m_J/\pTJ \le \tan(R/2) = R/2 + {\cal O}(R^3) $~\cite{Dasgupta:2012hg}.  Assuming a uniform energy distribution of particles within a circle of radius $R$ in $(\eta,\phi)$-space gives $m_J/\pTJ \le  R/\sqrt{2} + {\cal O}(R^3)$.  If we add a single massless particle at the center of this uniform distribution that carries a fraction $f$ of the total energy, then this gives $m_J/\pTJ \le  (1-f) R/\sqrt{2} + {\cal O}(R^3)$.   We will use  $m_J \lesssim \pTJ R/\sqrt{2}$ here, noting that even for $R=1.2$ the ${\cal O}(R^3)$ term gives only a $15\%$ correction. Near this boundary the jet mass spectrum has to fall off rapidly.

In between the nonperturbative region and fixed-order region is a perturbative region where resummation is important and power corrections are suppressed by $\sim \lqcd/\mu_S$, which we will refer to as the resummation or tail region.  Most of the differential jet mass cross section is in this region, in particular for gluon jets where the cross section peak is in the resummation region.  Transitions occur between this resummation region and the nonperturbative region, as well as between this resummation region and the fixed-order region, which must be handled smoothly.

To connect the peak, resummation, and fixed-order regions where the resummation must be handled
differently, we use $\Tau_i$-dependent scales, which are known as profile
functions~\cite{Ligeti:2008ac,Abbate:2010xh}. A transition between these three
regions is given by the following running scales for hard, jet, beam, and soft functions
%%%
\begin{align} \label{eq:runscales}
  \mu_H &= \mu
  \,, \\
  \mu_J(\tau_J) & =
  \Big[1 + e_J\, V(\tau_J,t_3) \Big]
  \sqrt{\mu\, \mu_\mathrm{run}(\delta_J\tau_J,\mu,1,\delta_J t_j)}
  \,, \nn \\
  \mu_{S_J}(\tau_J) & = \Big[1 + e_{S_J}\, V(\tau_J,t_3) \Big] 
        \mu_\mathrm{run}\Big(\tau_J,\mu,\delta_J^{1/2},t_j\Big)
  \,, \nn \\
  \mu_{B_i}(\tau_B) & =
  \Big[1 + e_B\, V(\tau_B,t_3') \Big]
  \sqrt{\mu\, \mu_\mathrm{run}(\delta_i \tau_B,\mu,1,\delta_i t_j')}
  \,, \nn \\
  \mu_{S_B}(\tau_B) & = \Big[1 + e_{S_B}\, V(\tau_B,t_3') \Big] \mu_\mathrm{run}(\tau_B,\mu,1,t_j')
 \,,\nn\end{align}
%%%
where the variables 
\begin{align}
 & \tau_J=m_J^2/(\pTJ Q_J) \,,\qquad
 \tau_B = \Tau^\cut/\pTJ \,,
\end{align} 
the fractions $\delta_J =Q_J/\pTJ$, $\delta_{a,b}=Q_{a,b}/(2\pTJ)$ and the function
\begin{align}
  V(\tau,t_3) &= \theta(t_3-\tau) \Big(1 - \frac{\tau}{t_3}\Big)^2  \,.
\end{align}
The function $\mu_\mathrm{run}(\tau,\mu,r_t,t_i)$ behaves as a constant in the nonperturbative and fixed-order regions, and as $\simeq \mu \tau r_t$ in the resummation region.  Since $\mu$ and $\tau$ are determined, it is choice for the dimensionless parameter $r_t$ that gives the slope for this region.  For this resummation region the choice of arguments in \eq{runscales} yields the desired canonical scalings given in \eq{canonicalscales}.  In the fixed-order region with large $\Tau^\cut$ we get $\mu_{S_B} = \mu_{B_i}= \mu_H$ and in the region with large $m_J$ we get $\mu_{S_J}=\mu_J=\mu_H$.  The expression for $\mu_\mathrm{run}$ can be found in \app{scales}, along with the central values used for the parameters $\mu$, $e_i$, $e_{S_i}$, $t_j$, $t_j'$, and details on the variations of these parameters that are used to estimate the perturbative uncertainties in our predictions.

To estimate the additional perturbative uncertainty associated with the
refactorization of the soft function in \subsec{sfact}, we reintroduce
correlations between the soft scales using a parameter $r$ satisfying $0 \leq r
\leq 1$,
%%%
\begin{align} \label{eq:softmix}
\mu_{S_J}^{(r)}&= (\bar \mu_S)^r\, (\mu_{S_J})^{1-r} 
\,, \quad
\mu_{S_B}^{(r)}= (\bar \mu_S)^r\, (\mu_{S_B})^{1-r}
\,, \nn \\
\ln \bar \mu_S &\equiv
 \frac{(T_a^2 + T_b^2) \ln \mu_{S_B} 
 + T_J^2 \ln \mu_{S_J} }{T_a^2 + T_b^2 + T_J^2} 
\,.\end{align}
%%%
Here $T_i^2=C_F$ for $i=q$ and $i=\bar q$, and $T_i^2=C_A$ for $i=g$.  For $r=0$
we have the original uncorrelated soft scales. By increasing $r$ the scales move
towards the ``color average" value $\bar \mu_S$. At $r=1$ they are all equal to this
average soft scale, so the refactorization is turned off (which as explained earlier
causes unphysical NGLs in the LL series).  To estimate the size of the freedom in the
refactorization we take $r=0.2$ as our default choice and include $r=0$ and
$r=0.4$ as separate scale variations in our uncertainty estimate.

The profiles for the SCET scales in \eq{runscales} are in distribution space for $m_J$ and cumulant space for $\Tau^\cut$, yielding the resummed $\df\sigma(\Tau^\cut)/\df m_J$.  To compute the $m_J$ distribution we use a derivative of the jet mass cumulant, utilizing the midpoint scale setting procedure discussed in Ref.~\cite{Abbate:2010xh}.  To compute the normalization $\sigma(m_J^\cut,\Tau^\cut)$ in \eq{norm} we then directly integrate our $m_J$ distribution result. This ensures that the normalized cross section $\df\hat\sigma(m_J^\cut, \Tau^\cut)/\df m_J$ integrates to $1$ over the desired range.

%===============================================================================
\subsection{Non-Global Logarithms}
\label{subsec:ngl}
%===============================================================================

If the NGLs are not large logarithms then they enter beyond NNLL order, and
should be of comparable size to other higher-order perturbative terms. This is
obviously only possible for some range of $m_J^2/(\pTJ\Tau^{\rm cut})$, 
which determines where our result is valid at NNLL order.
To determine this range we include the leading ${\cal
  O}(\alpha_s^2)$ NGL into our resummed calculation and compare the results with
and without this term for various parameter choices. In the factorized exclusive 1-jet cross
section all NGLs enter through the soft function $S$.  For simplicity we
restrict this analysis of the size of non-global logarithms to the $gg \to Hg$
channel, as the results for other channels are similar.

The leading NGL in the cumulant soft function is
\begin{align}\label{eq:S_NGL0}
  S_\text{NGL}(\{k_i^c\},\mu_S) 
  & = \prod_i \Big(\int_0^{k_i^c}\! \df k_i\Big) S_\text{NGL}(\{k_i\},\mu_S) 
  \\ &
  = -\frac{\al_s^2(\mu_S)C_A^2}{(2\pi)^2} 
  \sum_{i<j} G_{ij} \ln^2 \Big(\frac{k_i^c}{k_j^c}\Big)
\,.\nn
\end{align}
Here $G_{ij}$ is a geometric factor that depends on the boundaries of the jet
and beam regions.  Note the absence of explicit $\mu$-dependence in the NGLs.
These expressions for $S_\text{NGL}$ follow from the known result for
$e^+e^-\!\to 2$ jets~\cite{Dasgupta:2001sh,Dasgupta:2002dc,
  Kelley:2011ng,Hornig:2011iu}, by replacing the color factor $C_F C_A \to
C_A^2$. Unlike the global logarithms this contribution does not factor, so we
assign it a common soft scale which for our numerical analysis we take to be $\bar \mu_S$ given in
\eq{softmix}.

For the purpose of our numerical analysis we take $G_{ij}=\pi^2/3$, which is the
result for a hemisphere. This may be thought of as reasonable estimate and in reality
the values may differ by about 15\% to 30\%~\cite{Dasgupta:2012hg}. Converting the cumulant space result in
\eq{S_NGL0} into a full distribution yields
%%%
\begin{align} \label{eq:S_NGL}
  S_\text{NGL}(\{k_i\},\bar\mu_S) 
  &\simeq -\frac{\al_s^2(\bar\mu_S)C_A^2}{(2\pi)^2} \frac{\pi^2}{3}
  \Big[\sum_i \frac{4}{\mu'} \cL_1\Big(\frac{k_i}{\mu'}\Big)
  \nn \\ & \quad 
  -2\sum_{i<j} \frac{1}{\mu'} \cL_0\Big(\frac{k_i}{\mu'}\Big) \,
  \frac{1}{\mu'} \cL_0\Big(\frac{k_j}{\mu'}\Big) \Big]
\,,\end{align}
%%%
where the $\cL_n$ denote standard plus distributions as defined in \eq{Ln}.
Note that the $\mu'$ dependence cancels out explicitly between the terms, so the
choice of this scale is arbitrary and irrelevant. It is introduced for coding
purposes, since it is convenient to have the same type of $\cL_n$ distributions
as in the non-NGL part of the soft function.  When the NGLs are included in this
manner, via the soft function in the factorization, one automatically resums an
infinite series of global logarithms that multiply the NGL. In particular, this includes terms
that are schematically $[\alpha_s^2 \ln^2][\sum_k (\alpha_s\ln^2)^k]$ where the
first $\ln^2$ is non-global and the second $\ln^2$ is a large global logarithm.
The all-order structure of this series of terms is correctly predicted by the
factorization formula.

For our analysis we will mostly be interested in the normalized spectrum in
\eq{norm}. Here in the numerator the two jet veto variables are in cumulant space
and $m_J$ is in distribution space, while in the denominator all the variables
are in cumulant space. This result has two types of NGLs
\begin{align}
 \mathrm{i}) &\qquad  \alpha_s^2(\mu_S) \ln^2\biggl(\frac{m_J^{\rm  cut\,2}}{\pTJ\Tau^\cut}\biggr)
 \,, \\[1ex]
 \mathrm{ii}) & \qquad  \alpha_s^2(\mu_S) \frac{2}{\Tau^\cut}
 \cL_1\biggl(\frac{m_J^2}{\pTJ\Tau^\cut}\biggr)
  \,. \nn
\end{align}
For the denominator the relevant form of the NGL logarithms is as in \eq{S_NGL0},
yielding the terms i).  For the numerator the form of the NGL is as in ii). The
presence of two types of NGLs in the normalized spectrum implies a somewhat
different dependence than for the unnormalized cross section. The effect of
NGLs in these two cases are analyzed in detail in
\subsec{veto_ngls}.  There we will show that there is indeed a fairly large
range of $m_J$ and $\Tau^\cut$ values where the NGL terms in the exclusive jet cross
section are not large logarithms.

%%%%%%%%%%%%%%%%%%%%%%%%%%%%%%%%%%%%%%%%%%%%%%%%%%%%%%%%%%%%%%%%%%%%%%%%%%%%%%%%
\section{Results for Gluon and Quark Jets}
\label{sec:results_part}
%%%%%%%%%%%%%%%%%%%%%%%%%%%%%%%%%%%%%%%%%%%%%%%%%%%%%%%%%%%%%%%%%%%%%%%%%%%%%%%%

\begin{figure*}[t!]
\subfigure[Unnormalized jet mass spectrum for quark and gluon jets at NNLL. The uncertainties are sizable even at NNLL.]
{\label{fig:unnormuncert}\includegraphics[width=0.95\columnwidth]{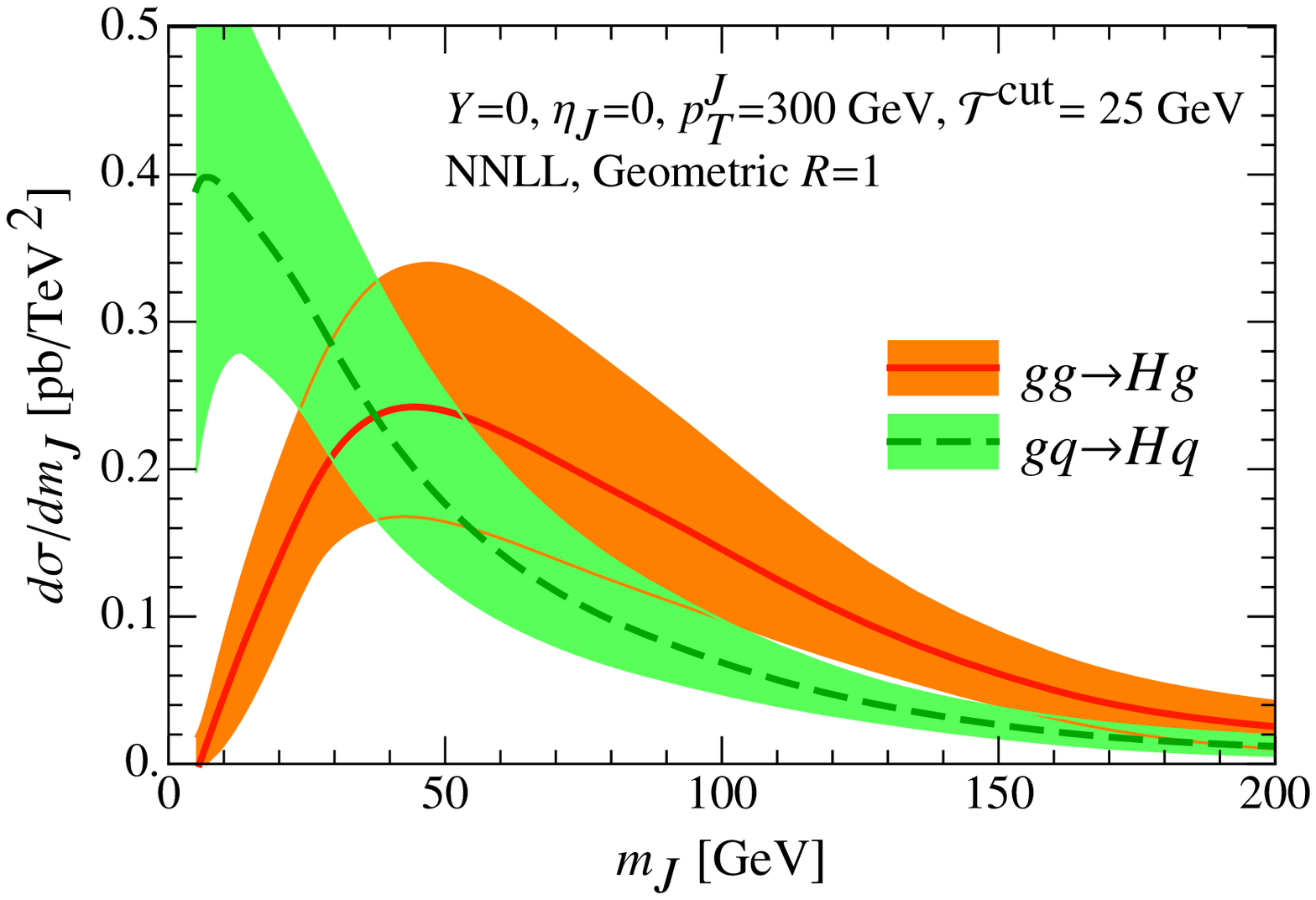}}%
\hfill%
\subfigure[Normalized jet mass spectrum for quark and gluon jets at NNLL. Compared to \fig{unnormuncert}, the normalization significantly reduces the perturbative uncertainties.]
{\label{fig:normuncert}\includegraphics[width=\columnwidth]{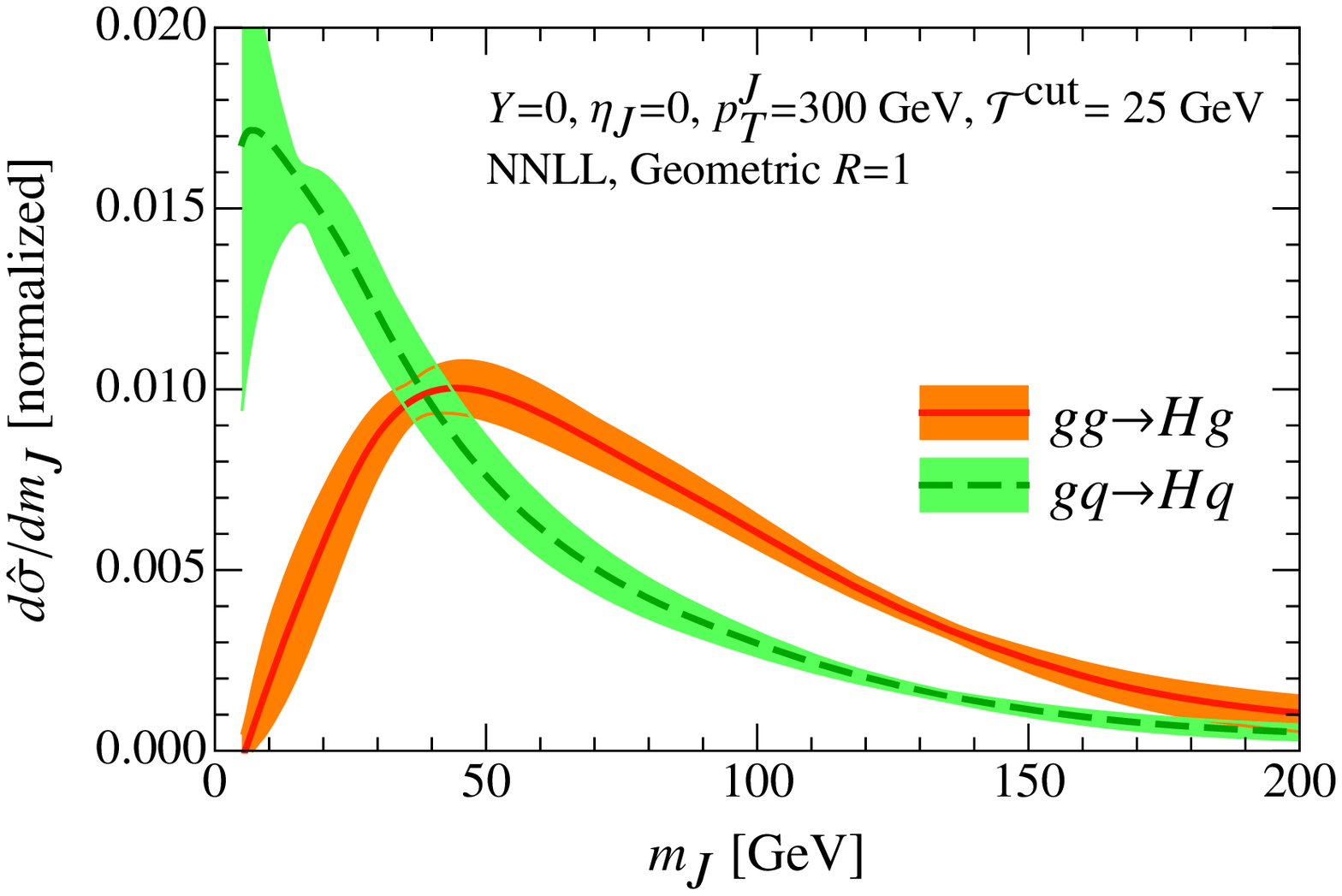}}%
\\
\subfigure[Convergence of the resummed calculation for gluon jets.]
{\label{fig:convergenceg}\includegraphics[width=\columnwidth]{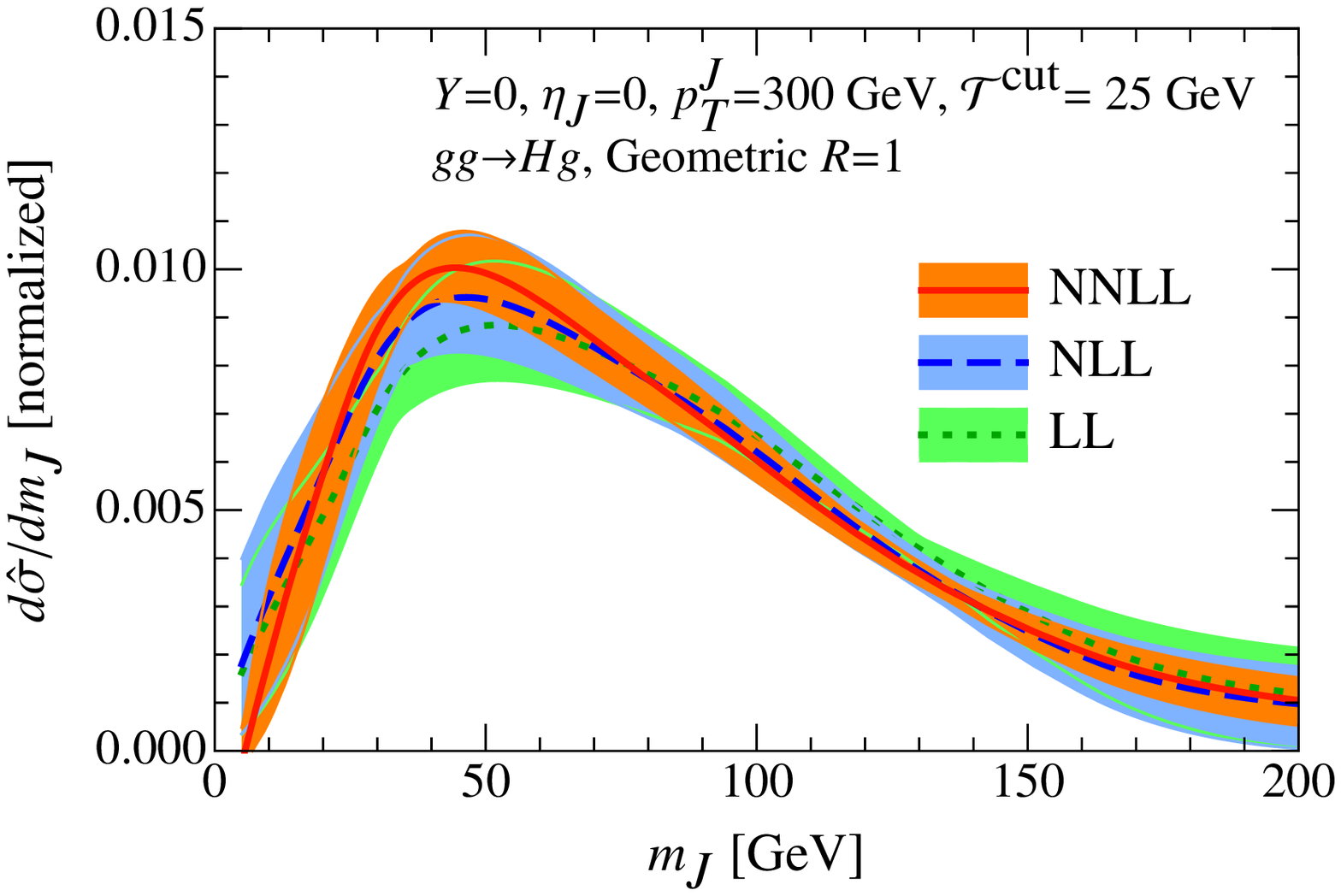}}%
\hfill%
\subfigure[Convergence of the resummed calculation for quark jets.]
{\label{fig:convergenceq}\includegraphics[width=\columnwidth]{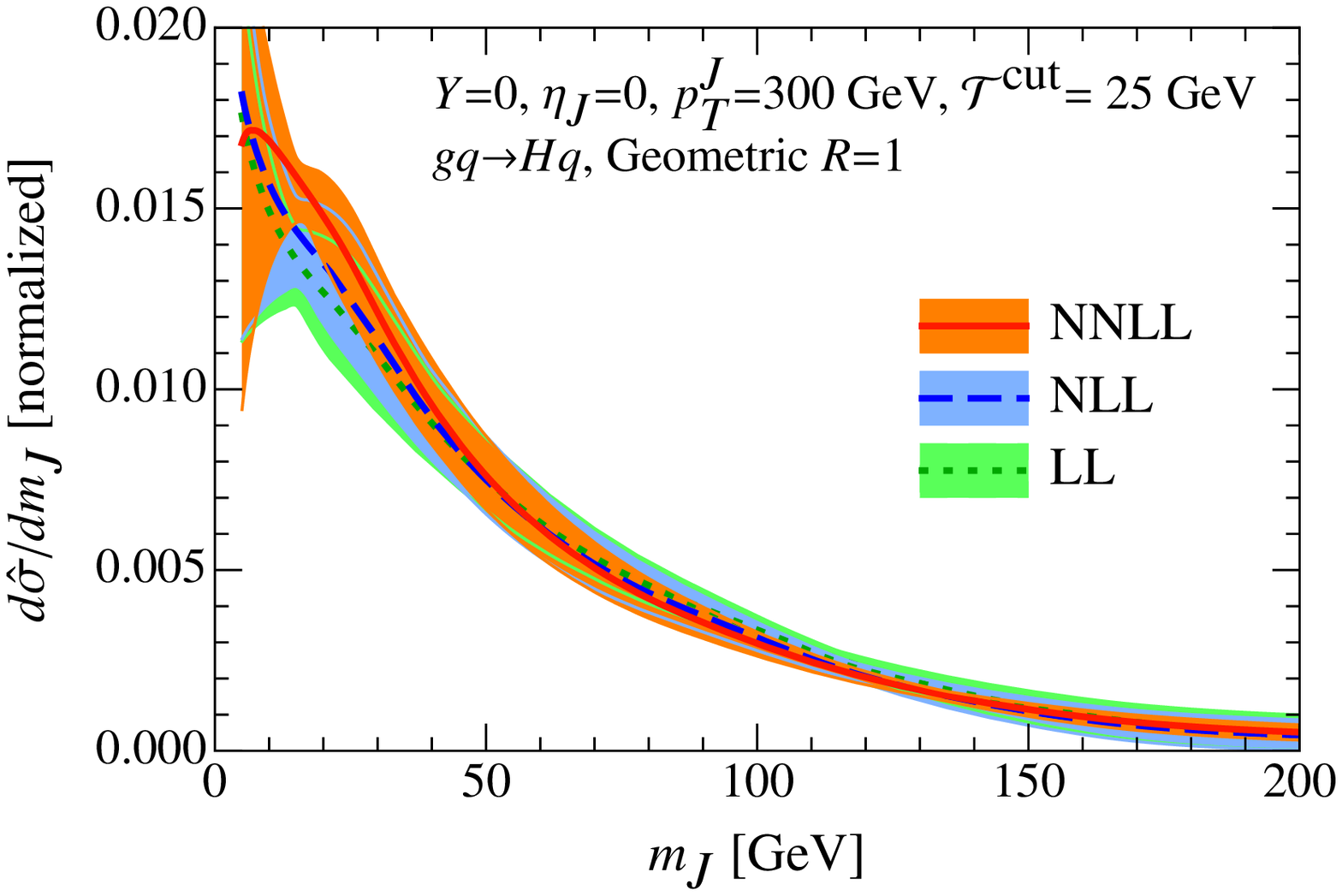}}%
\\
\subfigure[Individual scale variations that enter the uncertainty estimate for gluon jets at NNLL.
Shown are the variations relative to the central NNLL curve.]
{\label{fig:indivuncertg}\includegraphics[width=\columnwidth]{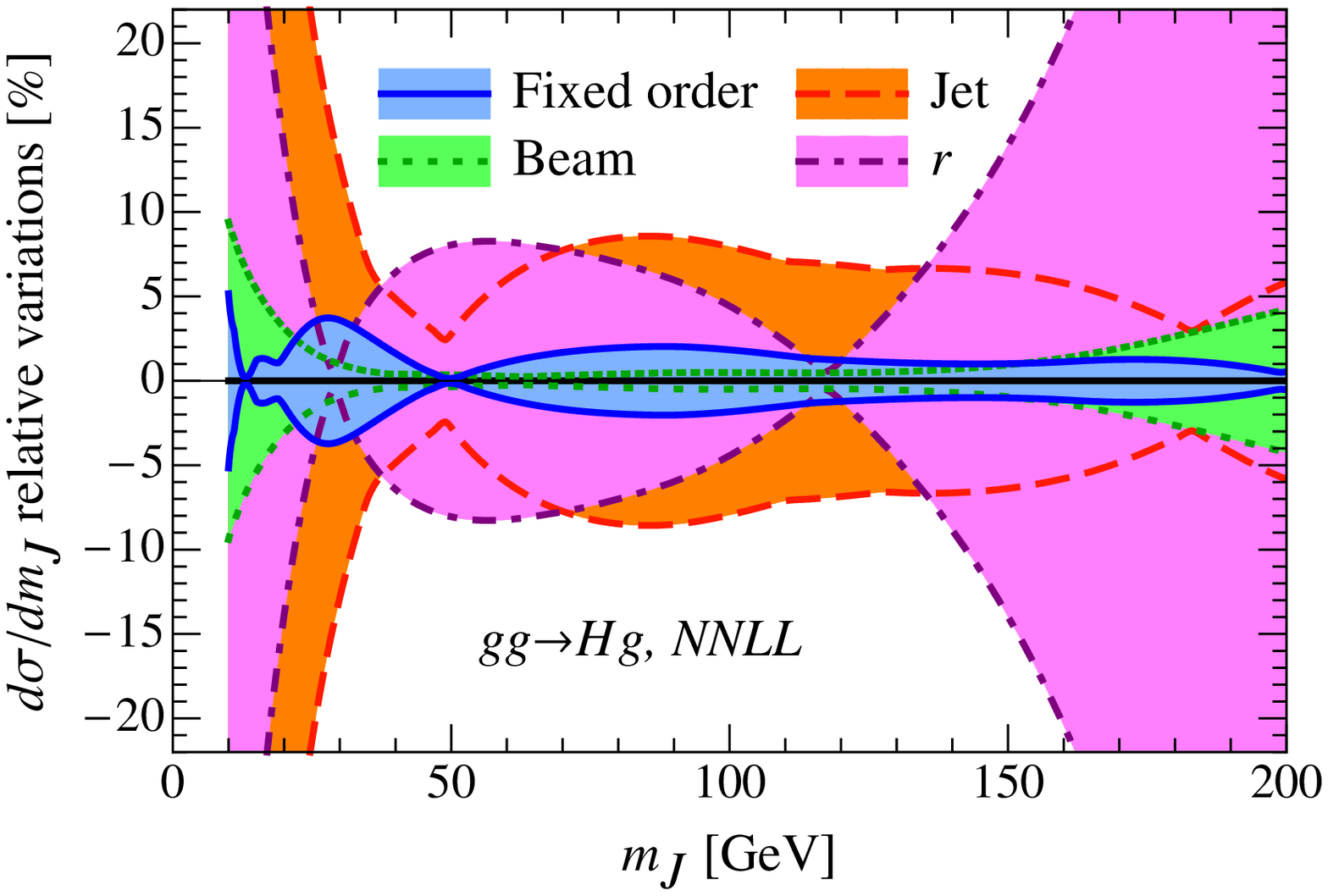}}%
\hfill%
\subfigure[Individual scale variations that enter the uncertainty estimate for quark jets at NNLL.
Shown are the variations relative to the central NNLL curve.]
{\label{fig:indivuncertq}\includegraphics[width=\columnwidth]{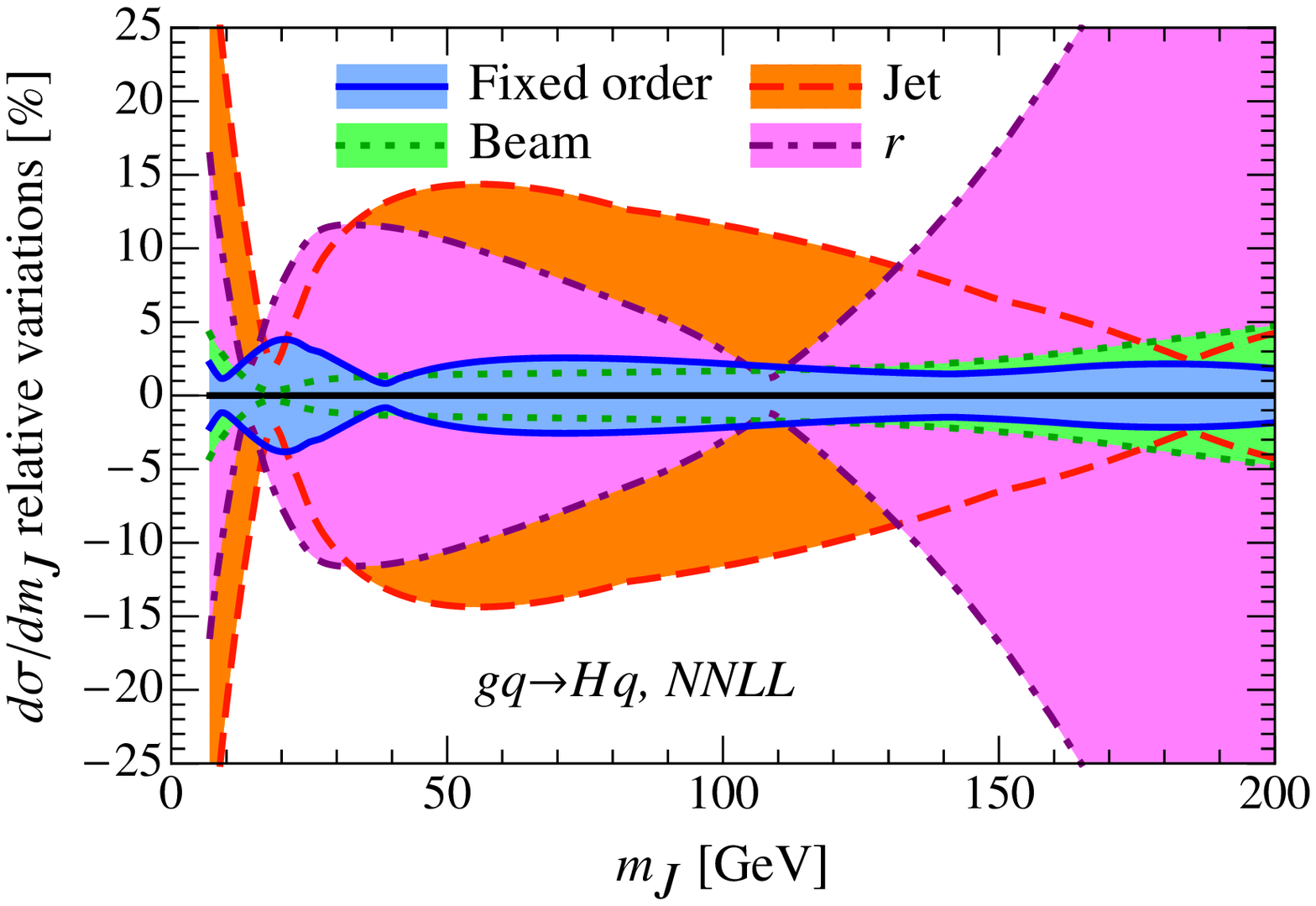}}%
\caption{Perturbative uncertainties and convergence for the jet mass spectrum in $gg\to Hg$ and $gq\to Hq$
with default inputs.}
\label{fig:uncert}
\end{figure*}

In this section we focus on the individual quark and gluon channels, leaving
results for $pp\to H+1$ jet to be discussed in \sec{results_Hj} below. We first study
the theoretical predictions for the $m_J$ spectrum with and without
normalization, and show that normalizing substantially reduces the perturbative
uncertainty. We also study the order-by-order convergence of this differential
cross section, and the size of various contributions to the perturbative
uncertainty bands.  Next, the dependence on the jet veto $\Tau^\cut$ is studied.
Finally, we investigate the size of non-global logarithms as a function of $m_J$
and $\Tau^\cut$.

%===============================================================================
\subsection{Default Parameter Choices}
%===============================================================================

Unless indicated otherwise we use the following default parameter choices for
all plots in Secs.~\ref{sec:results_part}, \ref{sec:results_Hj}, and \ref{sec:compare}.
For the Higgs mass we take $m_H = 125\,\GeV$~\cite{:2012gk,:2012gu}, and
for the LHC center-of-mass energy we take $\Ecm= 7\,{\rm TeV}$.  We always use
the MSTW NLO PDFs \cite{Martin:2009bu} with the corresponding value of
$\alpha_s(m_Z) = 0.1202$ for the strong coupling constant.  As our default we use
the geometric $R=1$ measure for defining the jets, $\Tau^\cut=25\,\GeV$ for the
jet veto, and $m_J^\cut=200\,\GeV$ for the normalization range.  Our default
hard kinematics are $\pTJ =300\,\GeV$, $\eta_J=0$, and $Y=0$.  Finally, for the
scale functions $\mu_H$, $\mu_{B_i}(\tau)$, $\mu_J(\tau)$, and $\mu_{S_i}(\tau)$
defined in \subsec{scales}, the central parameter values are given in
\app{scales}. There we also discuss the combination of scale variations used for
estimating the perturbative uncertainties.

%===============================================================================
\subsection{Normalization and Convergence}
%===============================================================================

The unnormalized jet mass spectrum at NNLL with our default inputs for the quark
and gluon channels are shown in \fig{unnormuncert}. As one expects, the gluon
jets peak at a much higher jet mass than the quark jets. We also see that the
perturbative uncertainties are quite sizable, even at NNLL.

\begin{figure*}[t!]
\includegraphics[width=0.95\columnwidth]{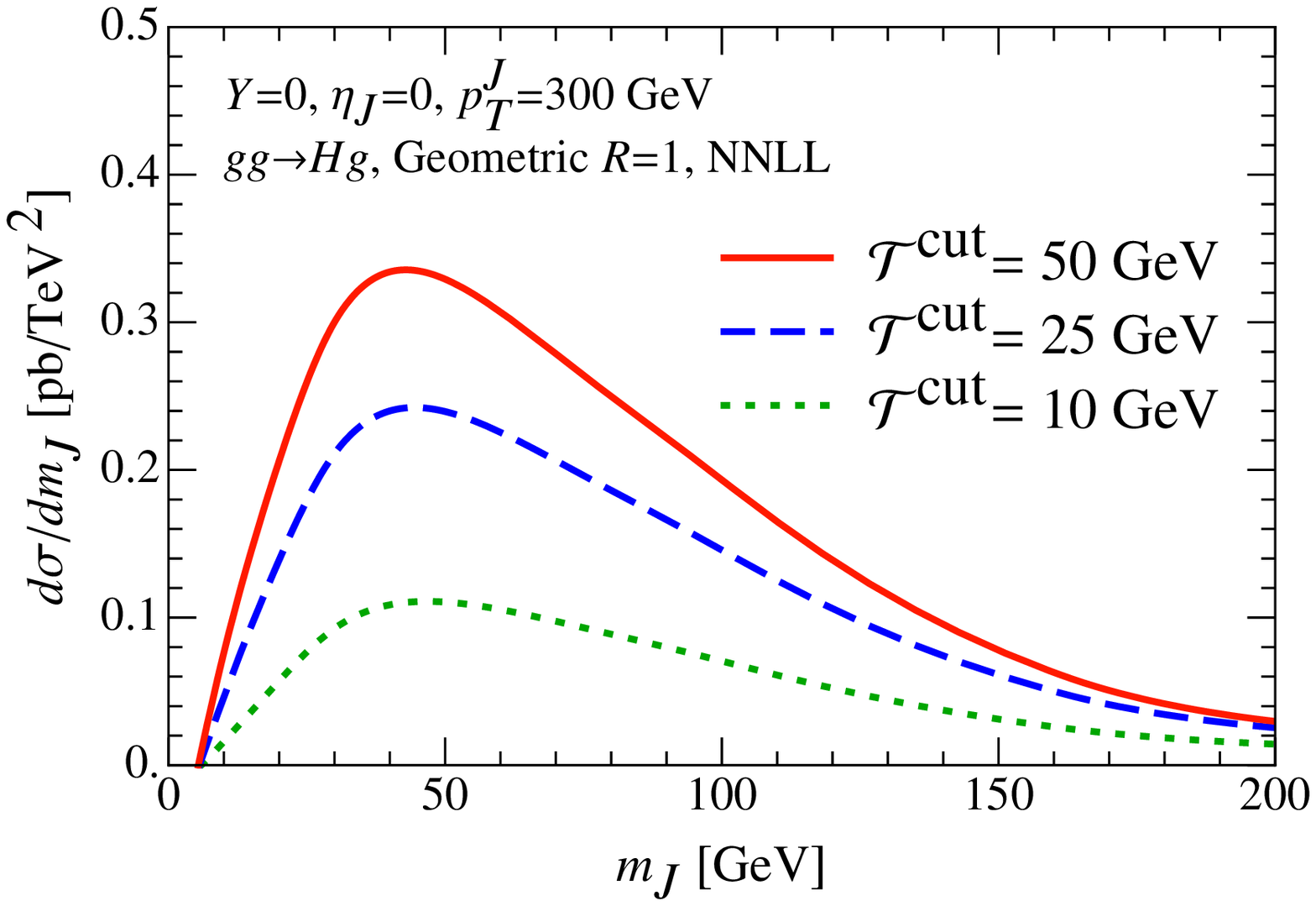}%
\hfill\includegraphics[width=\columnwidth]{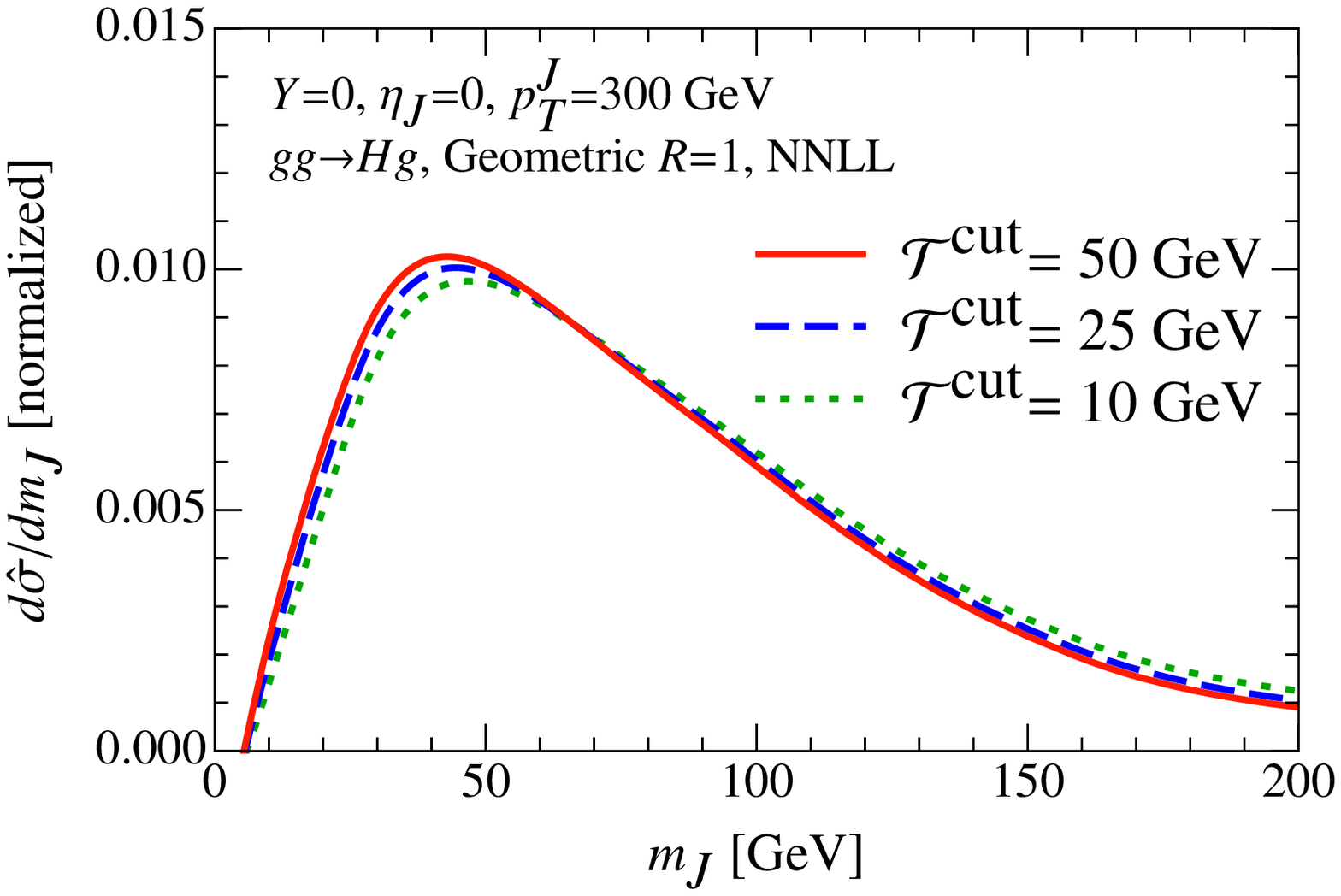}%
\caption{Effect of using different jet veto cuts on the jet mass spectrum for
  $gg\to Hg$.  While the unnormalized spectrum on the left is directly sensitive
  to the jet veto cut, this dependence almost completely cancels in the
  normalized spectrum on the right.  The same is true for the quark channel,
  $gq\to Hq$, and the sum over all partonic channels.}
\label{fig:TB}
\end{figure*}

Normalizing the jet mass spectrum allows one to study its shape without
contamination from the slow convergence of the integrated 1-jet
cross section, and also reduces the experimental uncertainties significantly. We
denote the normalized cross section as $\df\hat\sigma/\df m_J$ and calculate it
using \eq{norm} where we normalize over the range $0 \leq m_J \leq m_J^\cut$.

We first study the impact of normalization on the perturbative uncertainty.  To
preserve the normalization, we simultaneously vary the scales in the numerator
and denominator of \eq{norm}. Comparing the unnormalized cross section at NNLL
for the gluon and quark channels shown in \fig{unnormuncert} to the normalized
ones in \fig{normuncert}, we observe that a substantial portion of the
uncertainty is related to the integrated cross section rather than the shape. In
the resummation region of the $m_J$ spectrum, $30\,{\rm GeV}\le m_J\le 150\,{\rm GeV}$
the normalized cross sections have a quite reasonable remaining perturbative
uncertainty of $\simeq 6$--$9\%$ for gluons, and $\simeq 11$--$14\%$ for quarks.

A big part of the sizable uncertainty in the unnormalized 1-jet cross section is
due to the poor convergence of the hard function for $pp \to H+1$ jet, and thus
specific to the Higgs process. If we were to keep the hard function as an overall
multiplicative factor it would cancel exactly in the normalized
cross section for a given partonic channel and fixed phase space point.  As shown by the 
reduction in uncertainties seen in Fig.~\ref{fig:normuncert}, the majority of this
cancellation still takes place despite the fact that we are using an expanded hard function
as in \eq{Hmult}.  This cancellation also takes place approximately for the integrated
cross section summed over partonic channels as we show below in \subsec{integrate}. Our 
results with fixed kinematics are therefore representative of results integrated over the jet 
phase space.

The order-by-order convergence of our resummed jet mass calculation is displayed
in \figs{convergenceg}{convergenceq} for the gluon and quark jet channels, where
results at LL, NLL, and NNLL are shown. The various bands overlap with those of lower
orders, providing direct evidence that our scale variations yield a
reasonable estimate of the higher-order perturbative uncertainties.

There are several classes of perturbative scale uncertainties, the ``Fixed
Order'' scale variation that is correlated with the total cross section, the
``Beam'' scale variation from varying $\mu_{B_i}$ and $\mu_{S_B}$ that is related to
the presence of the jet veto, the ``Jet'' scale variation from varying $\mu_J$
and $\mu_{S_J}$ that is related to the jet mass measurement, and the uncertainty
from ``r'' that is related to the perturbative freedom in the refactorized
formula for the soft function.  For the NNLL results, these individual scale
variations are shown in \figs{indivuncertg}{indivuncertq} for gluon and quark
jets respectively. For simplicity we combined the uncertainty from varying the
jet scale $\mu_J$ and the scale of the jet part of the soft function $\mu_{S_J}$
by taking the envelope, and similarly for the beams. It is not too surprising
that the uncertainties associated with the hard and beam scale variations are
smaller, since they are mostly common to the numerator and denominator of the
normalized spectrum in \eq{norm}. It is also not surprising that the ``r" uncertainty 
dominates for large $m_J$ since in this region there is a hierarchy between $m_J^2$ 
and $\pTJ\Tau^\cut$, and the lack of resummation in this ratio shows up through this
uncertainty.  To obtain the total perturbative uncertainty we take the envelope of 
``Jet'', ``Beam'' and ``r'' uncertainties and combine it in quadrature with the 
``Fixed Order'' uncertainty. The total uncertainty in the jet mass spectrum is dominated 
by that of the jet and by the soft function refactorization.

%===============================================================================
\subsection{Jet Veto and Non-Global Logarithms}
\label{subsec:veto_ngls}
%===============================================================================

\begin{figure*}[t!]
\includegraphics[width=0.95\columnwidth]{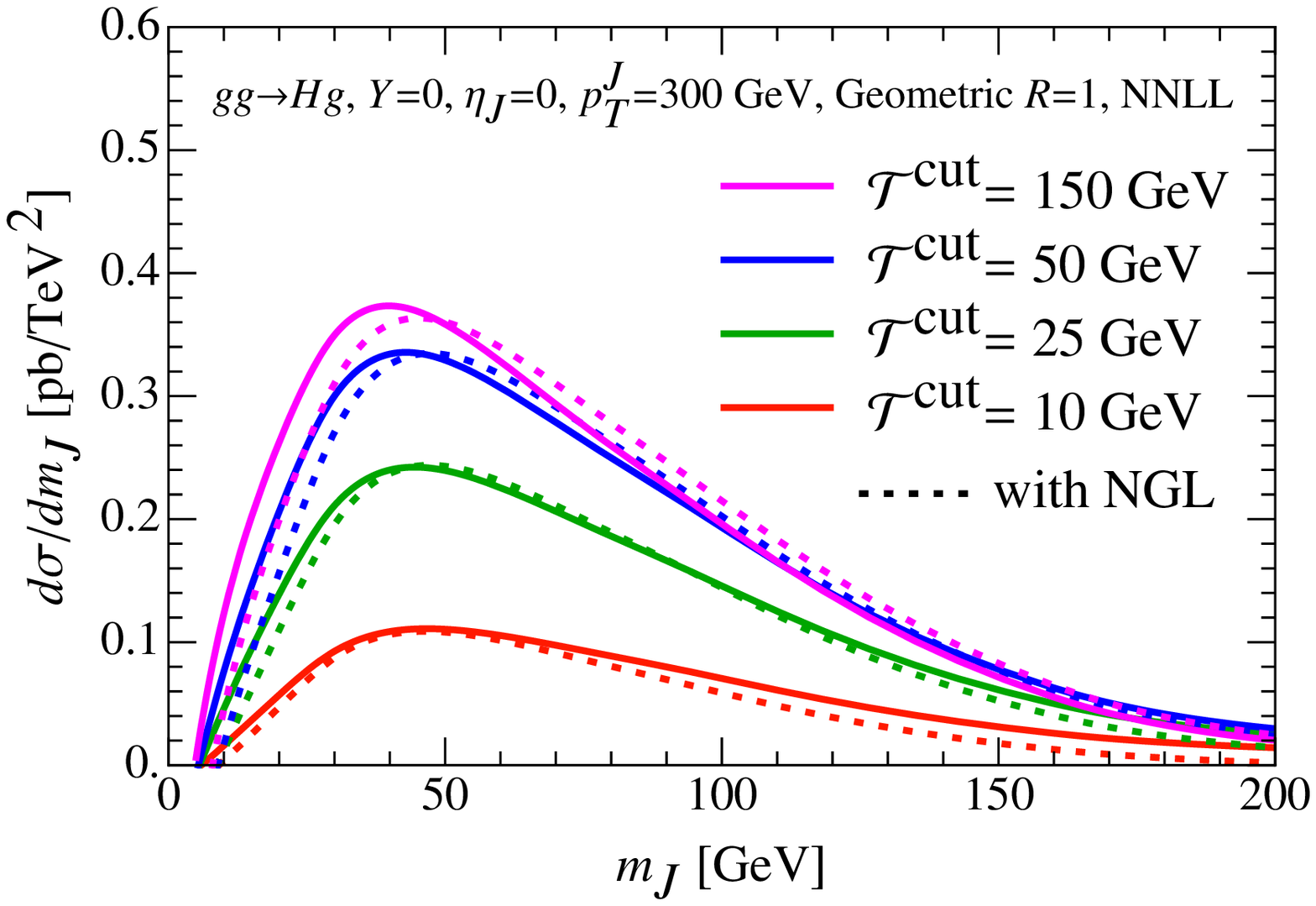}%
\hfill\includegraphics[width=\columnwidth]{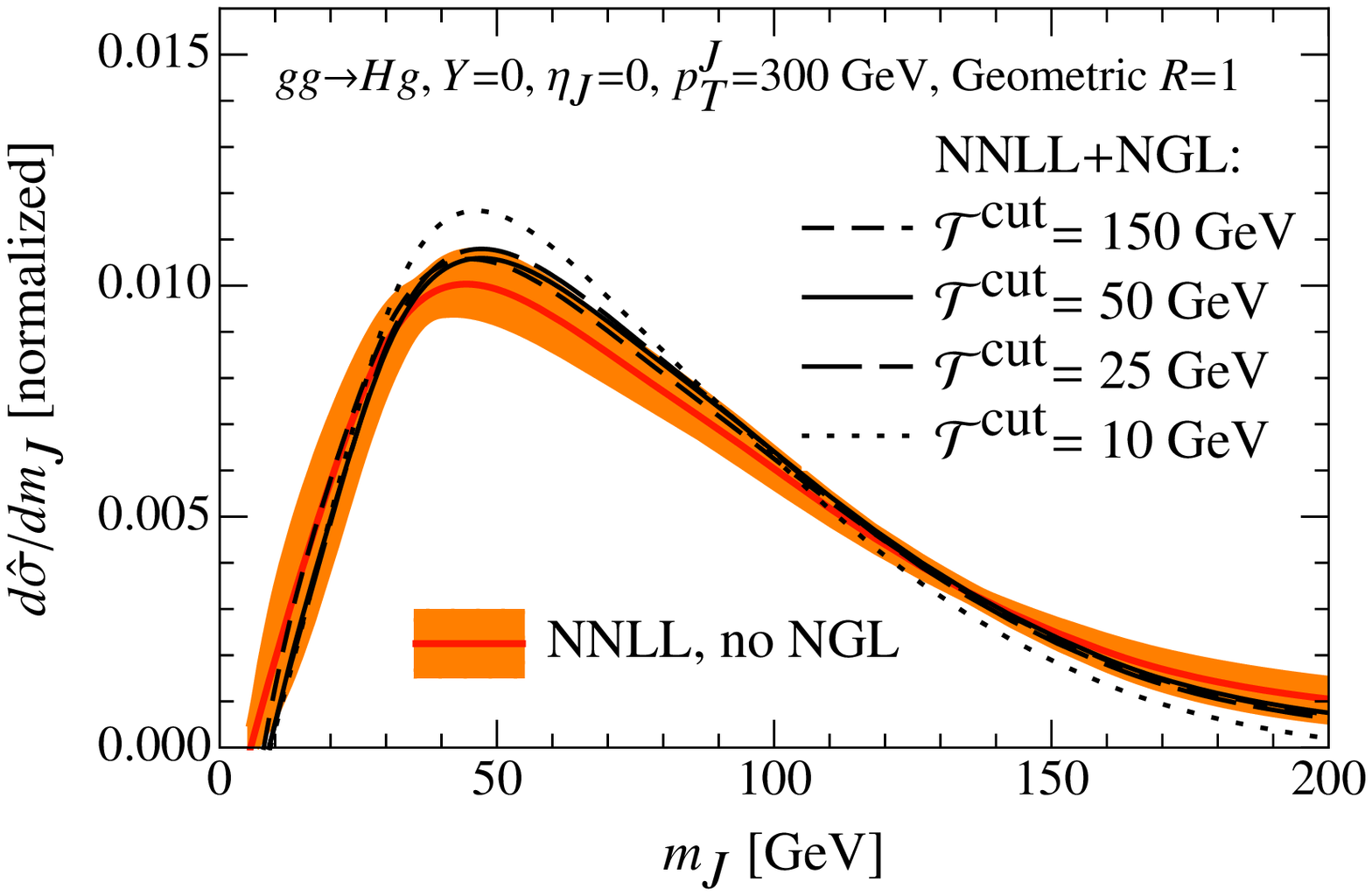}%
\caption{Effect of non-global logarithms on the NNLL jet mass spectrum for $gg\to Hg$ for different jet veto cuts.
Left panel: Including the leading NGLs (dashed lines) has a small effect on the
unnormalized spectrum, and is well within
the perturbative uncertainty for a wide range of jet veto cuts.
Right panel: 
  The effect of including the leading NGLs (black solid, dashed, and dotted curves) on the
  normalized NNLL spectrum (orange band) is still within the reduced perturbative uncertainty
  for a wide range of jet veto cuts, but start to become important for $\Tau^\cut=10\,{\rm GeV}$.}
\label{fig:ngls}
\end{figure*}
Next we discuss the effect of the jet veto on the jet mass spectrum. Our veto is
imposed through the variable $\Tau^\cut$, rather than a more traditional
$p_{TJ}^\cut$, since this simplifies the treatment of scales in the problem, and
allows us to make use of a known factorization theorem. This jet veto restricts
the initial and final-state collinear radiation as well as soft radiation. It
turns out that the \emph{normalized} jet mass spectrum is fairly insensitive to
the value of the jet veto cut.

We start by considering the effect of the jet veto on the \emph{unnormalized}
jet mass spectrum, as shown for $gg \to Hg$ in the left panel of \fig{TB}.
Decreasing $\Tau^\cut$ imposes a stronger restriction on the initial-state
radiation and reduces the unnormalized cross section. (This reduction is less
strong for $gq \to Hq$, because quarks radiate less than gluons.)  As the right
panel of \fig{TB} shows, the normalization removes the majority of the
$\Tau^\cut$ dependence.  Note that without the refactorization of the soft
function (see \subsec{sfact}) this cancellation would be spoiled by unphysical
logarithms. This strong cancellation is also the case for the other partonic
channels, as well as for their sum in $pp\to H+1$ jet. This insensitivity to
$\Tau^\cut$ also remains valid after integrating over the jet phase space, as we
show below in \fig{TB2}.
We have also studied the dependence on
$\Tau^\cut$ as well as a standard $p_{TJ}^\cut$ jet veto with \Pythia, where we
also find a similar insensitivity of the normalized jet mass spectrum to the
details of the used jet-veto variable and cut values.

Next we turn to our analysis of NGLs, both in the unnormalized and normalized
jet mass spectra.  As explained in \subsec{ngl}, we test for the size of the NGLs
by comparing the cross section with and without these terms. The leading NGL is
included in fixed-order perturbation theory in the soft function, on top of which 
we sum an infinite series of global logarithms through the factorization formula.

In the left panel of \fig{ngls} we show the unnormalized spectrum for
various $\Tau^\cut$ values at NNLL (solid lines) and the same spectra including
the NGL terms (dotted lines). As mentioned earlier, there is a point on the spectrum
where the NGLs exactly cancel. This point is at $m_J\simeq 50, 110, 165, 300$ for
$\Tau^\cut=10, 25, 50, 150\,{\rm GeV}$ respectively.  For all values of $m_J$
shown in this figure the effect of the NGL terms is well within the perturbative
uncertainty [cf.~the uncertainty bands shown in \fig{unnormuncert}].

When we normalize the spectrum we are dividing by the cumulant with $m_J^\cut$,
and the jet-veto dependence does not cancel out in the presence of the
non-global logarithms. There are two types of NGLs in the normalized result,
terms involving $\ln[m_J^2/(\pTJ \Tau^\cut)]$ from the numerator and terms
involving $\ln[m_J^{\cut\,2}/(\pTJ \Tau^\cut)]$ from the denominator. Therefore
for a fixed $\Tau^\cut$ there is no longer a value of $m_J$ where all the NGLs
will vanish. Results for the normalized spectrum with and without NGLs are shown
in the right panel of \fig{ngls}. The orange band shows the NNLL result without
NGLs along with its perturbative uncertainty, while the various black lines show
the central values for NNLL results that have the NGLs included. For the wide
range of values $25\,{\rm GeV}\le \Tau^\cut \le 150\,{\rm GeV}$ the effect of
the NGLs is of the same size as the reduced perturbative uncertainty in the
normalized spectrum.  This justifies our assertion that the NGLs do not have to
be considered as large logarithms for a significant range of cut values, so that
our NNLL result is complete at this order.  In the small $m_J$ region of the
spectrum the resummation of global logarithms on top of the NGL term provides an
appropriate Sudakov suppression in the the cross section.  For other $m_J$
values, and $25\,{\rm GeV}\le \Tau^\cut \le 150\,{\rm GeV}$, the argument of the
NGL remains between $1/8$ and $8$, which is the range over which we expect that
the NGLs do not dominate over nonlogarithmic corrections, as mentioned in the
introduction. On the other hand, for $\Tau^\cut =10\,{\rm GeV}$ one observes
that the NGLs become large enough that they are no longer contained within the
perturbative uncertainty, so this value is outside the range of validity of our
normalized NNLL results (though for this value the unnormalized results 
remain valid). For $\Tau^\cut =10\,{\rm GeV}$  the argument of the NGL involving
$m_J^\cut$ becomes $\simeq 13$, which is outside of the range mentioned above.

Although we have only explored the $gg\to Hg$ channel at a fixed kinematic point
in this section, we have also checked explicitly that the same conclusions about
NGLs hold when integrating over a kinematic range, and when considering quark
jets from $gq\to H q$.

%%%%%%%%%%%%%%%%%%%%%%%%%%%%%%%%%%%%%%%%%%%%%%%%%%%%%%%%%%%%%%%%%%%%%%%%%%%%%%%%
\section{\boldmath Results for $pp\to H+1$ jet}
\label{sec:results_Hj}
%%%%%%%%%%%%%%%%%%%%%%%%%%%%%%%%%%%%%%%%%%%%%%%%%%%%%%%%%%%%%%%%%%%%%%%%%%%%%%%%

In this section we show results for the $pp\to H+1$ jet cross section at NNLL,
summing the contributions from the various partonic channels: $gg \to Hg$, $gq
\to Hq$, and the (small) $q \bar q \to Hg$. We present results for the
dependence of the jet mass spectrum on the jet kinematics, on the choice of jet
definition which affects the shape of the jets, and on the jet size $R$. We also
compare the $m_J$ spectrum obtained for a fixed point in the jet kinematics
to that obtained from integrating over a range of jet momenta.

%%%
\begin{figure*}[t!]
\subfigure[The cross section decreases with increasing $\pTJ$.]
{\label{fig:varypT_UnNorm}\includegraphics[width=0.95\columnwidth]{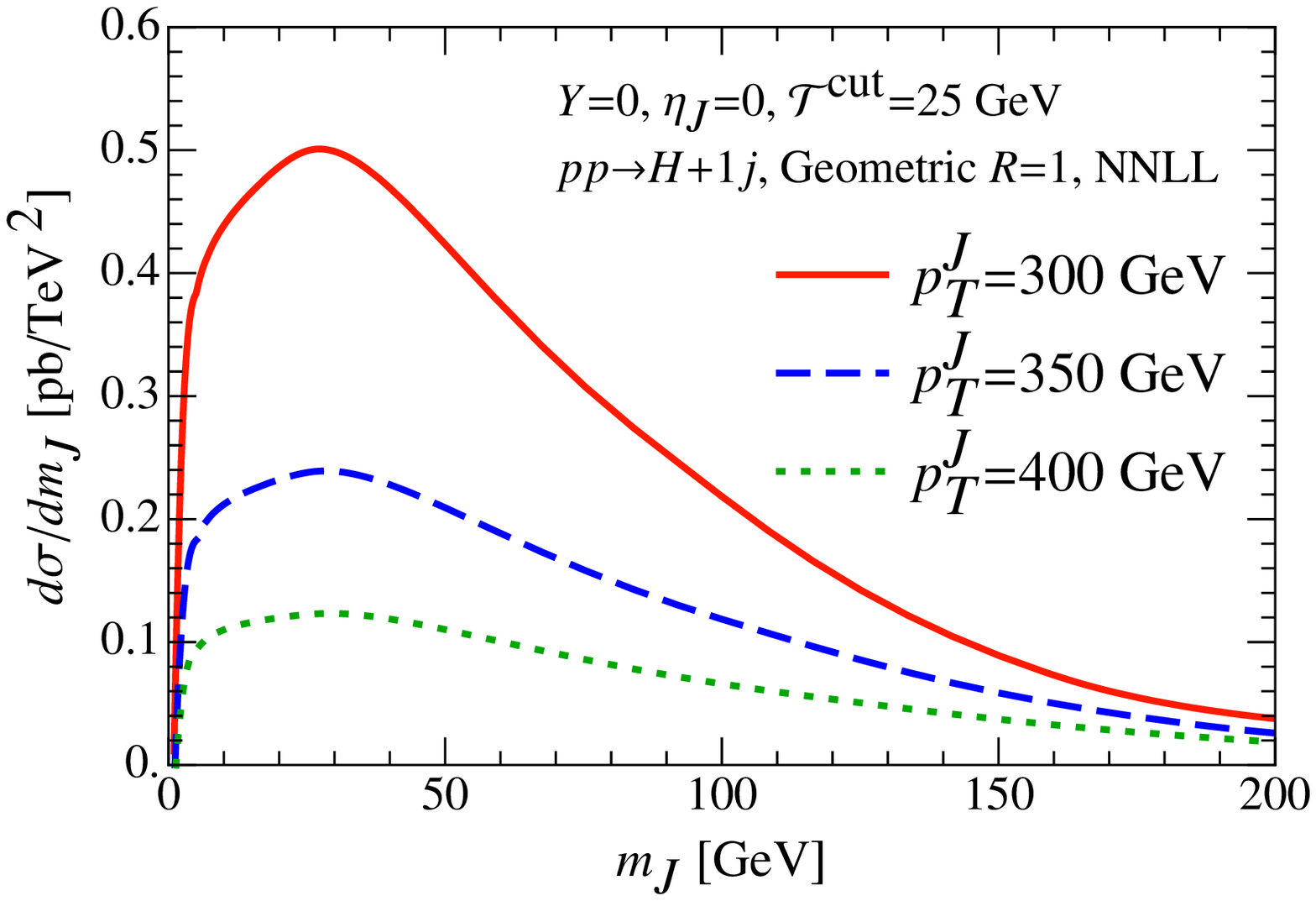}}%
\hfill%
\subfigure[For $pp\to H+1$ jet the peak position remains stable and the spectrum slightly broadens with increasing $\pTJ$.]
{\label{fig:varypT}\includegraphics[width=\columnwidth]{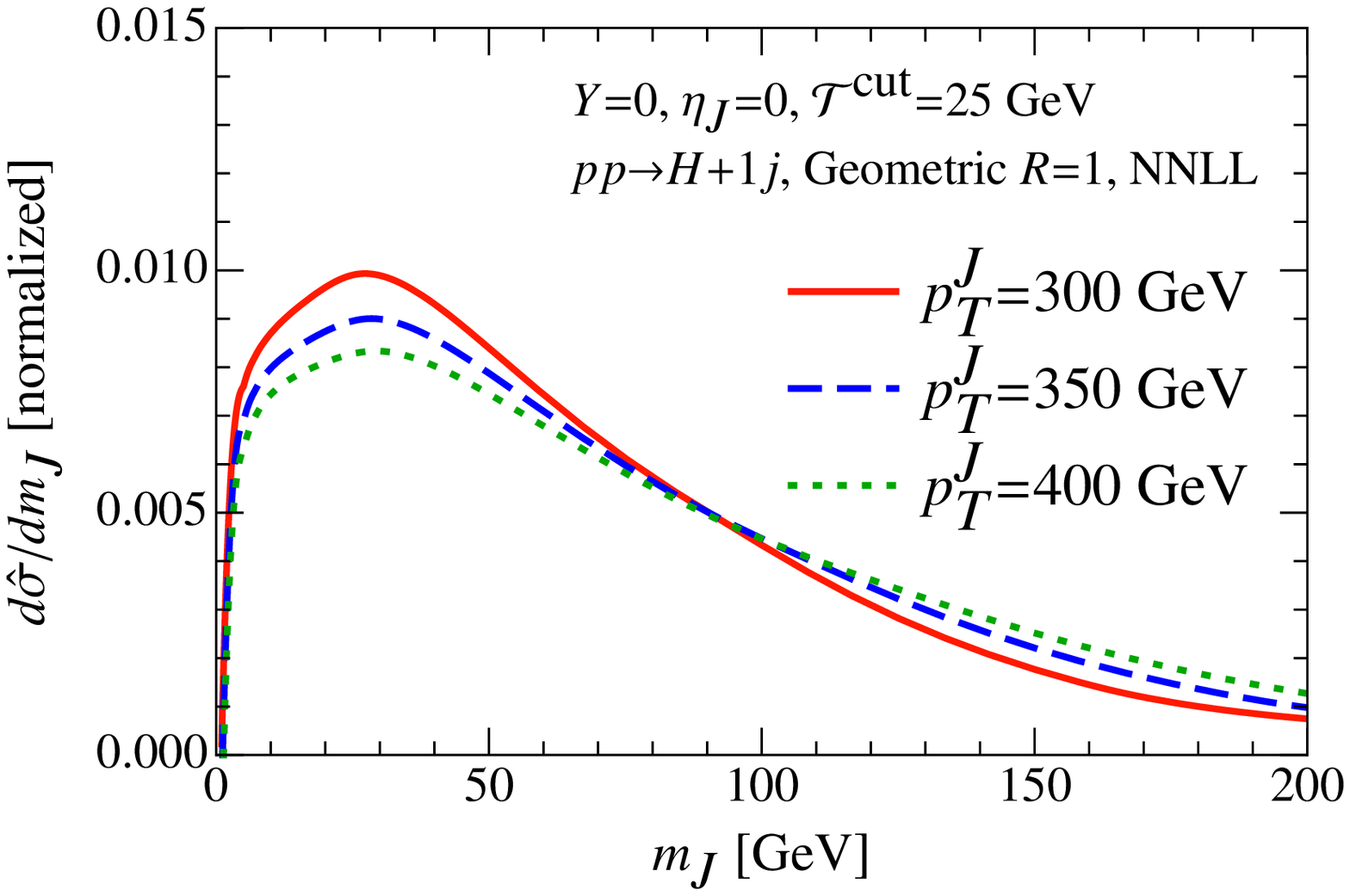}}
\\
\subfigure[The cross section decreases with increasing $\eta_J$.]
{\label{fig:varyEta_UnNorm}\includegraphics[width=0.95\columnwidth]{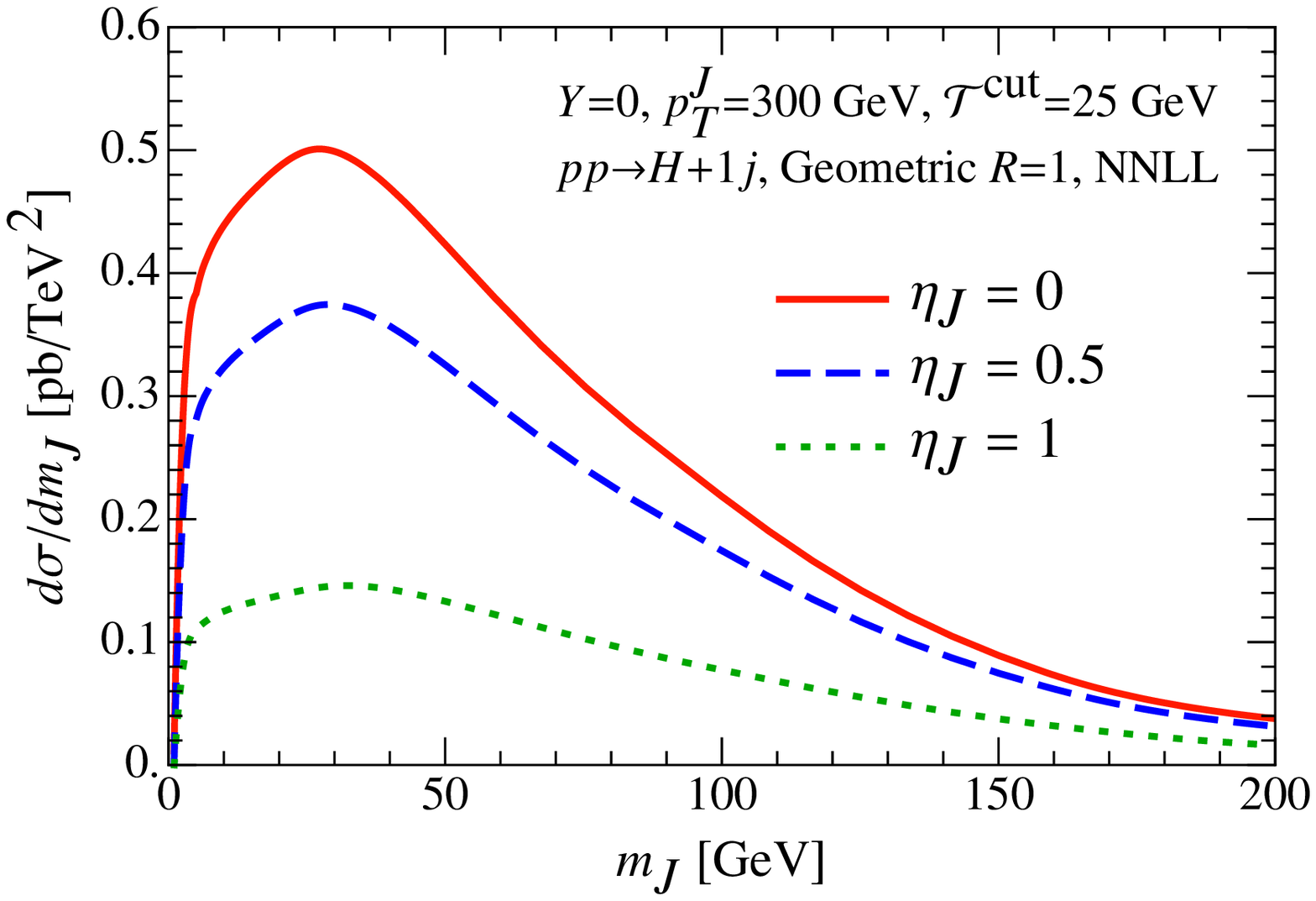}}%
\hfill%
\subfigure[For $pp\to H+1$ jet the peak position shifts slightly and the spectrum slightly broadens with increasing $\eta_J$.]
{\label{fig:varyEta}\includegraphics[width=\columnwidth]{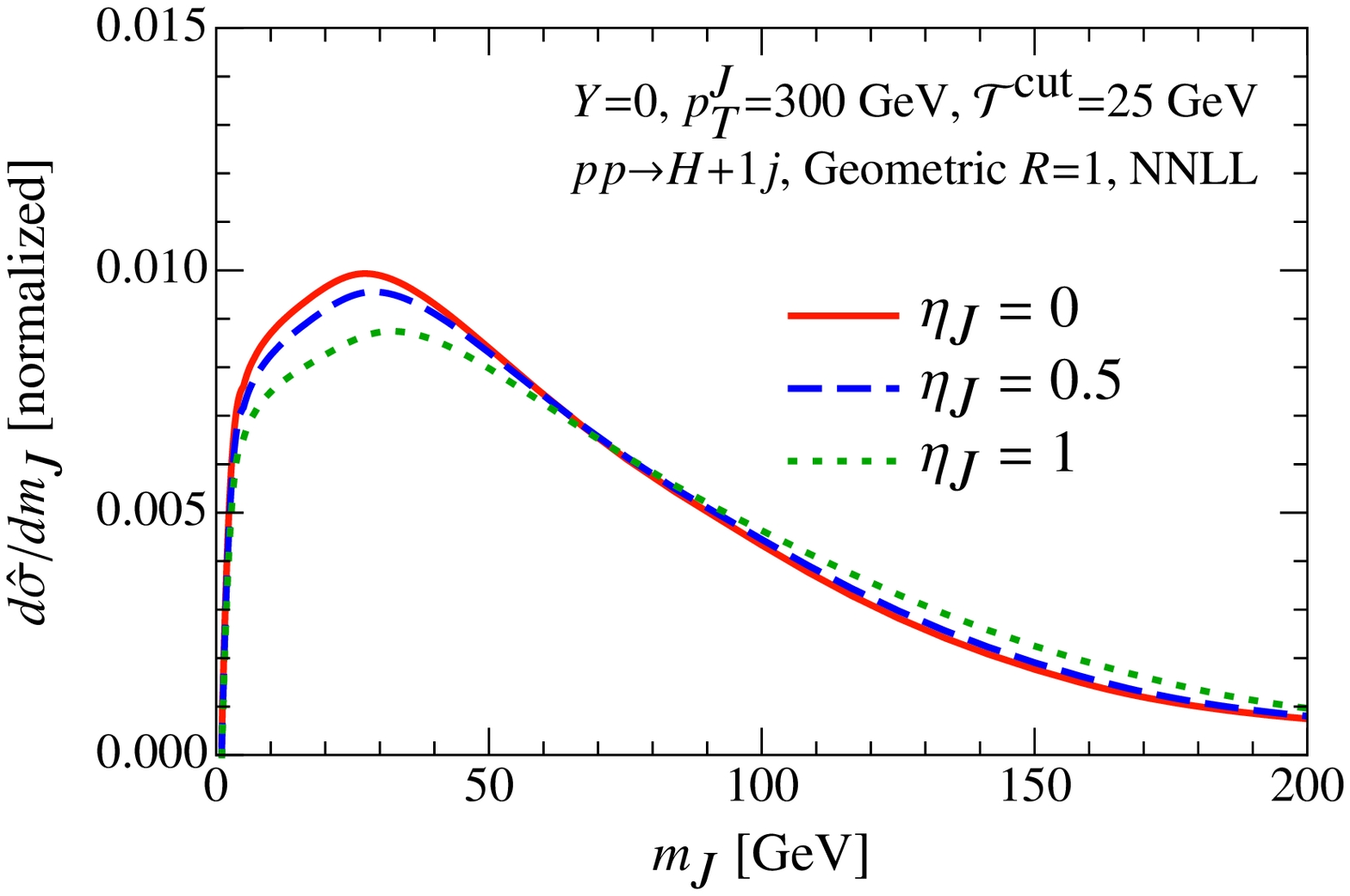}}
\\
\subfigure[The cross section quickly decreases for larger $Y$.]
{\label{fig:varyY_UnNorm}\includegraphics[width=0.95\columnwidth]{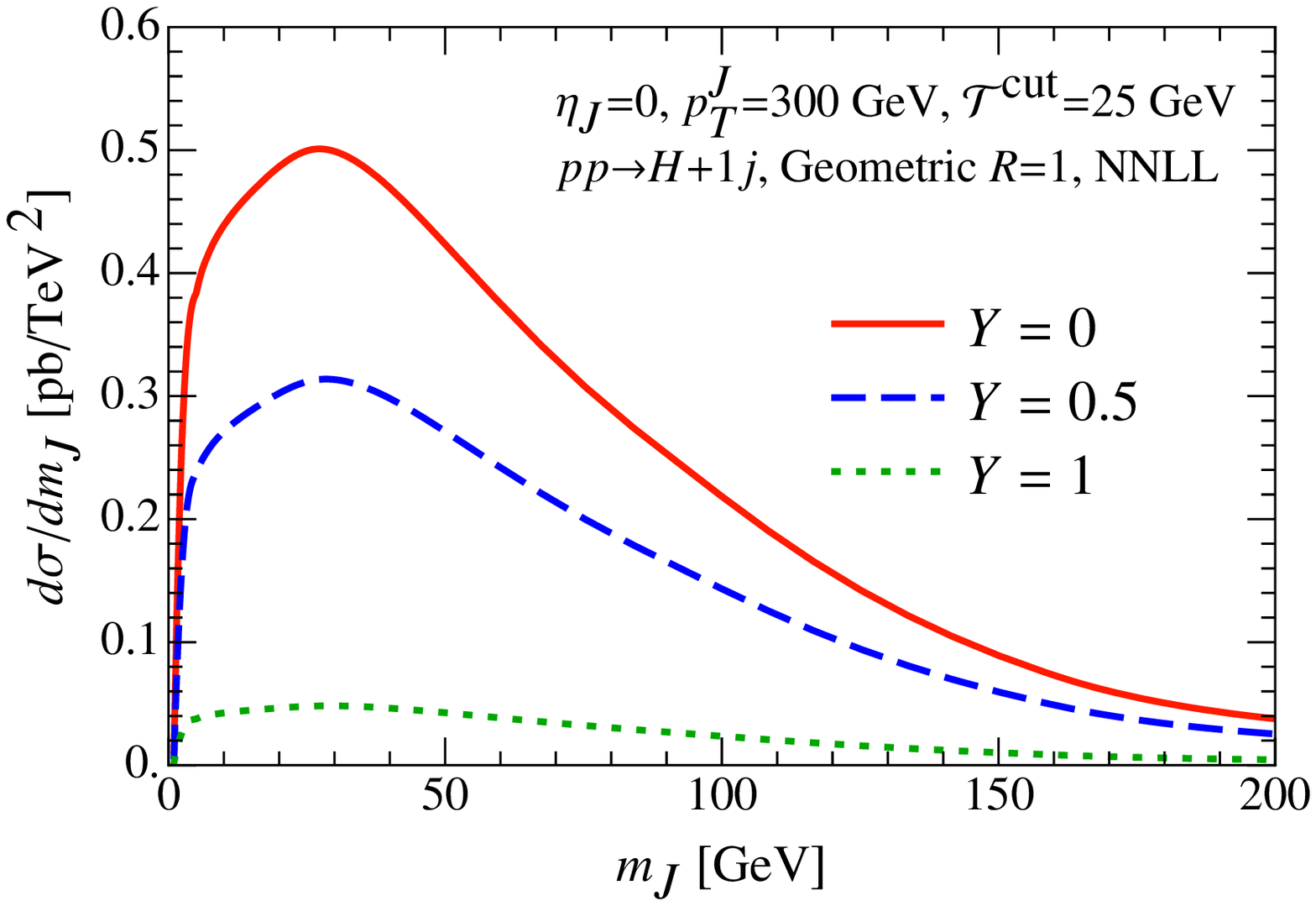}}%
\hfill%
\subfigure[For $pp\to H+1$ jet the peak position remains stable and the spectrum slightly broadens with increasing $Y$.]
{\label{fig:varyY}\includegraphics[width=\columnwidth]{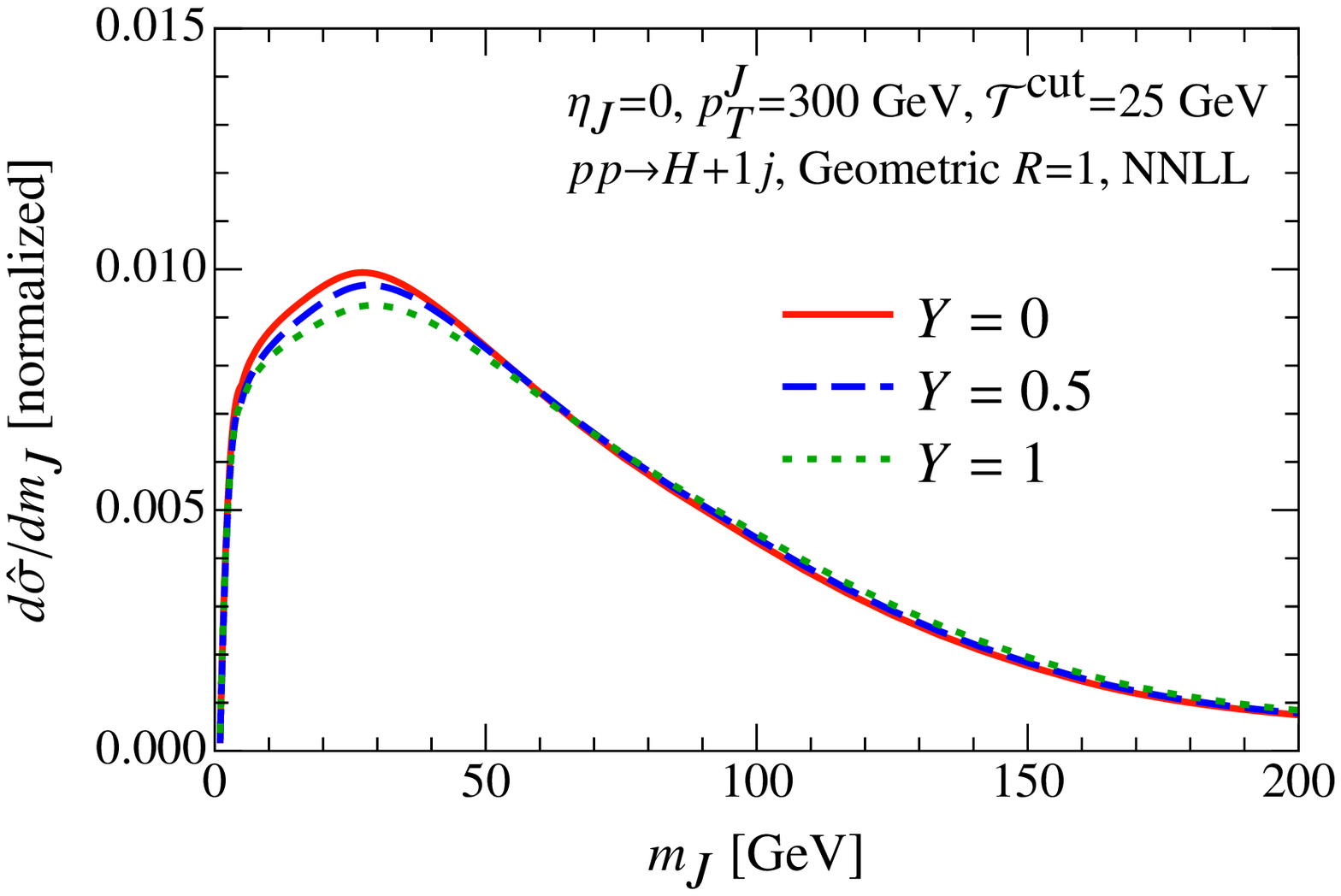}}
\caption{Dependence on the kinematic variables $p_T^J$, $\eta_J$, and $Y$ for
  the unnormalized and normalized NNLL jet mass spectra for $pp\to H+1$ jet. }
\label{fig:kinematics}
\end{figure*}
%%%

\begin{figure*}[t!]
\includegraphics[width=\columnwidth]{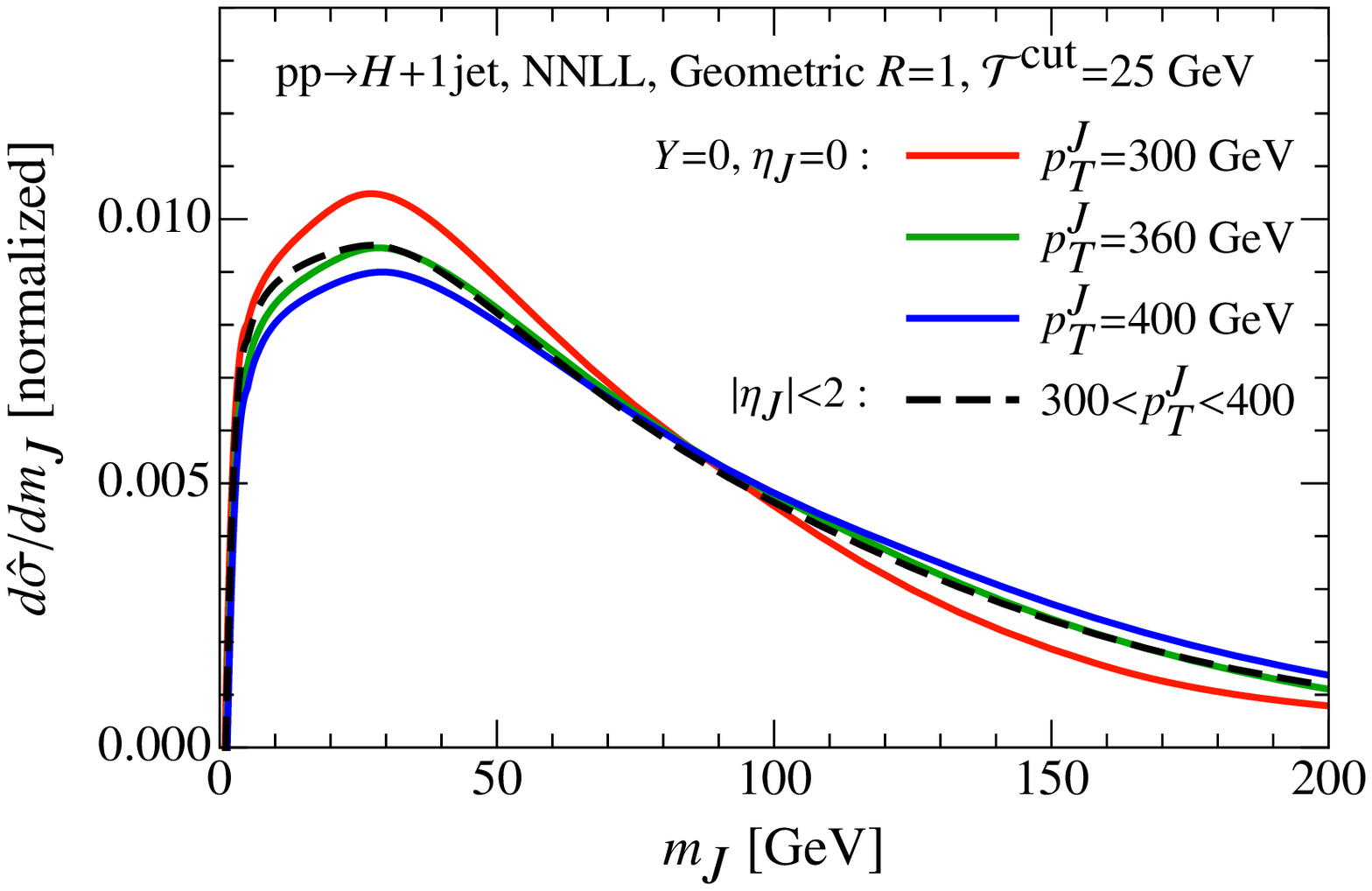}%
\hfill\includegraphics[width=\columnwidth]{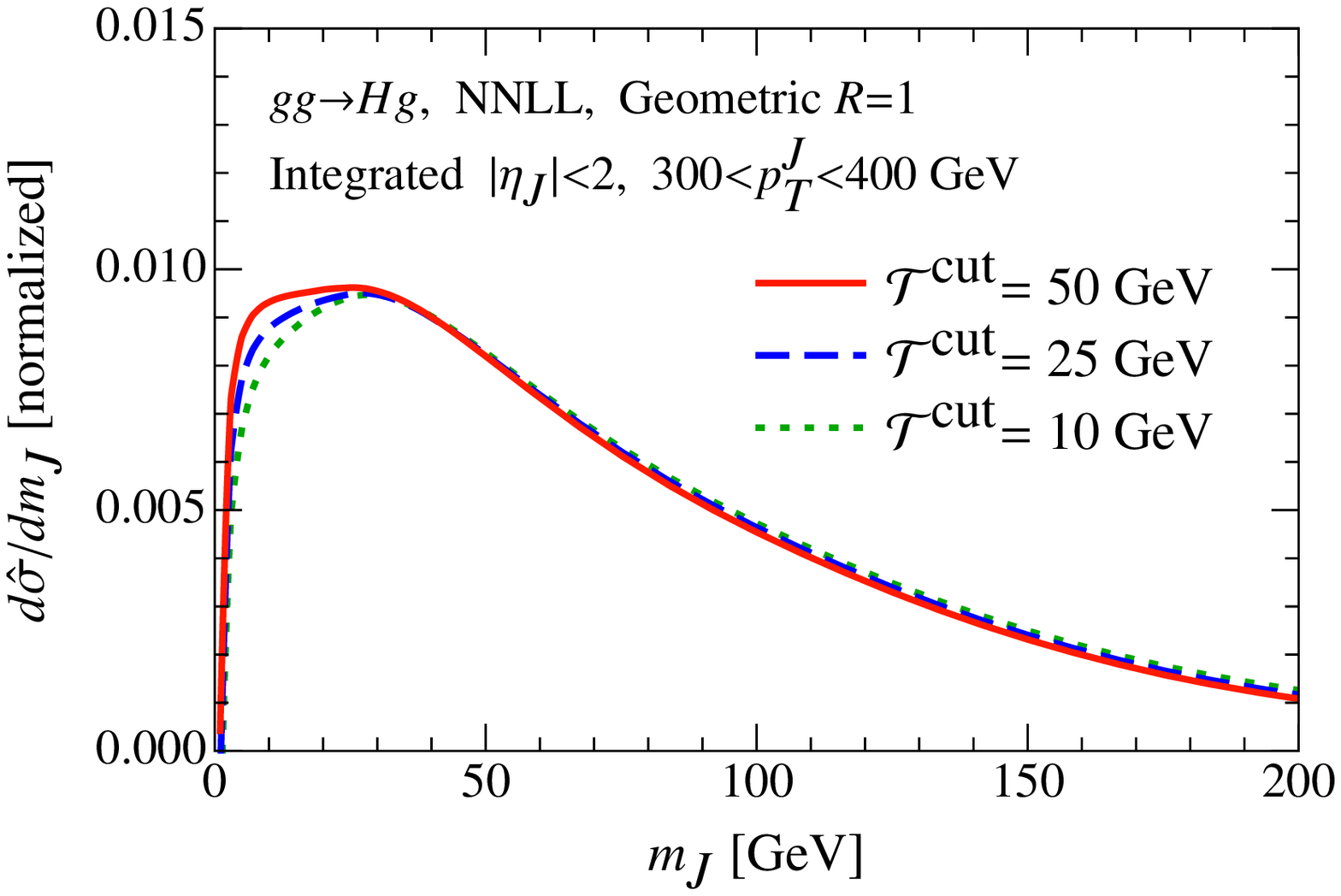}%
\caption{Results for the normalized jet mass spectrum at NNLL for $pp\to H+1$
  jet after integrating over $300\,{\rm GeV}<\pTJ<400\,{\rm GeV}$, $|\eta_J|<2$,
  and all $Y$. The left panel compares the spectrum for integrated kinematics
  (dashed line) to those for fixed kinematics with $Y=\eta_J=0$ and $\pTJ=300,
  360, 400\,{\rm GeV}$ (solid lines from top to bottom at the peak).  The right
  panel shows the impact of ${\cal T}^\cut$ on the normalized spectrum for
  integrated kinematics, which is the analog of the comparison in the right
  panel of \fig{TB} for fixed kinematics.}
\label{fig:TB2}
\end{figure*}

%===============================================================================
\subsection{Dependence on Kinematics}
%===============================================================================

For $pp\to H+1$ jet there are three nontrivial kinematic variables: the
transverse momentum of the jet $p_T^J$, rapidity of the jet $\eta_J$, and the
total rapidity $Y$ of the combined Higgs+jet system.  We show how each of these variables
affect both the unnormalized and normalized jet mass spectrum, which allows us
to separate the impact of kinematics on the normalization and the shape.

The falloff of the PDFs at larger $x$ values causes the cross section to
strongly decrease for increasing $\pTJ$ and for increasing $|\eta_J|$ (for
$Y=0$). This is shown in \figs{varypT_UnNorm}{varyEta_UnNorm}. The dependence on
$\pTJ$ and $\eta_J$ in the corresponding normalized spectra are shown in
\figs{varypT}{varyEta}.  Here we see that there is a decrease in the height of
the peak and a compensating increase in the tail height as $\pTJ$ or $|\eta_J|$
are increased.  Note that for these variables there is a marked difference
between the total $pp \to H+1$ jet process compared to the individual partonic
channels (which are not shown).  For each partonic channel the peak position of
the jet mass spectrum increases as $m_J^{\rm peak}\propto \sqrt{\smash\pTJ \vphantom{p_T}\rule[2ex]{0pt}{0pt}}$ and also
increases with increasing $|\eta_J|$.  However, at the same time the
contribution of $gq \to Hq$ relative to $gg \to Hg$ is enhanced, and the peak of
the jet mass spectrum is at lower values for quark jets than for gluon jets [see
\fig{normuncert}].  These two effects largely cancel for $pp\to H+1$ jet, such
that the peak position is practically unchanged with increasing $\pTJ$, whereas
for increasing $\eta_J$ a small net increase in the peak position remains.  

Note that our ability to calculate the $\eta_J$ dependence implies that it is
trivial to impose rapidity cuts in our framework.  This is an advantage of calculating
the jet mass spectrum for an exclusive jet sample, where the jet veto controls radiation 
in the out-of-jet region.

The main dependence on the total system rapidity $Y$ enters through the shape of
the PDFs, causing the cross section to strongly decrease with increasing $|Y|$,
as \fig{varyY_UnNorm} shows. (This is also the reason for taking central jets
with $Y=0$ for our default value when using a single phase space point.)  The
value of $Y$ also affects the shape of the jet mass spectrum, as can be seen in
\fig{varyY}. The jet rapidity relative to the partonic center of mass is
$\eta_J-Y$, so one would expect the shape change as function of $Y$ to be similar
to that as function of $\eta_J$, shown in \fig{varyEta}. The agreement is close
but not exact because the $Y$ dependence induced by the shape of the PDFs
differs channel by channel, and thus affects their relative weight in the sum
over channels.

%===============================================================================
\subsection{Integrated Kinematics}
\label{subsec:integrate}
%===============================================================================

So far we have shown the $m_J$ spectra for a fixed point in
$\pTJ$, $\eta_J$, and $Y$. We now consider the impact of integrating the kinematic
variables over a bin with $|\eta_J|<2$, $300\,{\rm GeV}< \pTJ < 400\,{\rm GeV}$,
and any $Y$. These kinematic ranges are realistic experimentally for jets at the LHC.

In the left panel of \fig{TB2} the jet mass spectrum for integrated kinematics
is shown by a black dashed line, and is compared to three spectra with fixed
kinematics shown by solid lines (with $Y=\eta_J=0$ and $\pTJ=300,360,400\,{\rm
  GeV}$ from top to bottom at the peak of the spectrum). One observes that the
$m_J$ spectrum in the integrated bin is very close to the $m_J$ spectrum with
$Y=\eta_J=0$ and near the center of the $\pTJ$ bin.  Thus our conclusions made
from studies of a single kinematic point directly carry over to the results
obtained by integrating over a phase space bin.

The one situation where this is not immediately obvious is the dependence of the
normalized cross section on the jet-veto cut, ${\cal T}^\cut$, shown for fixed
kinematics in \fig{TB}. When we integrate over the kinematic bin the hard
function, including its Sudakov form factor depending on ${\cal T}^\cut$, no
longer exactly cancels between the numerator and denominator. Nevertheless,
comparing the spectra for integrated kinematics and different values of ${\cal
  T}^\cut$, shown in \fig{TB2}, we see that the normalized spectrum is still
very insensitive to the details of the jet veto also after summing over partonic
channels and integrating over a range of kinematics. (We have also confirmed
that upon phase space integration the size of the NGL effect remains the same as
shown in \fig{ngls}.)

%===============================================================================
\subsection{Jet Definitions and Jet Area}
%===============================================================================

%%%
\begin{figure*}[t!]
\includegraphics[width=0.32\textwidth]{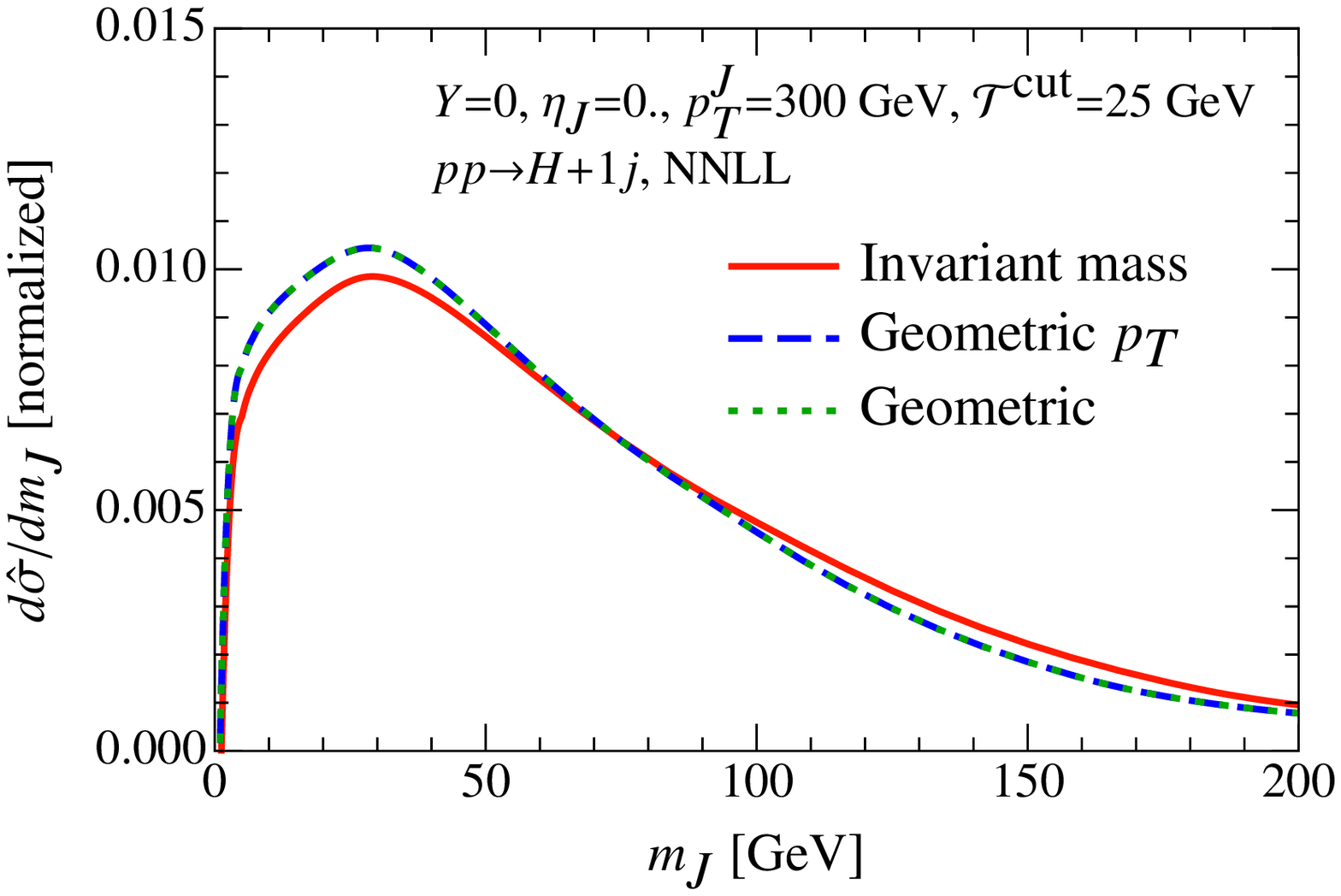}%
\hfill%
\includegraphics[width=0.32\textwidth]{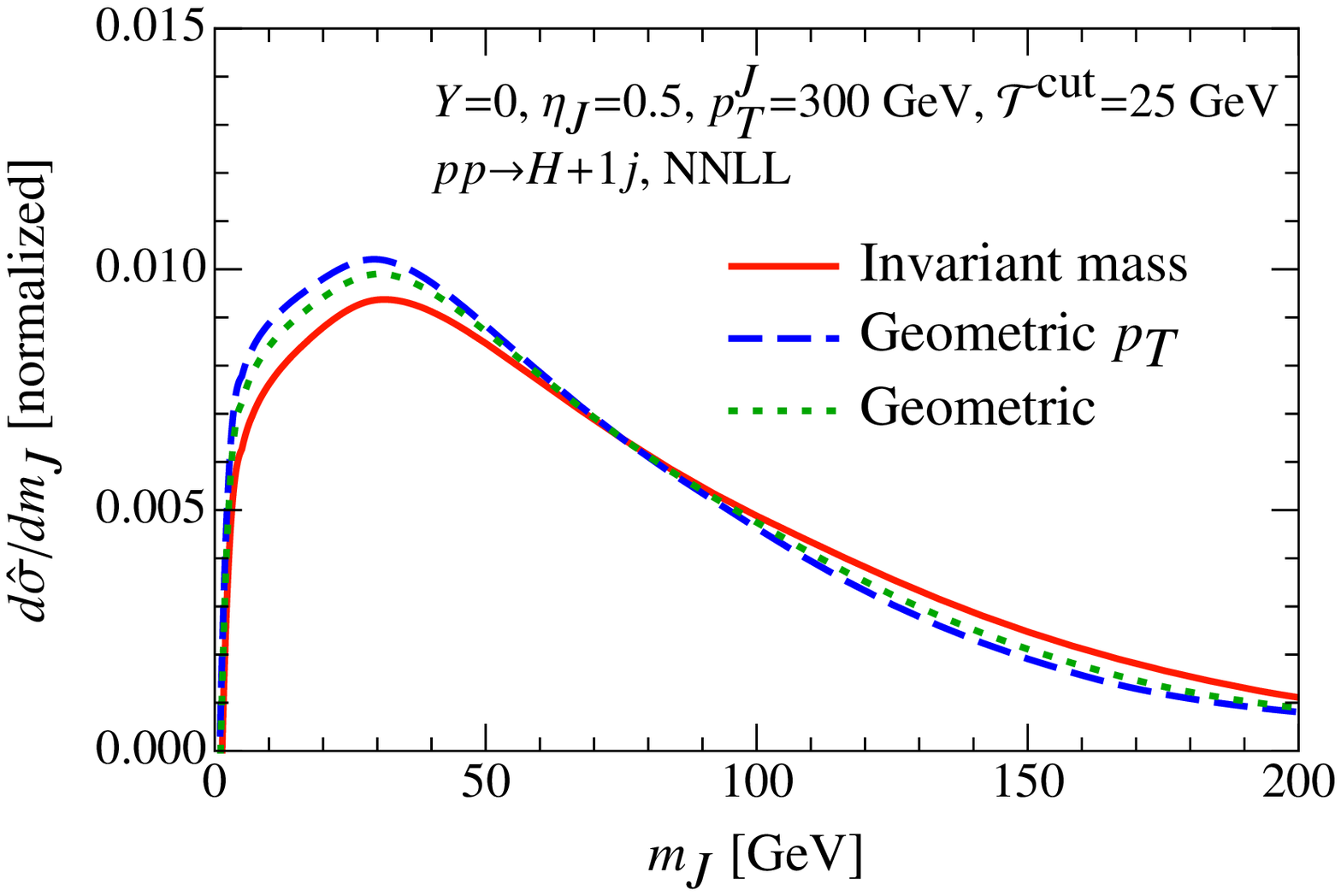}%
\hfill%
\includegraphics[width=0.32\textwidth]{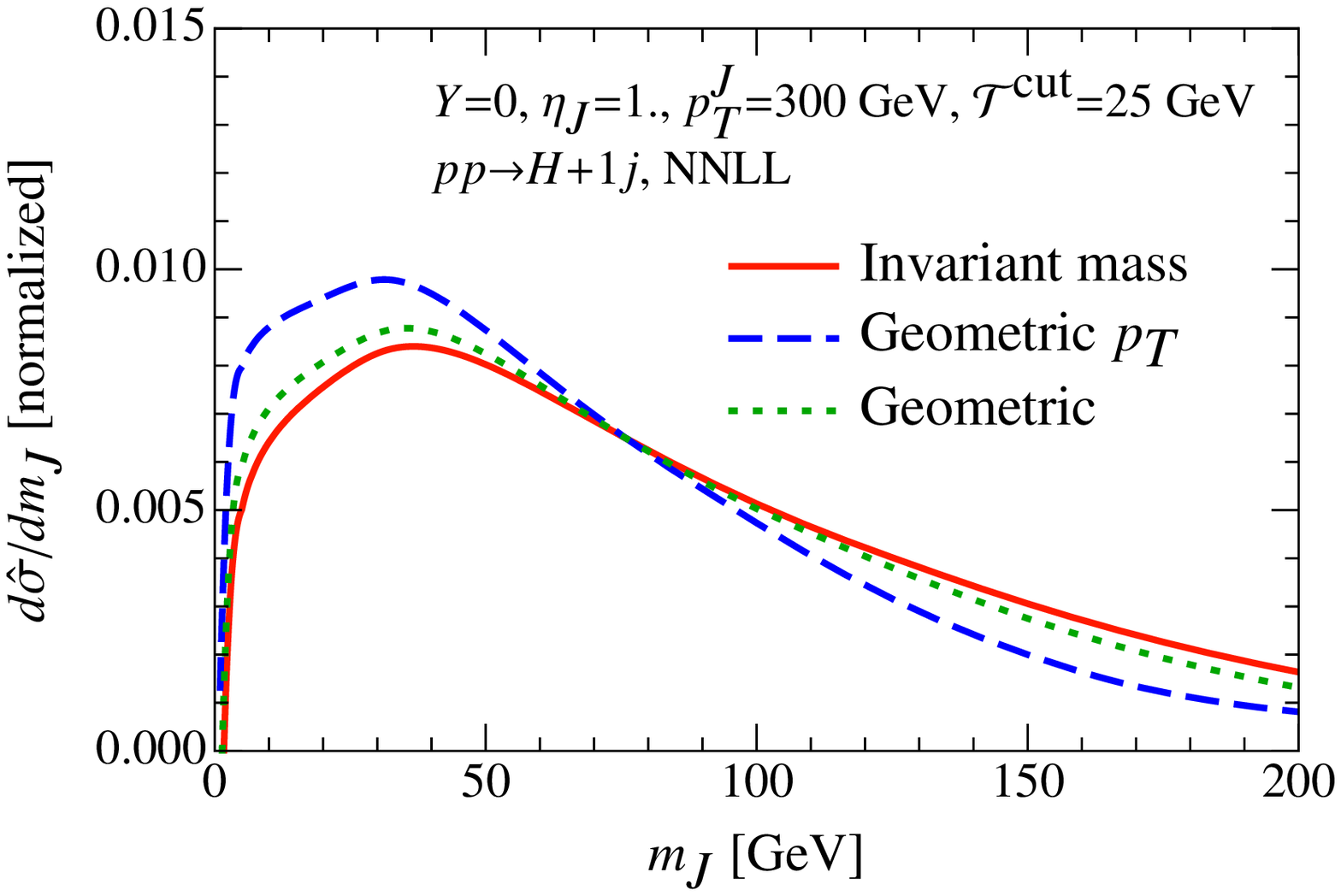}%
\caption{Dependence of the NNLL jet mass spectrum for $pp\to H+1$ jet on the $N$-jettiness measure used to define the jets.}
\vspace{-1ex}
\label{fig:jetdefinition}
\end{figure*}
%%%

%%%
\begin{figure}[t!]
 \includegraphics[width=\columnwidth]{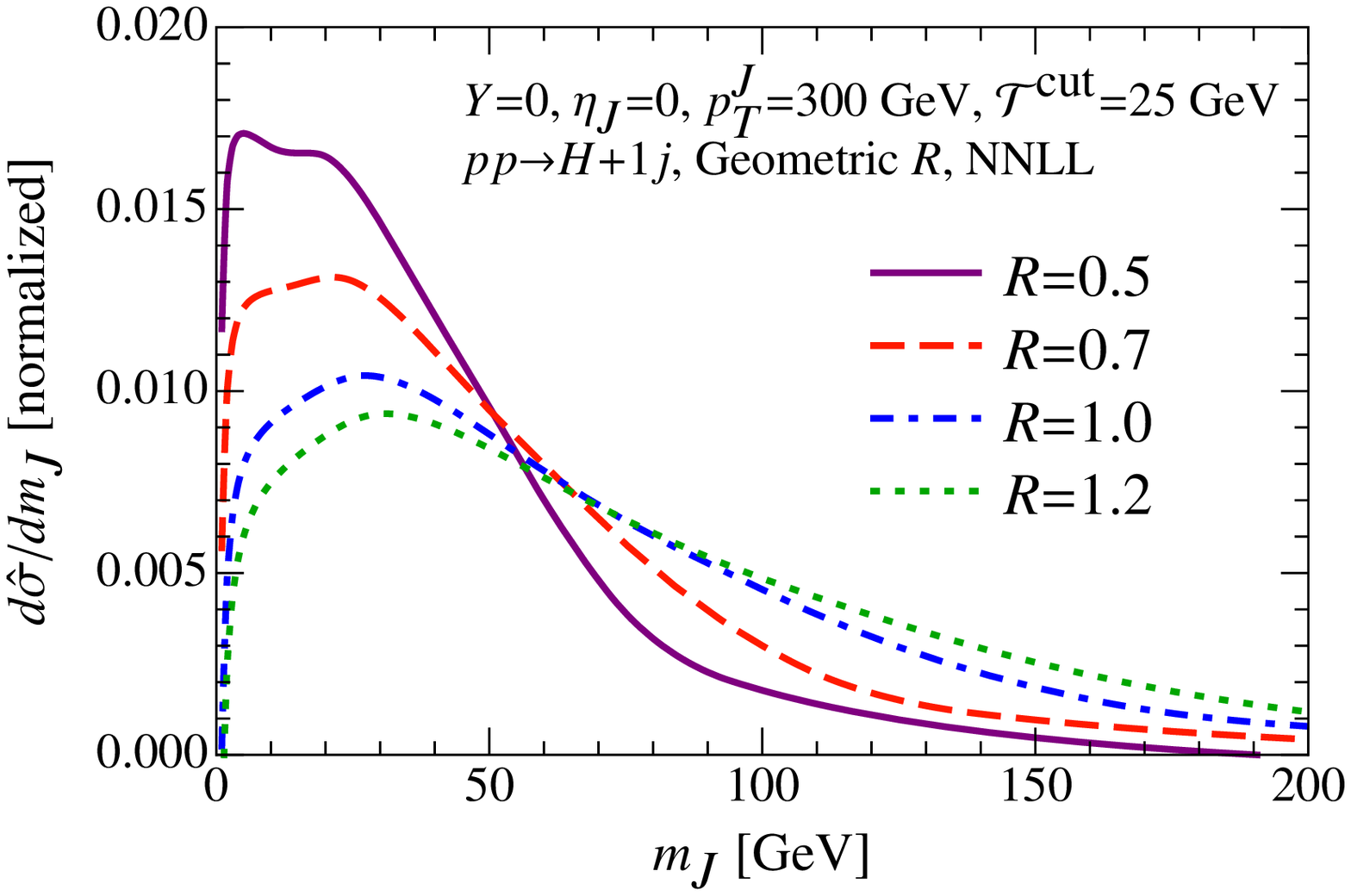}
 \caption{Dependence of the NNLL jet mass spectrum for the geometric $R$ measure
   on the jet radius $R$. Only the $R$ dependence from singular terms in the
   factorization formula is shown here.}
 \label{fig:jetr}
\end{figure}
%%%

In \sec{kin} we discussed the various $N$-jettiness measures (defined by the
$Q_i$) and illustrated the corresponding size and shape of the jet regions for
the geometric cases. An illustration of the more irregular regions that appear
for the invariant mass measure can be found in Ref.~\cite{Jouttenus:2011wh}.  We
now study how the jet mass spectrum is affected by these different jet
definitions as well as by their jet area ($R$ dependence). We start by noting
that in the $N$-jettiness factorization only the soft function is sensitive to
the boundaries of the jet regions.  Up to NLL the only jet algorithm dependence 
enters through the arguments of the logarithms, such as $\ln[m_J^2/(Q_J\pTJ)]$. 
More complicated dependence on the boundaries enters through the soft function 
starting at NLO, which appears in our NNLL results.  The nontrivial jet radius and jet algorithm 
dependence in the singular terms in the factorization theorem is formally enhanced for $m_J\ll \pTJ$ 
over the dependence on the jet algorithm and jet area in the power-suppressed nonsingular terms that 
are not part of \eq{sigmaTau1evo}.

In \fig{jetdefinition} we compare the invariant mass, geometric \geopT, and
geometric measures for three different kinematic configurations with $\eta_J = \{0, 0.5,
1\}$, $Y=0$, and $p_T^J = 300$ GeV.  We fix $\rho =0.834$ for the two geometric measures 
(which corresponds to $R=1$ for the geometric measure at $\eta_J=0$). When increasing $\eta_J$, 
all three measures show a mild decrease in the peak height and mild increase in the tail.  For 
$\eta_J=0$ the dependence of the jet mass on the jet definition is quite mild (for jets of similar area):
 the invariant mass measure is very similar to the geometric measures, and the two geometric 
measures agree exactly as we saw already in \fig{geopte}.  For more forward jet rapidities the 
two geometric measures start to progressively differ, with the geometric measure being closer to the 
invariant mass result.

In \fig{jetr} we show the jet mass spectrum for the geometric $R$ measure for
various values of the jet radius $R$. A smaller jet radius translates into a higher peak
and shorter tail.  (The small bump at the top of the $R=0.5$ peak is not significant
within our uncertainties.) Indeed, one of the most significant  effects on the jet mass spectra for
different values of $R$ is the fact that the size of the jet puts an effective upper boundary 
on its mass $m_J \lesssim p_T^J R/\sqrt{2}$.  At this boundary the jet mass spectrum has 
to fall off rapidly. This boundary is seen in \Pythia and LHC data and is incorporated into
our resummation by determining the point where we transition from the resummation region 
to the fixed-order region. Since this decreases the size of the tail of the jet mass spectrum
there must be a corresponding increase to the peak to ensure the result remains normalized.
Note that the precise form of the jet mass spectrum near $m_J \sim p_T^J R/\sqrt{2}$ is not
fully predicted by our calculation, because we have not yet incorporated the nonsingular
contributions to the cross section. These are important for making accurate predictions
in this part of the tail of the distribution, where their size is not fully captured by our
perturbative uncertainty estimates.

%%%%%%%%%%%%%%%%%%%%%%%%%%%%%%%%%%%%%%%%%%%%%%%%%%%%%%%%%%%%%%%%%%%%%%%%%%%%%%%%
\section{Monte Carlo Comparisons}
\label{sec:compare}
%%%%%%%%%%%%%%%%%%%%%%%%%%%%%%%%%%%%%%%%%%%%%%%%%%%%%%%%%%%%%%%%%%%%%%%%%%%%%%%%

In this section we study various aspects of the jet mass spectrum in \Pythia.
Although formally the perturbative accuracy of \Pythia is significantly lower
than that of our NNLL calculation, it is also well known that after sufficient
tuning \Pythia is able to reproduce the shape of many jet observables. Here we
are particularly interested in testing the impact on the jet mass spectrum from
using different hard processes, using different jet algorithms, and from adding
hadronization and underlying event (the latter being described by \Pythia's
multi-parton interaction model).  We also perform a comparison between our
calculation and \Pythia for the same geometric $R=1$ $N$-jettiness jets used in
our analysis. Finally we compare our exclusive 1-jet $m_J$ calculation with the
inclusive jet mass spectrum measured in $pp\to$ jets by ATLAS. We always use
\textsc{Pythia}8 with its default tune 5 (``Tune 4C''), which as we will see provides a
good description of the ATLAS jet mass data.

%===============================================================================
\subsection{Hard Process and Jet Algorithm Dependence in PYTHIA}
\label{subsec:pythia}
%===============================================================================

\begin{figure}
 \includegraphics[width=\columnwidth]{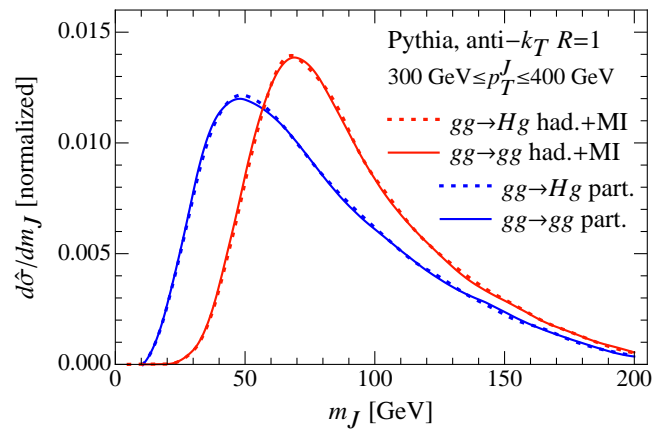}
 \caption{The gluon jet mass spectrum in \Pythia does not depend on the underlying
   hard process producing the jets. This is true both for partons (left peaks)
   and with hadronization and underlying event (right peaks).}
 \label{fig:pythia_gluon}
\end{figure}

We start by investigating to what extent the jet mass spectrum depends on the
underlying hard process in \Pythia. In \fig{pythia_gluon} we show the spectrum
for a gluon jet from $gg \to gg$ (solid) and from $gg \to Hg$ (dotted),
demonstrating that in \Pythia there is essentially negligible process dependence
for individual partonic channels. This is true both at the partonic level (blue
curves with peak on the left) and after including hadronization and multiple
interactions (red curves with peak on the right).  In reality one expects some
differences from the hard process due to the additional soft radiation produced
with more available colored particles, and from the different color flow, where
in particular $gg \to gg$ involves a matrix of color channels with nontrivial
interference. These effects may not be sufficiently described by \Pythia so one
should not conclude that the hard process dependence on the jet mass spectrum is
as small as is shown.

%%%
%%%

%%%
\begin{figure}
 \includegraphics[width=\columnwidth]{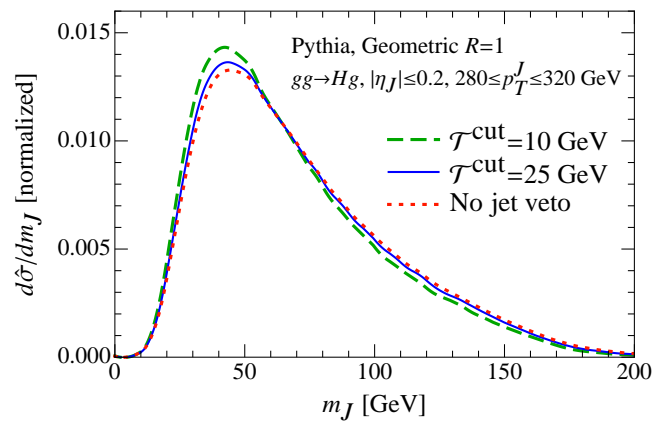}
 \caption{Comparison of the normalized jet mass spectra for exclusive and
   inclusive jet samples in \Pythia.}
 \label{fig:pythia_inex}
\end{figure}
%%%

%%%
\begin{figure}
 \includegraphics[width=\columnwidth]{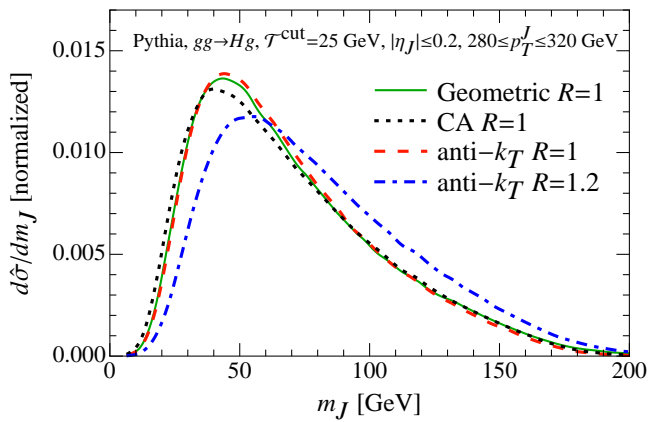}
 \caption{Comparison of the anti-$k_T$, CA, and geometric $R$ jet algorithms
   in \Pythia.}
 \label{fig:pythia_alg}
\end{figure}
%%%

Next, we look at the difference in \Pythia between the jet mass for exclusive and
inclusive jet production. We use the process $gg \to Hg$, imposing the jet veto
$\Tau^\cut = 10, 25$ GeV to obtain two exclusive samples, and using no jet veto
for our inclusive sample. The resulting normalized jet mass spectra are shown in
\fig{pythia_inex}. The difference between $\Tau^\cut=25$ GeV (our default value)
and the inclusive case is small, allowing our calculation to be compared to
inclusive spectra. The difference is slightly larger for $\Tau^\cut = 10$ GeV
and increases significantly for smaller values of $\Tau^\cut$. However, we will
not consider such strong jet vetos, as they lead to large NGLs (see
\subsec{veto_ngls}).

In \fig{pythia_alg} we compare the jet mass spectrum from \Pythia for
different jet algorithms, specifically our 1-jettiness $R=1$-algorithm,
Cambridge-Aachen with $R=1$, and anti-\kT with $R=1$ and
$R=1.2$~\cite{Cacciari:2011ma}. To stay close to a calculation for a single
phase space point, we restrict the jet to a narrow $p_T$ and rapidity bin, and
impose a veto using $\Tau^\cut=25$ GeV. The differences between the $R=1$ curves
are within the size of the uncertainty band from our NNLL calculation in the
same phase space bin. This result agrees with the small differences observed in
each of the panels of \fig{jetdefinition} from comparing different jet measures
for 1-jettiness jets.  The difference between $R=1$ and $R=1.2$ for anti-\kT is
a bit larger than that observed in our calculation using geometric $R$ jets in
\fig{jetr}.  In \Pythia the difference between $R=1$ and $R=1.2$ becomes smaller
when $\Tau^\cut$ is decreased, since with a stronger jet veto less additional
radiation is present that would be absorbed by larger jets. To be specific, the
15\% difference in the peak heights for anti-\kT with $R=1$ and $R=1.2$ for
$\Tau^\cut=25$ GeV reduces to 7\% for $\Tau^\cut =5$ GeV.

%===============================================================================
\subsection{Comparison of NNLL with  PYTHIA}
%===============================================================================

\begin{figure}
 \includegraphics[width=\columnwidth]{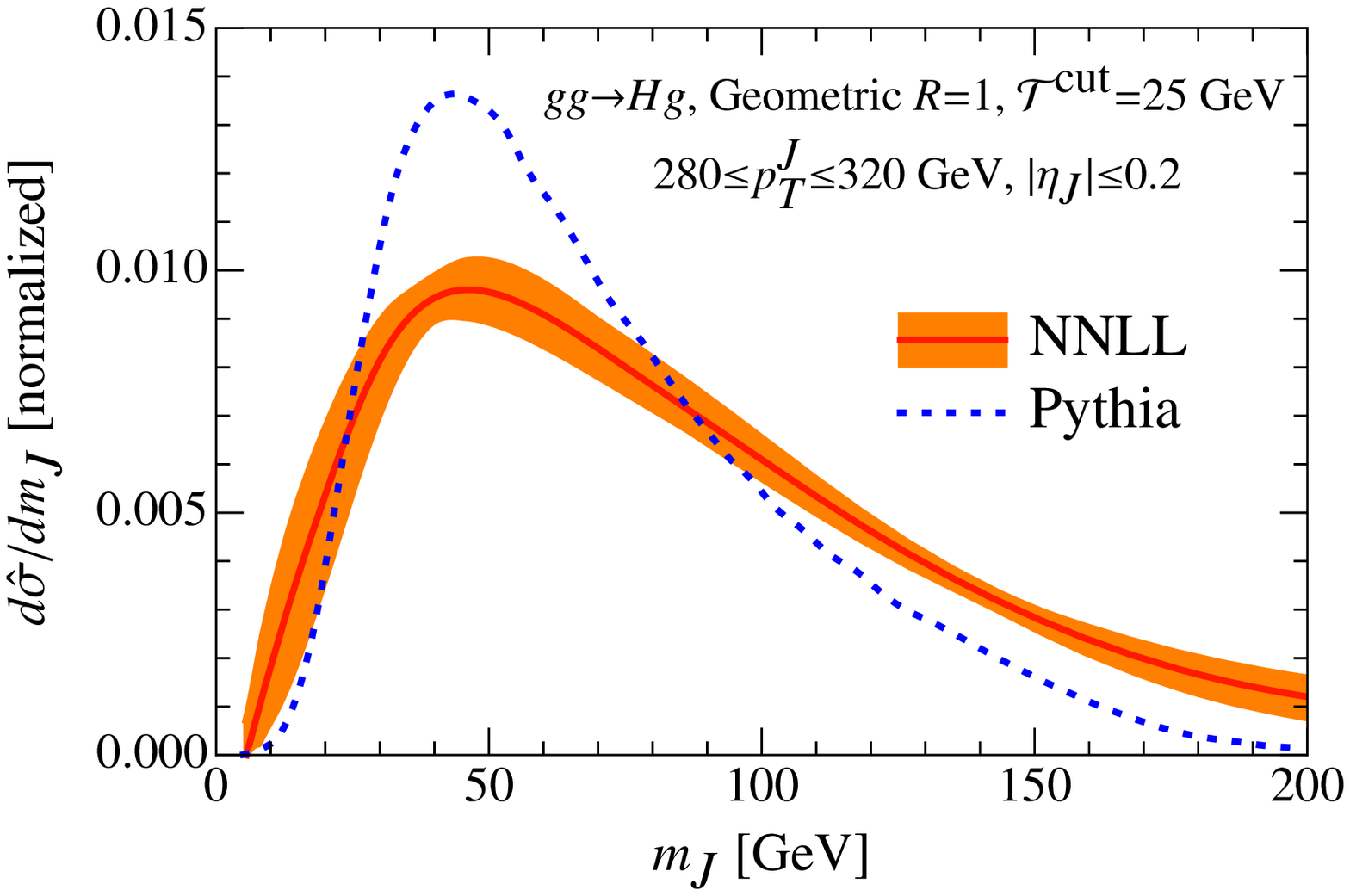} 
  \includegraphics[width=\columnwidth]{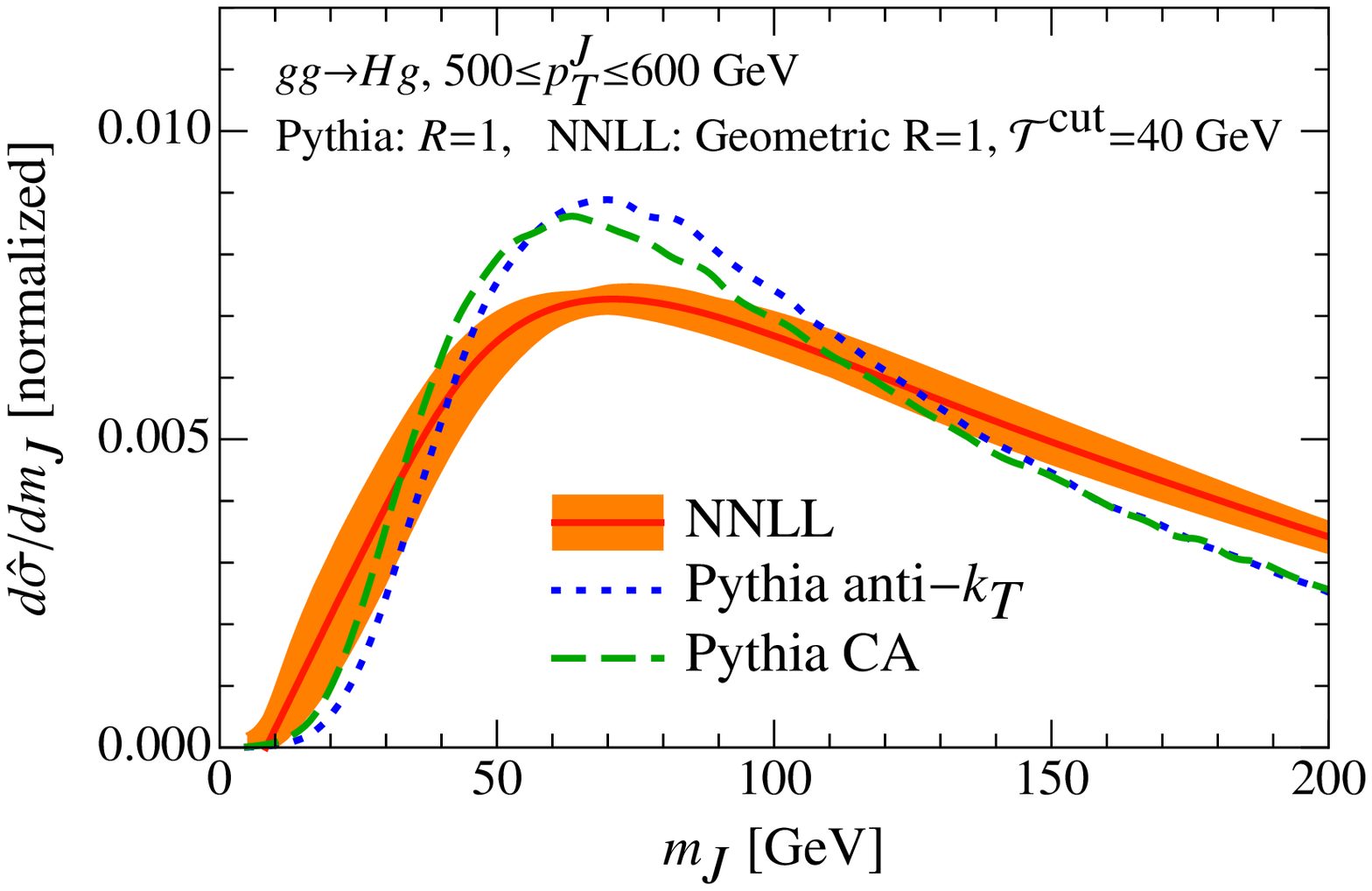} 
 \caption{Comparison between our NNLL calculation and partonic \Pythia for the
   $gg\to Hg$ channel. Both results use geometric $R=1$ jets and the same
   kinematic cuts.}
 \label{fig:pythia_nnll}
\end{figure}

A comparison between our NNLL calculation and partonic \Pythia results for
$gg\to Hg$ are shown in the two panels of \fig{pythia_nnll}.  

%%%
%%%

In the top panel of \fig{pythia_nnll} we show results for a 
narrow $\pTJ$ bin about $\pTJ=300\,{\rm GeV}$ and use the geometric $R=1$ jet 
definition for both \Pythia and the NNLL results. The peak positions in both cases agree very
well. To ensure that this is not an accident and that the peak position in
\Pythia does not depend on the PDF set used by our default tune, we checked that
an alternative tune (number 10, which is based on our default \Pythia tune but uses MSTW2008
LO PDFs) only shifts the peak by a small amount, similar to the small difference in
peak positions between \Pythia and our NNLL calculation.  However, as seen in
\fig{pythia_nnll}, the NNLL calculation has a somewhat lower peak and a correspondingly
higher tail.  Since the spectrum is normalized these two effects are related,
namely higher values in the tail must be compensated by a lower peak.  There are
several possibilities that may account for this difference.  Due to the
stability of our order-by-order results in \fig{convergenceg} it is unlikely to
be related to the lower order accuracy of \Pythia's LL parton shower
resummation. Most likely the differences are related to the fact that we have
not yet included nonsingular contributions to the spectrum which are important
in the fixed-order region, in particular for the spectrum to fall off rapidly enough.
Due to the fact that the results are normalized, this mismatch in the tail then
also leads to a disagreement of the peak heights. Thus we expect that the
inclusion of the nonsingular contributions will reduce this difference.  Note that an estimate for 
the size of these nonsingular terms is not included in our perturbative uncertainty bands.

In the bottom panel of \fig{pythia_nnll} we compare results at larger $\pTJ$ bin, $500\le \pTJ \le 600\,{\rm GeV}$, again normalizing
both the \Pythia and NNLL results over the same $m_J=0$--$200\,{\rm GeV}$ range. For a common
jet radius $R=1$ there is mild dependence on the jet algorithm as explored earlier, and we show the \Pythia results for 
anti-$k_T$ and CA.  Here there is an improved agreement between our NNLL results and \Pythia, with the largest effect again being the higher tail.

%===============================================================================
\subsection{Hadronization in PYTHIA}
%===============================================================================

%%%
\begin{figure}[t!]
 \includegraphics[width=\columnwidth]{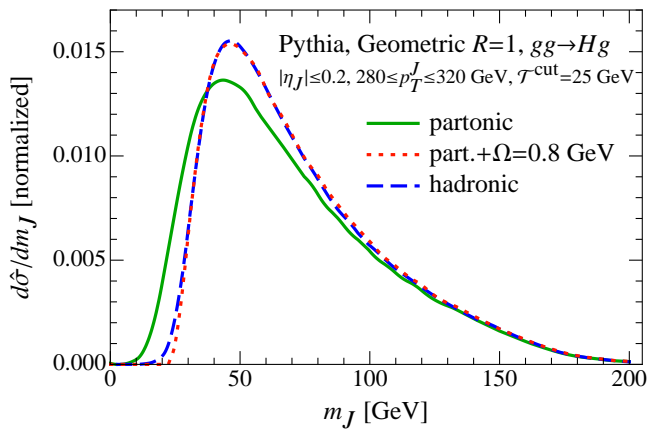}
 \caption{The nonperturbative hadronization correction in \Pythia is well described by a shift in $m_J^2$.}
 \label{fig:pythia_hadr}
\end{figure}
%%%

%%%
\begin{figure*}[t!]
\includegraphics[width=\columnwidth]{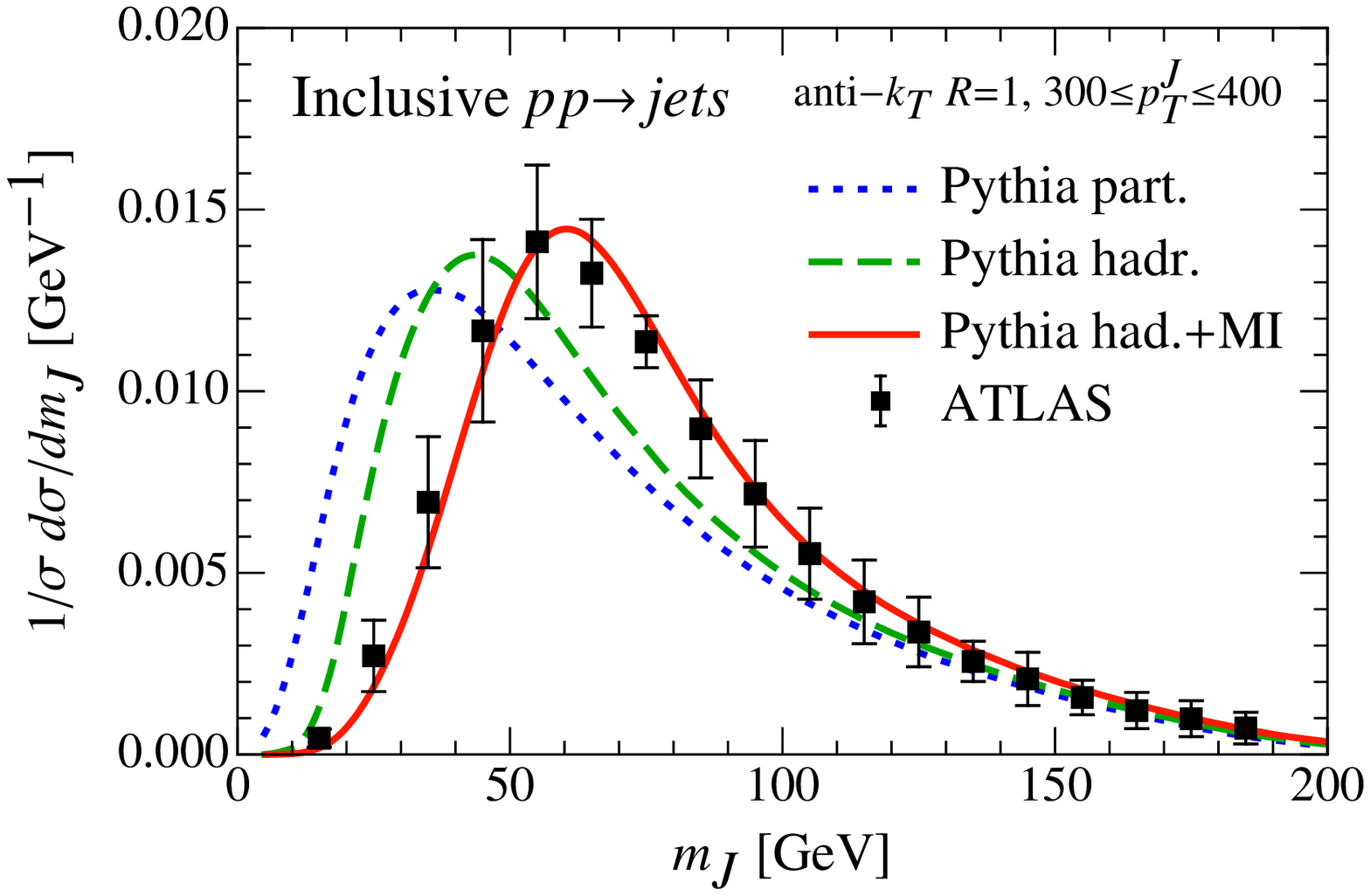}%
\hfill%
\includegraphics[width=\columnwidth]{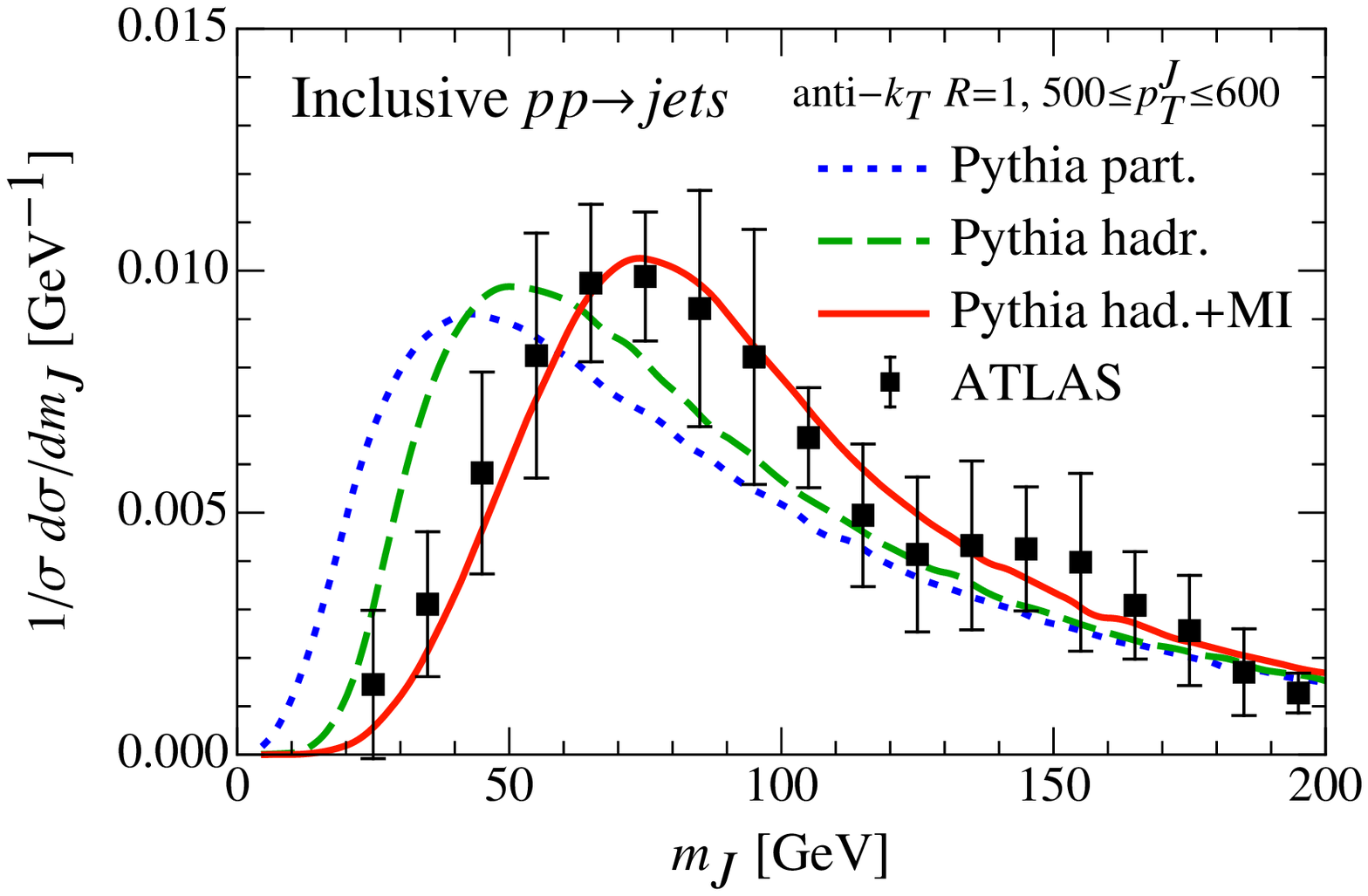}%
\caption{Comparison of the \Pythia jet mass spectrum for inclusive $pp\to $ jets to the corresponding ATLAS data~\cite{ATLAS:2012am}.
\Pythia results are shown at parton level (dotted), including hadronization (dashed), and including hadronization and multiple interactions (solid).
The final \Pythia results reproduce the data well.}
\label{fig:UE}
\end{figure*}
%%%

We now explore the effect of hadronization on the jet mass spectrum using
\Pythia.  In the factorization formula the hadronization is encoded through
nonperturbative corrections in the soft function $S$ at a scale $\sim
\Lambda_{\rm QCD}$, which must be separated from perturbative corrections at the
soft scale $\mu_S\sim m_J^2/\pTJ$.  For $e^+e^-\!\to2$ jets there is an analytic
understanding of the analogous nonperturbative corrections originating in
Refs.~\cite{Dokshitzer:1995qm, Dokshitzer:1995zt, Dokshitzer:1997iz,
  Salam:2001bd} as well as a modern understanding in terms of field theory
operators~\cite{Korchemsky:1999kt, Lee:2006fn, Hoang:2007vb, Mateu:2012nk}.  For
these processes, as soon as the relevant soft scale $\mu_S$ is perturbative, the
nonperturbative corrections can be power expanded in $\Lambda_{\rm QCD}/\mu_S$,
and the dominant power correction simply shifts the event shape distribution,
$e\to e-\Omega_e/Q$. In the case at hand, the nonperturbative soft function is
built from more than two Wilson lines, so the description of the power
corrections becomes more complicated. Nevertheless, for a given kinematic
configuration we still expect that the dominant effect will be described by a
shift involving a parameter $\Omega\sim\Lambda_{\rm QCD}$. For a jet mass $m_J^2 \simeq
p_J^+ p_J^-$ this shift occurs due to nonperturbative soft radiation causing a
shift in the small momentum $p_J^+$, so it takes the form
%%%
\begin{align} \label{eq:hadr} 
  m_J^2 \to m_J^2 - 2\W \, p_T^J\, R 
  \,.
\end{align}
The factor of $R$ accounts for the fact that there is a decreased amount of soft
momentum contamination in the jet for decreasing $R$~\cite{Dasgupta:2007wa}. It
is straightforward to test whether this shift agrees with the hadronization
model in \Pythia, by comparing the results with and without hadronization. As
demonstrated in \fig{pythia_hadr}, a shift with the choice $\W=0.8$ GeV works
very well, in reasonable agreement with the $\W=1.0$~GeV found earlier in
Ref.~\cite{Dasgupta:2012hg} for the inclusive $\ge 1$ jet cross section.

%%%
\begin{figure*}[t!]
\includegraphics[width=\columnwidth]{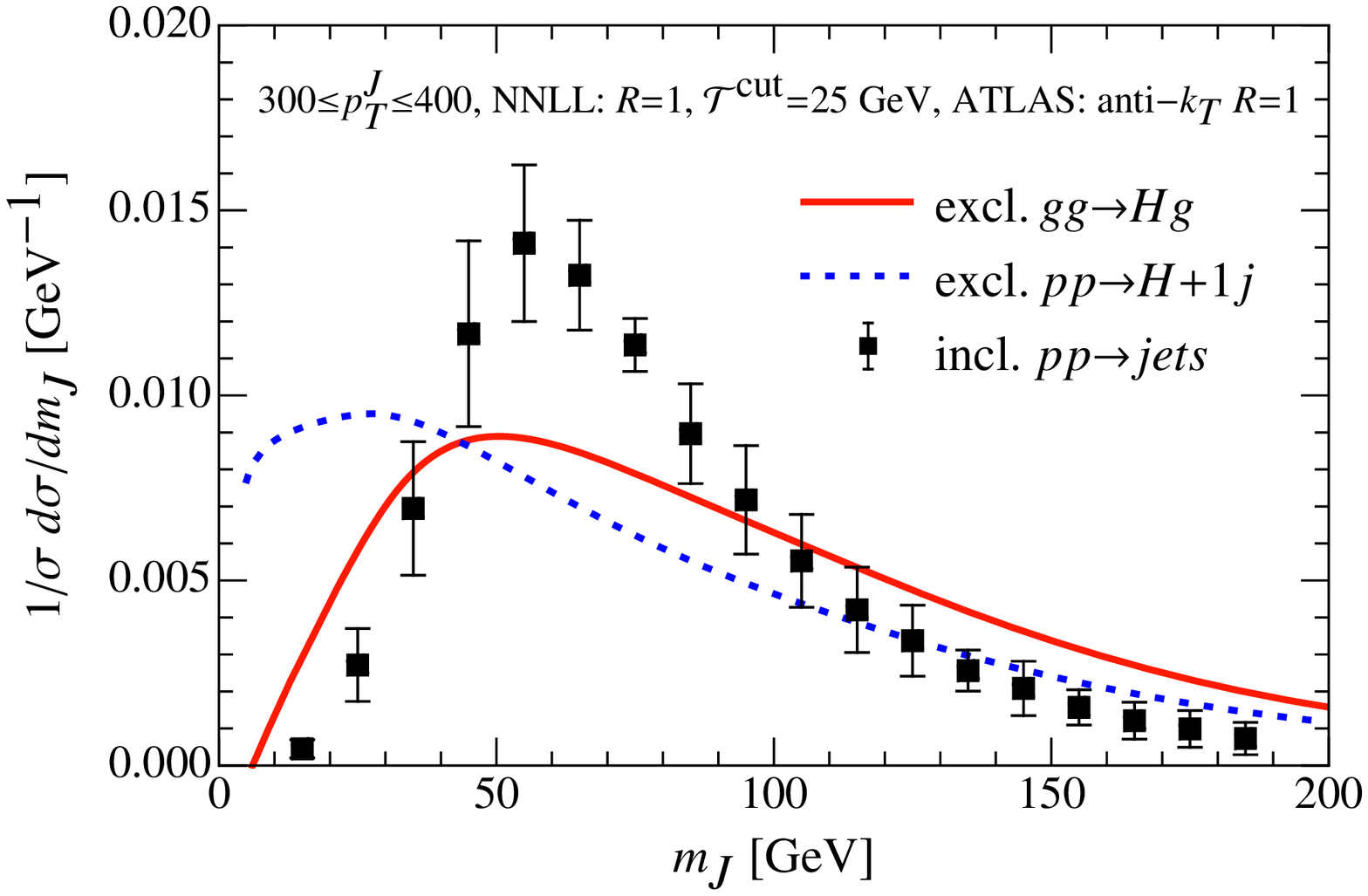}%
\hfill%
\includegraphics[width=\columnwidth]{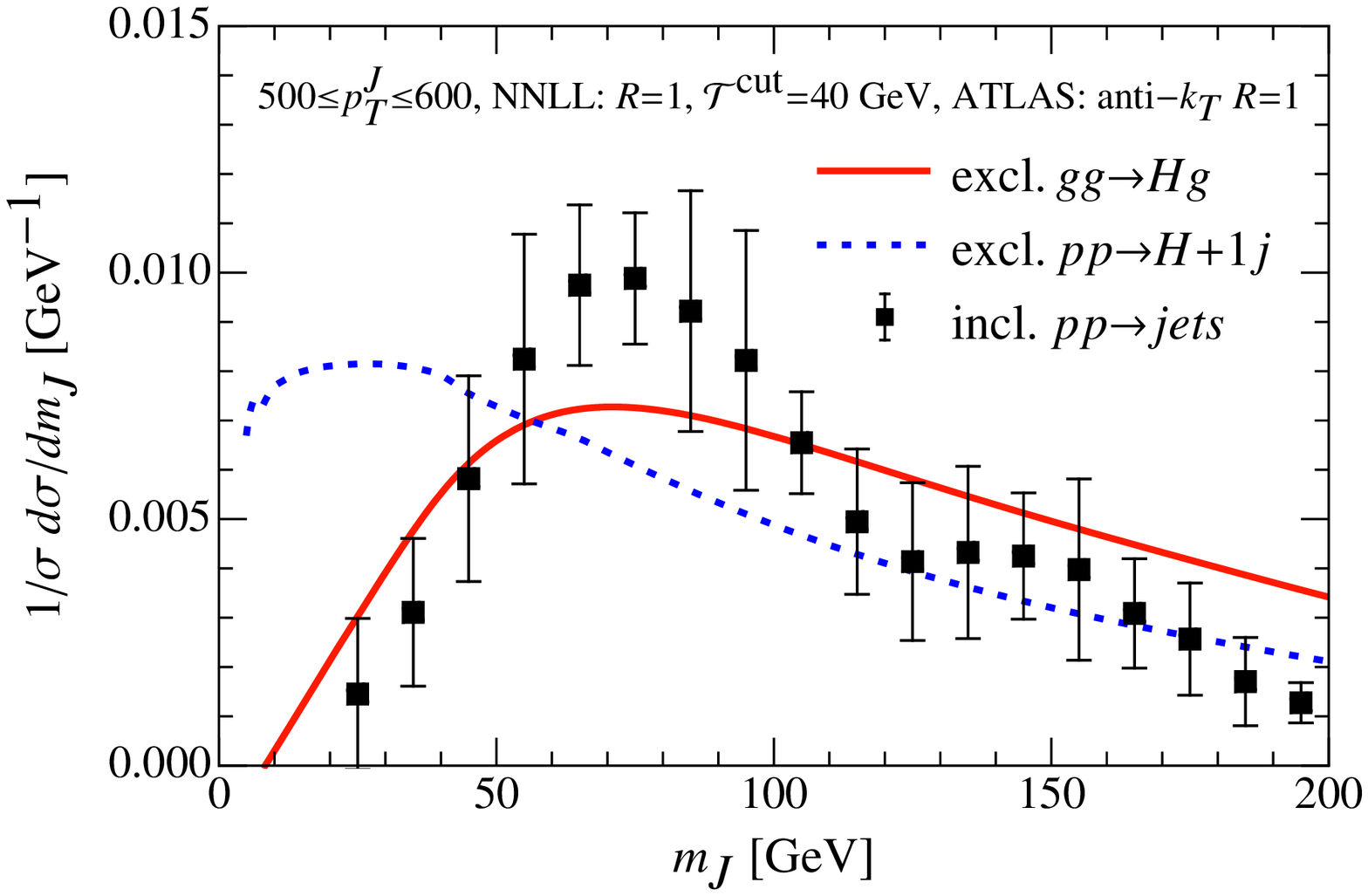}%
\caption{Comparison of our exclusive NNLL calculation with ATLAS inclusive jet
  mass data~\cite{ATLAS:2012am}.  The peak position of our gluon jets from
  $gg\to Hg$ agrees remarkably well with the inclusive dijet data. For the ATLAS
  date there is presumably a shift to lower values due to quark jets which is
  compensated by a shift to higher values due to hadronization and multiple
  interactions.}
 \label{fig:atlas}
\end{figure*}
%%%

%===============================================================================
\subsection{Underlying Event and ATLAS Data}
%===============================================================================

In \Pythia the effect of the underlying event is modeled by multiple partonic interactions,
and its effect on the jet mass spectrum is more pronounced than that of
hadronization. This is shown in \fig{UE} where we plot the jet mass spectrum for
inclusive $pp\to $ jets from \Pythia at parton level, including hadronization, and
including hadronization and multiple interactions. Also shown are the corresponding ATLAS data from
Ref.~\cite{ATLAS:2012am}, where the uncertainty bars are from linearly combining the
statistical and systematic uncertainties. This channel is dominated by the copious
$pp \to $ dijet production at the LHC. We use the same inputs and
cuts as ATLAS, namely $\Ecm=7\,{\rm TeV}$, anti-\kT jets with $R=1$,
$|\eta_J| \le 2$, and consider both $300\,{\rm GeV}\le \pTJ\le 400\,{\rm GeV}$ and
$500\,{\rm GeV}\le \pTJ\le 600\,{\rm GeV}$.
The shift to the peak location from
hadronization is of similar magnitude as that for $gg\to Hg$ in
\fig{pythia_hadr}, namely $\simeq 3.0\,{\rm GeV}$ for $gg\to Hg$ compared to
$\simeq 8.0\,{\rm GeV}$ for the $300\,{\rm GeV}\le \pTJ\le 400\,{\rm GeV}$
inclusive jets which have a slightly larger average $\pTJ$. For the inclusive
$pp\to $ jets in \Pythia the additional shift to the peak location from the
underlying event is $\simeq 17.4\,{\rm GeV}$. The final \Pythia results agree well
with the ATLAS data for both $\pTJ$ bins. In a NNLL calculation the effect of
hadronization and part of the effect of the underlying event will be captured by
corrections to the soft function, but it is not clear if hadronic corrections in
the multi-jet soft function will fully capture the effect of the underlying event.

Given that \Pythia agrees well with the ATLAS inclusive dijet spectrum, one
might wonder what the purpose of a higher-order NNLL dijet calculation
would be.  An advantage of our calculational framework
over \Pythia is that it follows from first principles and does not
involve the modeling and tuning present in \Pythia.  Specifically, the input to
our calculation is limited to $\al_s(m_Z)$, the parton distributions functions,
and simple soft function parameters like $\W$ for the hadronic effects.
Furthermore, we have a rigorous estimate of the higher-order perturbative
uncertainty from scale variation, as well as from order-by-order convergence,
which enable us to fully assess the reliability of the result. Finally, it should
be emphasized that our calculation is fully analytic (up to the numerical
convolution with the PDFs) and hence provides an analytic QCD calculation of an
LHC spectrum for jets.

To the extent that the normalized jet mass spectrum is independent of the hard
process and independent of using an inclusive or exclusive jet sample, which \Pythia
seems to suggest in \figs{pythia_gluon}{pythia_inex}, a comparison between jet
mass spectra involving different hard processes and with and without jet veto
cuts is appropriate. The approximate hard process independence only holds
separately for gluon or quark jets, which themselves have fairly different jet
mass spectra, see \fig{normuncert}. Therefore when varying the hard process we
expect the dominant change in the jet mass spectrum to be related to the process
dependent fraction of quark and gluon jets produced.

In \fig{atlas} we compare our NNLL result for $pp\to H+1$ jet and for $gg \to Hg$ to
the ATLAS data for $pp\to$ jets.  Recall that the peak location of the NNLL
$H+1$ jet calculation matches well with that from \Pythia, see
\fig{pythia_nnll}. Because of the significant contribution from quark jets the
$H+1$ jet spectrum peaks to the left of the spectrum from dijets. On the other
hand, the peak location with pure gluon jets ($gg \to Hg$) agrees quite
well with the data on dijets, particularly for the larger $\pTJ$ bin. From the results 
already obtained above, we expect only small differences (comparable to the ATLAS 
error bars) for effects related to the choice of the jet algorithm, the choice of 
inclusive versus exclusive jets, or the choice of looking at gluon jets in dijets or 
in Higgs production.  On the other hand there will be a more significant shift of 
the spectrum to the left from quark channels in the dijet production, and a shift to 
the right from adding hadronization and underlying event, neither of which is included 
in the solid red curve. The agreement between peak locations seems to indicate that
these two effects largely compensate for one another. Finally, there will be an
effect related to the fact that there are nontrivial color correlations in
$gg\to gg$ which are not present in $gg\to Hg$ (these effects are not apparent
in \Pythia, see \fig{pythia_gluon}).

One may also look at the peak heights in \fig{atlas}, for which the agreement is
not as good. As described earlier, this effect is related to the fact that we
have not yet included nonsingular corrections. These corrections are known to
decrease the tail to enable it to rapidly fall off by $m_J^2 \sim \pTJs R^2/2$,
and they also affect the peak directly through the normalization. Since with
additional work these can be included in future results the difference in peak
heights is not of too much concern.

Finally one may also compare the results in \figs{UE}{atlas} for the $300\,{\rm
  GeV}\le \pTJ\le 400\,{\rm GeV}$ and $500\,{\rm GeV}\le \pTJ\le 600\,{\rm GeV}$
bins. For dijets the peak location moves to higher $m_J$ with increased $\pTJ$, unlike for
$pp\to H+1$ jet, again indicating that gluon jets likely dominate. The
conclusions from the comparison with \Pythia and the contrast to our NNLL
calculation remains the same for these two ranges of $\pTJ$.

%%%%%%%%%%%%%%%%%%%%%%%%%%%%%%%%%%%%%%%%%%%%%%%%%%%%%%%%%%%%%%%%%%%%%%%%%%%%%%%%
\section{Conclusions}
\label{sec:conc}
%%%%%%%%%%%%%%%%%%%%%%%%%%%%%%%%%%%%%%%%%%%%%%%%%%%%%%%%%%%%%%%%%%%%%%%%%%%%%%%%

In this paper we calculated the jet mass spectrum for $pp\to H+1$ jet to NNLL order.
For this exclusive 1-jet cross section we veto additional jets with the 1-jettiness event shape,
and used the 1-jettiness factorization formula in terms of hard, beam, jet, and
soft functions to obtain our results.  For the \emph{normalized} jet mass
spectrum the remaining higher-order perturbative uncertainties from scale
variation are at the $\simeq 6-14\%$ level at NNLL order, being on the smaller side of this range 
for gluon jets. In addition, our results exhibit excellent order-by-order convergence.

The normalized NNLL spectrum is quite insensitive to the jet veto over a wide
range of values, even when accounting for non-global logarithms. Thus in our
framework non-global logarithms can be accurately treated as fixed-order
contributions to the soft function, upon which additional global logarithms are
automatically resummed in the factorization framework. An essential ingredient
in the resummation of the global logarithms was the refactorization of the soft
function, which we demonstrate is required to avoid introducing spurious leading
logarithms in certain regions of phase space. Our treatment of the NNLL
exclusive cross section with a jet veto has significantly smaller non-global
logarithmic terms when compared to the size of these terms observed in the
earlier inclusive NLL analysis in Ref.~\cite{Banfi:2010pa}, and the earlier
inclusive partial NNLL analysis in Ref.~\cite{Kelley:2012kj}. Finally, we note
that in \Pythia the inclusive jet mass spectrum and the exclusive jet mass
spectrum with our default jet veto are essentially identical.

Utilizing our calculation we investigated the dependence of the jet mass
spectrum on various parameters of the exclusive jet cross section. Part of the
power of our framework is that the factorization formula is fully differential
in the jet kinematics ($\pTJ$, $\eta_J$, and $Y$), allowing us to vary the
definition of the jets and the jet area, and can be easily separated into quark
jet and gluon jet channels. As expected we find that the spectrum peaks at
larger $m_J$ values for gluon jets than for quark jets. For a given partonic
channel the factorization framework predicts little sensitivity to the
underlying hard process, and this result is also found to be the case in
\Pythia. The main process dependence is therefore the relative mix of quark and
gluon jets.  The peak of our NNLL $m_J$ spectrum moves to the right for larger
$\pTJ$ and for larger $|\eta_J|$, but more so for the individual partonic
channels than for $pp\to H+$1 jet, where the change to the mix of quarks and
gluons provides a compensating effect. The complete description of the various
kinematic variables also makes it trivial to implement rapidity cuts.  For a bin
$|\eta_J|<2$ and a not too large bin in $\pTJ$, we find that the integrated NNLL
result is very consistent with the NNLL result for fixed kinematic variables
taken at the center of the bin.

Varying the jet definition with fixed jet area leads to small changes in
the jet mass spectrum, both for various jet definitions in our NNLL result and
for anti-$k_T$, CA, and geometric-$R$ jets in \Pythia. This suggests that there
are only small differences between the spectrum for 1-jettiness jets and
traditional jet algorithms. On the other hand, both \Pythia and our NNLL results exhibit a larger
dependence on the jet radius $R$.   In the fixed-order region near the jet boundary
$m_J \sim \pTJ R/\sqrt{2}$ there  are nonsingular terms that become important that have 
not been included in our analysis here. The absence of these terms likely leads to
a larger tail in our NNLL spectrum than in \Pythia, and correspondingly a
smaller peak height in the normalized NNLL result. On the other hand, the peak
location agrees very well between our NNLL calculation and \Pythia. An analysis of
these additional nonsingular terms will be carried out in the future.

We investigated the dependence of the jet mass spectrum on hadronization and
underlying event using \Pythia. Hadronization is very well described by a shift
to the mass spectrum, $m_J^2\to m_J^2 - (2R\pTJ) \Omega $ with
$\Omega\sim\Lambda_{\rm QCD}$, which is the anticipated result from
nonperturbative soft gluon contributions in our factorization formula's soft
function. In \Pythia the underlying event is modeled by multiple partonic interactions
and leads to a somewhat larger shift to the spectrum than for hadronization. It
plays an important role in obtaining agreement with the ATLAS jet mass
results for inclusive dijets. Comparing our results to ATLAS we find that the
NNLL $pp\to H+$1 jet spectrum peaks to the left of the dijet data, whereas the
NNLL $gg\to Hg$ spectrum peaks in the same location.  The comparison made so far
with the ATLAS data is promising. The extension of our NNLL calculation to
$pp\to $ dijets is completely feasible using 2-jettiness, and it will be interesting
to see to what extent the contributions from quark channels, color mixing, and
hadronization and underlying event will affect this comparison with the data.
Theoretically, the only remaining challenge to a complete comparison appears to be
incorporating the effect of the underlying event from first principles rather than
relying on its modeling via Monte Carlo.

%%%%%%%%%%%%%%%%%%%%%%%%%%%%%%%%%%%%%%%%%%%%%
\vspace{-1ex}
\begin{acknowledgments}
\vspace{-1ex}
This work was supported in part by the Office of Nuclear Physics of the U.S.
Department of Energy under Grants No.  DE-FG02-94ER40818 and No. DE-FG02-90ER40546,
and by the DFG Emmy-Noether grant TA 867/1-1.  T.J.  was also supported by a 
LHC-TI grant under the NSF grant PHY-0705682.
\end{acknowledgments}

%%%%%%%%%%%%%%%%%%%%%%%%%%%%%%%%%%%%%%%%%%%%%

\appendix

%%%%%%%%%%%%%%%%%%%%%%%%%%%%%%%%%%%%%%%%%%%%%%%%%%%%%%%%%%%%%%%%%%%%%%%%%%%%%%%%
\section{Perturbative Inputs}
\label{app:inputs}
%%%%%%%%%%%%%%%%%%%%%%%%%%%%%%%%%%%%%%%%%%%%%%%%%%%%%%%%%%%%%%%%%%%%%%%%%%%%%%%%

In this section we collect the fixed-order ingredients and evolution kernels for evaluating the jet mass cross section for $pp \to H+1j$ in \eqs{sigmaTau1}{sigmaTau1evo} at NNLL order. We first give expressions for the hard, jet, beam and soft functions at next-to-leading order. This is followed by the evolution kernels and the coefficients that they depend on.

%===============================================================================
\subsection{Hard Function}
\label{app:hard}
%===============================================================================

The hard functions $H_\kappa$ for the various partonic channels $\kappa$ that contribute to $pp \to H+1$ jet can be obtained from the finite part of the helicity amplitudes $A$ determined in Ref.~\cite{Schmidt:1997wr} following the procedure of Ref.~\cite{Stewart:2012yh},
\begin{widetext}
%%%
\begin{align} \label{eq:hard}
H_{ggg}(\{q_i^\mu\}, \mu_H) 
&= \frac{16\al_s(\mu_H)^3 C_A^2 C_F}{9\pi v^2} \frac{1}{[2(N_c^2-1)]^2}
\Big[|A(1_g^+,2_g^+,3_g^+;4_H)|^2
+|A(1_g^+,2_g^+,3_g^-;4_H)|^2
\nn \\ & \quad
+|A(1_g^+,3_g^+,2_g^-;4_H)|^2
+|A(3_g^+,2_g^+,1_g^-;4_H)|^2\Big]
\,,\nn \\
H_{g\bq\bq}(\{q_i^\mu\}, \mu_H) 
&= \frac{8\al_s(\mu_H)^3 C_A C_F}{9\pi v^2} \frac{1}{2N_c} \frac{1}{2(N_c^2-1)}
\Big[|A(1_g^+,2_q^+,3_\bq^-;4_H)|^2
+|A(1_g^-,2_q^+,3_\bq^-;4_H)|^2 \Big]
\,,\nn \\
H_{\bar q g\bar q}(\{q_i^\mu\}, \mu_H) 
&= \frac{8\al_s(\mu_H)^3 C_A C_F}{9\pi v^2} \frac{1}{2N_c} \frac{1}{2(N_c^2-1)}
\Big[|A(2_g^+,1_q^+,3_\bq^-;4_H)|^2
+|A(2_g^-,1_q^+,3_\bq^-;4_H)|^2 \Big]
\,,\nn\\
H_{q\bq g}(\{q_i^\mu\}, \mu_H) 
&= \frac{8\al_s(\mu_H)^3 C_A C_F}{9\pi v^2} \frac{1}{(2N_c)^2}
\Big[|A(3_g^+,2_q^+,1_\bq^-;4_H)|^2
+|A(3_g^-,2_q^+,1_\bq^-;4_H)|^2 \Big]
\,,\nn\\
H_{gqq}(\{q_i^\mu\}, \mu_H) &= H_{g\bq\bq}(\{q_i^\mu\}, \mu_H)
\,,\qquad
H_{qgq}(\{q_i^\mu\}, \mu_H) = H_{\bar q g\bar q}(\{q_i^\mu\}, \mu_H) 
 \,,\nn\\*
H_{\bar q qg}(\{q_i^\mu\}, \mu_H) &=H_{q\bq g}(\{q_i^\mu\}, \mu_H)
 \,.
\end{align}
%%%
The factors of $1/(2N_c)$ and $1/[2(N_c^2-1)]$ arise from averaging over the
spins and colors of the colliding quarks and gluons. The arguments of a helicity
amplitude $A$ have the form $i_t^h$, where $i$ denotes the momentum $q_i^\mu$,
$t$ denotes the parton type, and $h$ denotes the helicity of this particle.
\emph{Only in the helicity amplitudes} will we use an outgoing convention for
all these quantities, to make crossing symmetry direct. This implies that if we
want to convert to the convention used in the main text, then the $s_{ij}$'s in
the helicity amplitudes below will pick up additional minus signs if one of the
particles $i$ and $j$ is in and the other is out. The amplitudes that enter in
\eq{hard} are given by
%%%
\begin{align} \label{eq:ampl}
A(1_g^+,2_g^+,3_g^+;4_H)
&= \frac{m_H^4}{\sqrt{2|s_{12} s_{13} s_{23}|}}  \biggl\{
1 \!+\! \frac{\al_s(\mu_H)}{4\pi}\Big[
f(s_{12}, s_{13}, s_{23}, m_H^2, \mu_H)
\!+\! \frac{1}{3}(C_A \!-\! 2 T_F n_f)\,\frac{s_{12} s_{13}\!+\! s_{12} s_{23}\!+\! s_{13} s_{23}}{m_H^4} \Big]\biggr\}
\,,\nn\\
A(1_g^+,2_g^+,3_g^-;4_H)
&= \frac{s_{12}^2}{\sqrt{2|s_{12} s_{13} s_{23}|}}
\biggl\{
1 + \frac{\al_s(\mu_H)}{4\pi}\Big[
f(s_{12}, s_{13}, s_{23}, m_H^2, \mu_H)
+ \frac{1}{3}(C_A - 2 T_F n_f)\,\frac{s_{13} s_{23}}{s_{12}^2} \Big]\biggr\}
\,,\nn\\
A(1_g^+;2_q^+,3_\bq^-;4_H)
&= \frac{s_{12}}{\sqrt{2|s_{23}|}}
\biggl\{
1 + \frac{\al_s(\mu_H)}{4\pi}\Big[
   g(s_{12}, s_{13}, s_{23}, m_H^2, \mu_H) + (C_F - C_A)\, \frac{s_{23}}{s_{12}} \Big]\biggr\}
\,,\nn\\
A(1_g^-;2_q^+,3_\bq^-;4_H)
&= \frac{s_{13}}{\sqrt{2|s_{23}|}}
\biggl\{
1 + \frac{\al_s(\mu_H)}{4\pi}\Big[
   g(s_{12}, s_{13}, s_{23}, m_H^2, \mu_H) + (C_F - C_A)\, \frac{s_{23}}{s_{13}} \Big]\biggr\}
\,,\nn\\
f(s_{12}, s_{13}, s_{23}, m_H^2, \mu_H)
&= -C_A \biggl[\frac{1}{2} (L_{12}^2+L_{13}^2+L_{23}^2)
 + L_{12/H} L_{13/H} + L_{12/H} L_{23/H} + L_{13/H} L_{23/H}
\nn \\ & \quad
+ 2\Li_2\Bigl(1-\frac{s_{12}}{m_H^2}\Bigr) + 2\Li_2\Bigl(1-\frac{s_{13}}{m_H^2}\Bigr)
+ 2\Li_2\Bigl(1-\frac{s_{23}}{m_H^2}\Bigr) - 5 - \frac{3\pi^2}{4} \biggr] - 3 C_F
\,,\nn\\
g(s_{12}, s_{13}, s_{23}, m_H^2, \mu_H)
&= C_A\biggl[-\frac{1}{2}(L_{12}^2 + L_{13}^2 - L_{23}^2) + L_{12/H} L_{13/H}
  - (L_{12/H} + L_{13/H}) L_{23/H}
  -2\Li_2\Bigl(1-\frac{s_{23}}{m_H^2}\Bigr)
\nn \\ & \quad
  + \frac{22}{3} +\frac{\pi^2}{4} \biggr]
+ C_F\biggl[ -L_{23}^2 + 3L_{23} - 2 L_{12/H} L_{13/H}
  - 2\Li_2\Bigl(1-\frac{s_{12}}{m_H^2}\Bigr) - 2\Li_2\Bigl(1-\frac{s_{13}}{m_H^2}\Bigr)
\nn \\ & \quad
  - 11 + \frac{\pi^2}{2} \biggr]
+ \beta_0 \Bigl(-L_{23} + \frac{5}{3}\Bigr)
\,.\end{align}
\end{widetext}
%%%
Here we use the shorthand notation
%%%
\begin{align}
L_{ij} &= \ln\Bigl(-\frac{s_{ij}}{\mu_H^2} - \img 0 \Bigr)
\,, \nn \\
L_{ij/H} &= \ln\Bigl(-\frac{s_{ij}}{\mu_H^2} - \img 0 \Bigr) - \ln\Bigl(-\frac{m_H^2}{\mu_H^2} - \img 0 \Bigr)
\,.\end{align}
%%%
Explicit values for the $s_{ij}$ follow once we identify $q_i$ and $q_j$ as corresponding to the jet or a beam region. In particular here
\begin{align} \label{eq:sij}
  s_{ab} &= Q^2 \,, 
 & s_{a1} &= -Q\, \pTJ\, e^{Y-\eta_J} \,,
 & s_{b1} &= -Q\, \pTJ\, e^{\eta_J-Y} \,.
\end{align}
In contrast, the convention used in the main text is $s_{ij}>0$.

%===============================================================================
\subsection{Jet Functions}
\label{app:jet}
%===============================================================================

The one-loop jet functions are given by~\cite{Bauer:2003pi, Fleming:2003gt, Becher:2009th}
%%%
\begin{align} \label{eq:jetf}
   J_q(s,\mu_J) &= \de(s) + \frac{\al_s(\mu_J) C_F}{2\pi} \Big[ \frac{2}{\mu_J^2} \cL_1\Big(\frac{s}{\mu_J^2}\Big)
   \nn \\
  & \quad
    - \frac{3}{2 \mu_J^2} \cL_0\Big(\frac{s}{\mu_J^2}\Big) - \Big(\frac{\pi^2}{2} - \frac{7}{2} \Big) \de(s) \Big]
 \,, \nn \\
   J_g(s,\mu_J) &= \de(s) + \frac{\al_s(\mu_J)}{2\pi} \Big\{\frac{2 C_A}{\mu_J^2} \cL_1\Big(\frac{s}{\mu_J^2}\Big) - \frac{\bt_0}{2\mu_J^2} \cL_0\Big(\frac{s}{\mu_J^2}\Big) 
   \nn \\ 
  & \quad   
   + \Big[\Big(\frac{2}{3} - \frac{\pi^2}{2} \Big)C_A + \frac{5}{6} \bt_0 \Big] \de(s)  \Big\}
  \,,
\end{align}
%%%
where the plus distributions $\cL_n$ are defined as
%%%
\begin{align} \label{eq:Ln}
\cL_n(x)
&\equiv \biggl[ \frac{\theta(x) \ln^n x}{x}\biggr]_+
   \\* &
 = \lim_{\beta \to 0} \biggl[
  \frac{\theta(x- \beta)\ln^n x}{x} +
  \delta(x- \beta) \, \frac{\ln^{n+1}\!\beta}{n+1} \biggr]
\,. \nn
\end{align}
%%%
The ${\cal L}_n(x)$ integrate to zero if the range in $x$ is $[0,1]$.

%===============================================================================
\subsection{Beam Functions}
\label{app:beam}
%===============================================================================

The beam functions can be expressed in terms of standard gluon and quark PDFs
using an operator product expansion~\cite{Fleming:2006cd,Stewart:2009yx},
%%%
\begin{align} \label{eq:B_OPE}
B_i(t,x,\mu_B)
&= \sum_{j = \{g,q,\bar{q}\}} \int_x^1 \! \frac{\df \xi}{\xi}\, \cI_{ij}\Bigl(t,\frac{x}{\xi},\mu_B \Bigr) f_j(\xi, \mu_B)
\nn \\ & \quad \times
  \biggl[1 + \ORd{\frac{\lqcd^2}{t}}\biggr]
\,.\end{align}
%%%
The one-loop matching coefficients are~\cite{Stewart:2010qs,Berger:2010xi} are
%%%
\begin{widetext}
\begin{align} \label{eq:Iij}
\cI_{qq}(t,z,\mu_B)
&= \delta(t)\, \delta(1 - z)
  + \frac{\alpha_s(\mu_B)C_F}{2\pi}\, \theta(z) \biggl\{
  \frac{2}{\mu_B^2} \cL_1\Bigl(\frac{t}{\mu_B^2}\Bigr) \delta(1 - z) +
  \frac{1}{\mu_B^2} \cL_0\Bigl(\frac{t}{\mu_B^2}\Bigr) P_{qq}(z)
  \nn \\ & \quad
  + \delta(t) \biggl[
  \cL_1(1 - z)(1 + z^2) - P_{qq}(z) \ln z
  -\frac{\pi^2}{6}\, \delta(1 - z)
  + \theta(1 - z)(1 - z) \biggr]
  \biggr\}
  \,, \nn\\\nn
\cI_{qg}(t,z,\mu_B)
 &= \frac{ \alpha_s(\mu_B) T_F }{2\pi}\, \theta(z) \biggl\{
\frac{1}{\mu_B^2} \cL_0\Bigl(\frac{t}{\mu_B^2}\Bigr) P_{qg}(z)
 + \delta(t) \biggl[P_{qg}(z)\Bigl(\ln\frac{1-z}{z} - 1\Bigr) +  \theta(1-z) \biggr]
\biggr\}
\nn \\
\cI_{gg}(t,z,\mu_B)
&= \delta(t)\,\delta(1-z)
   + \frac{\alpha_s(\mu_B)C_A}{2\pi}\, \theta(z)
  \biggl\{ \frac{2}{\mu_B^2} \cL_1\Bigl(\frac{t}{\mu_B^2}\Bigr) \delta(1-z)
  + \frac{1}{\mu_B^2} \cL_0\Bigl(\frac{t}{\mu_B^2}\Bigr) P_{gg}(z)
  \nn \\ & \quad  
  + \delta(t)\, \Big[\cL_1(1-z) \frac{2(1-z + z^2)^2}{z} - P_{gg}(z) \ln z - \frac{\pi^2}{6} \delta(1-z)\Big] \biggr\}
\,, \nn \\
\cI_{gq}(t,z,\mu_B)
&= \frac{\alpha_s(\mu_B)C_F}{2\pi}\, \theta(z)
   \biggl\{\frac{1}{\mu_B^2} \cL_0\Bigl(\frac{t}{\mu_B^2}\Bigr) P_{gq}(z)
   + \delta(t)\, \Big[P_{gq}(z)\ln \frac{1-z}{z} + \theta(1-z) z\Big]\biggr\}
\,.\end{align}
\end{widetext}
%%%
The splitting functions in this equation are defined as
%%%
\begin{align} \label{eq:P_def}
P_{qq}(z)
&= \cL_0(1-z)(1+z^2)
\,,\nn\\
P_{qg}(z) &= \theta(1-z)\bigl[(1-z)^2+ z^2\bigr]
\,,\nn \\
P_{gg}(z)
&= 2 \cL_0(1-z)z + 2\theta(1-z)\Bigl[\frac{1-z}{z} +  z(1-z)\Bigr]
\,,\nn\\
P_{gq}(z) &= \theta(1-z)\, \frac{1+(1-z)^2}{z}
\,.\end{align}
%%%

%===============================================================================
\subsection{Factorized Soft Function}
\label{app:soft}
%===============================================================================

We now give expressions for the $N$-jettiness soft function, showing explicitly how the factorization in \eq{nlofact} is implemented. We remind the reader that there is some freedom in this refactorization, and that the corresponding uncertainty is probed by varying the parameter $r$ in \eq{softmix}.

Up to NLO the $1$-jettiness soft function is given by
%%%
\begin{align} 
S_\kappa(\{k_i\}, \{\mu_{S_i}\})
= \prod_{i=a,b,J} S_i(k_i,\{\hat q_i^\mu\},\mu_{S_i})
  + \ord{\al_s^2} 
\,.\end{align}
%%%
\begin{widetext}
From the NLO calculation in Ref.~\cite{Jouttenus:2011wh} we obtain
%%%
\begin{align}
 S_i(k_i,\{\hat q_i^\mu\},\mu_{S_i}) &= \id\, \de(k_i) + \frac{\alpha_s(\mu_{S_i})}{\pi} 
 \sum_{j \neq i} 
\bigg\{
\bT_i\cdot \bT_j  \Bigl[\frac{2}{\sqrt{\hs_{ij}}\, \mu_{S_i}}\cL_1\Bigl(\frac{k_i}{\sqrt{\hs_{ij}}\,\mu_{S_i}}\Bigr) - \frac{\pi^2}{24}\,\delta(k_i)\Bigr] 
 \\ & \quad
+ \sum_{m \neq i,j} \bigg[
 \Big\{ \bT_i \cdot \bT_j \,
  I_0\Bigl(\frac{\hs_{jm}}{\hs_{ij}}, \frac{\hs_{im}}{\hs_{ij}} \Bigr)
  - \bT_m \cdot\bT_j\,
  I_0\Bigl(\frac{\hs_{ij}}{\hs_{mj}}, \frac{\hs_{im}}{\hs_{mj}} \Bigr)
 \Big\}
\frac{1}{\mu}\cL_0\Bigl(\frac{k_i}{\mu_{S_i}}\Bigr)
\nn \\ & \quad +
\frac{1}{6} \Big\{\bT_i\cdot \bT_j \Big[I_0\Bigl(\frac{\hs_{jm}}{\hs_{ij}}, \frac{\hs_{im}}{\hs_{ij}} \Bigr) \ln\frac{\hs_{jm}}{\hs_{ij}} + I_1\Bigl(\frac{\hs_{jm}}{\hs_{ij}}, \frac{\hs_{im}}{\hs_{ij}} \Bigr) \Big] + \text{5 permutations of } (i, j, m) \Big\}  \delta(k_i) \bigg] \bigg\}
\,.\nn
\end{align}
%%%
Here $\hat s_{ij} = | s_{ij} /(Q_i Q_j) |$ with the $s_{ij}$ from \eq{sij}, the two integrals are
\begin{align} \label{eq:I01_def}
I_0(\alpha, \beta)
&=  \frac{1}{\pi}\int_{-\pi}^{\pi}\!\df\phi \int\!\frac{\df y}{y}\,
\theta\bigl(y - \sqrt{\beta/\alpha}\bigr)\,
\theta\bigl(1/\alpha - 1 - y^2 + 2 y\cos\phi\bigr)
\,,\nn\\
I_1(\alpha, \beta)
&= \frac{1}{\pi}\int_{-\pi}^{\pi}\!\df\phi \int\!\frac{\df y}{y}\,
\ln(1 + y^2 - 2y\cos\phi\bigr)\,
\theta\bigl(y - \sqrt{\beta/\alpha}\bigr)\, \theta\bigl(1/\alpha - 1 - y^2 + 2y\, \cos\phi\bigr)
\,,\end{align}
\end{widetext}
and the various color factors are
%%%
\begin{align} \label{eq:titj}
  gg\to Hg: \quad & \bT_a^2 = \bT_b^2 = \bT_J^2 = C_A
 \,, \nn \\ 
  & \bT_a \cdot \bT_b = \bT_a \cdot \bT_J = \bT_b \cdot \bT_J = -\frac{C_A}{2}
 \,, \nn \\
  gq\to Hq: \quad & \bT_a^2 = C_A\,, \ \bT_b^2 = \bT_J^2 = C_F
 \,, \nn \\ 
  & \bT_a \cdot \bT_b = \bT_a \cdot \bT_J = -\frac{C_A}{2}
 \,, \nn \\ 
  & \bT_b \cdot \bT_J = \frac{C_A}{2} - C_F
\,.\end{align}
%%%

%===============================================================================
\subsection{Evolution Factors}
\label{app:evo}
%===============================================================================

Following the discussion in \subsec{sfact}, we give expressions for the factorized evolution of the hard function,
%%%
\begin{align} \label{eq:Hrun}
& H_\kappa(\{q_j^\mu\}, \{\mu_i\}) 
  = H_\kappa(\{q_i^\mu\}, \mu_H)\!\! \prod_{i=a,b,J} \!\!
 U_{H_{\kappa_i}}\!(\{q_j^\mu\},\mu_H, \mu_i)
\,, \nn\\
& U_{H_{\kappa_i}}(\{q_j^\mu\}, \mu_H, \mu_i) 
  = \bigg\lvert e^{K_H^{i}} \prod_{j \neq i} 
  \Bigl(\frac{-s_{ij} - \img 0}{\mu_H^2}\Bigr)^{\bT_i \cdot \bT_j \eta_H}\bigg\rvert
\,,\nn \\
& K_H^{i}(\mu_H,\mu_i) 
  = -2 K_{\Ga^{\kappa_i}}(\mu_H,\mu_i)+ K_{\ga_H^{\kappa_i}}(\mu_H,\mu_i) 
\,, \nn \\
& \eta_H(\mu_H,\mu_i) 
  = -\frac{\eta_{\Ga^q}(\mu_H,\mu_i)}{C_F} = -\frac{\eta_{\Ga^g}(\mu_H,\mu_i)}{C_A}\!
\,.\end{align}
%%%
Here the products over $i$ and $j$ run over all colored particles, with
corresponding flavor $\kappa_i$ and $\kappa_j$. For each channel contributing to
$pp \to H+1j$ there is only one color structure so $\bT_i \cdot \bT_j$ is simply
a number [see \eq{titj}]. The functions $K_\Gamma$, $\eta_\Gamma$ and $K_\gamma$ are given below
in \eq{Keta}.

The solution of the RG evolution of the jet function is given by~\cite{Balzereit:1998yf, Neubert:2004dd, Fleming:2007xt, Ligeti:2008ac}
%%%
\begin{align} 
J_{\kappa_i}(s,\mu) & =  \int\! \df s'\, J_{\kappa_i}(s - s',\mu_J)\, U_{J_{\kappa_i}}(s',\mu_J, \mu)
\,, \nn \\
U_{J_{\kappa_i}}(s, \mu_J, \mu) &= \frac{e^{K_J^i -\gamma_E\, \eta_J^i}}{\Gamma(1+\eta_J^i)}\,
\biggl[\frac{\eta_J^i}{\mu_J^2} \cL^{\eta_J^i} \Bigl( \frac{s}{\mu_J^2} \Bigr) 
  + \delta(s) \biggr]
\,, \nn \\
K_J^i(\mu_J,\mu) &= 4 K_{\Gamma^{\kappa_i}}(\mu_J,\mu) 
  + K_{\gamma_J^{\kappa_i}}(\mu_J,\mu)
\,, \nn \\
\eta_J^i(\mu_J,\mu) &= -2\eta_{\Gamma^{\kappa_i}}(\mu_J,\mu)
\,.\end{align}
%%%
The plus distribution $\cL^\eta$ is defined as
%%%
\begin{align} \label{eq:Leta}
\cL^\eta(x)
&\equiv \biggl[ \frac{\theta(x)}{x^{1-\eta}}\biggr]_+
\nn \\ &
 = \lim_{\beta \to 0} \biggl[
  \frac{\theta(x - \beta)}{x^{1-\eta}} +
  \delta(x- \beta) \, \frac{x^\eta - 1}{\eta} \biggr]
\,.\end{align}
%%%
General relations for the rescaling and convolutions of $\cL_n(x)$ in \eq{Ln}
and $\cL^\eta(x)$ can be found in App.~B of Ref.~\cite{Ligeti:2008ac}.  The
renormalization group evolution of the beam functions is
identical~\cite{Stewart:2010qs} and can be obtained from the above expressions
by replacing $J_i(s,\mu) \to B_i(t,x,\mu)$.  We do not give the evolution of the
soft function, as it is not needed for evaluating \eq{sigmaTau1evo}. It can be
obtained from the evolution of the hard function and beam function by using the
$\mu$-independence of the cross section.

The functions $K_\Gamma(\mu_0, \mu)$, $\eta_\Gamma(\mu_0, \mu)$,
$K_\gamma(\mu_0, \mu)$ in the above RGE solutions at NNLL are given by,
%%%
\begin{widetext}
\begin{align} \label{eq:Keta}
K_\Gamma(\mu_0, \mu) &= -\frac{\Gamma_0}{4\beta_0^2}\,
\biggl\{ \frac{4\pi}{\alpha_s(\mu_0)}\, \Bigl(1 - \frac{1}{r} - \ln r\Bigr)
   + \biggl(\frac{\Gamma_1 }{\Gamma_0 } 
   - \frac{\beta_1}{\beta_0}\biggr) (1-r+\ln r)
   + \frac{\beta_1}{2\beta_0} \ln^2 r
\nn\\ & \quad
+ \frac{\alpha_s(\mu_0)}{4\pi}\, \biggl[
  \biggl(\frac{\beta_1^2}{\beta_0^2} 
 - \frac{\beta_2}{\beta_0} \biggr) \Bigl(\frac{1 - r^2}{2} + \ln r\Bigr)
  + \biggl(\frac{\beta_1\Gamma_1 }{\beta_0 \Gamma_0 } 
  - \frac{\beta_1^2}{\beta_0^2} \biggr) (1- r+ r\ln r)
  - \biggl(\frac{\Gamma_2 }{\Gamma_0} 
  - \frac{\beta_1\Gamma_1}{\beta_0\Gamma_0} \biggr) \frac{(1- r)^2}{2}
     \biggr] \biggr\}
\,, \nn\\
\eta_\Gamma(\mu_0, \mu) &=
 - \frac{\Gamma_0}{2\beta_0}\, \biggl[ \ln r
 + \frac{\alpha_s(\mu_0)}{4\pi}\, \biggl(\frac{\Gamma_1 }{\Gamma_0 }
 - \frac{\beta_1}{\beta_0}\biggr)(r-1)
 + \frac{\alpha_s^2(\mu_0)}{16\pi^2} \biggl(
    \frac{\Gamma_2 }{\Gamma_0 } - \frac{\beta_1\Gamma_1 }{\beta_0 \Gamma_0 }
      + \frac{\beta_1^2}{\beta_0^2} -\frac{\beta_2}{\beta_0} \biggr) \frac{r^2-1}{2}
    \biggr]
\,, \nn\\
K_\gamma(\mu_0, \mu) &=
 - \frac{\gamma_0}{2\beta_0}\, \biggl[ \ln r
 + \frac{\alpha_s(\mu_0)}{4\pi}\, \biggl(\frac{\gamma_1 }{\gamma_0 }
 - \frac{\beta_1}{\beta_0}\biggr)(r-1) \biggr]
\,.\end{align}
\end{widetext}
%%%
Here, $r = \alpha_s(\mu)/\alpha_s(\mu_0)$ and the running coupling at the scale
$\mu$ is given in terms of that at the reference scale $\mu_0$ by the three-loop
expression
%%%
\begin{align} \label{eq:alphas}
\frac{1}{\alpha_s(\mu)} &= \frac{X}{\alpha_s(\mu_0)}
  +\frac{\beta_1}{4\pi\beta_0}  \ln X
  + \frac{\alpha_s(\mu_0)}{16\pi^2} \biggr[
  \frac{\beta_2}{\beta_0} \Bigl(1-\frac{1}{X}\Bigr)
  \nn \\ & \quad
  + \frac{\beta_1^2}{\beta_0^2} \Bigl( \frac{\ln X}{X} +\frac{1}{X} -1\Bigr) \biggl]
\,,\end{align}
%%%
where $X\equiv 1+\alpha_s(\mu_0)\beta_0 \ln(\mu/\mu_0)/(2\pi)$.

%===============================================================================
\subsection{RGE Coefficients}
%===============================================================================

Up to three loops, the coefficients of the beta function~\cite{Tarasov:1980au,
  Larin:1993tp} and cusp anomalous dimension~\cite{Korchemsky:1987wg,
  Moch:2004pa} in $\overline{\mathrm{MS}}$ are
%%%
\begin{align} 
\beta_0 &= \frac{11}{3}\,C_A -\frac{4}{3}\,T_F\,n_f
\,,\\
\beta_1 &= \frac{34}{3}\,C_A^2  - \Bigl(\frac{20}{3}\,C_A\, + 4 C_F\Bigr)\, T_F\,n_f
\,, \nn\\
\beta_2 &=
\frac{2857}{54}\,C_A^3 + \Bigl(C_F^2 - \frac{205}{18}\,C_F C_A
 - \frac{1415}{54}\,C_A^2 \Bigr)\, 2T_F\,n_f
 \nn \\ & \quad
 + \Bigl(\frac{11}{9}\, C_F + \frac{79}{54}\, C_A \Bigr)\, 4T_F^2\,n_f^2
\,,\nn \\ 
\Gamma^q_0 &= 4C_F
\,,\nn\\
\Gamma^q_1 &= 4C_F \Bigl[\Bigl( \frac{67}{9} -\frac{\pi^2}{3} \Bigr)\,C_A  -
   \frac{20}{9}\,T_F\, n_f \Bigr]
\,,\nn\\
\Gamma^q_2 &= 4C_F \Bigl[
\Bigl(\frac{245}{6} -\frac{134 \pi^2}{27} + \frac{11 \pi ^4}{45}
  + \frac{22 \zeta_3}{3}\Bigr)C_A^2
  \nn \\ & \quad
  + \Bigl(- \frac{418}{27} + \frac{40 \pi^2}{27}  
  - \frac{56 \zeta_3}{3} \Bigr)C_A\, T_F\,n_f
\nn\\ & \quad
  + \Bigl(- \frac{55}{3} + 16 \zeta_3 \Bigr) C_F\, T_F\,n_f
  - \frac{16}{27}\,T_F^2\, n_f^2 \Bigr]
\,,  \nn \\ 
\Gamma^g_n &= \frac{C_A}{C_F}\, \Gamma^q_n  \quad \text{for } n \leq 2
  \nn
\,.\end{align}
%%%

Up to two loops, the $\overline{\mathrm{MS}}$ non-cusp anomalous dimension for
the hard function~\cite{Idilbi:2006dg, Becher:2006mr} and jet and beam
functions~\cite{Becher:2006qw, Becher:2009th, Stewart:2010qs, Berger:2010xi} are
%%%
\begin{align} \label{eq:gaHexp}
\gamma_{H\,0}^{q} &= -6 C_F
\,,\\
\gamma_{H\,1}^{q}
&= - C_F \Bigl[
  \Bigl(\frac{82}{9} - 52 \zeta_3\Bigr) C_A
+ (3 - 4 \pi^2 + 48 \zeta_3) C_F
\nn \\ & \quad
+ \Bigl(\frac{65}{9} + \pi^2 \Bigr) \beta_0 \Bigr]
\,, \nn \\
\ga_{H\,0}^{g} &= -2 \bt_0
\,,\nn\\
\ga_{H\,1}^{g}
&= \Big(-\frac{118}{9} + 4\zeta_3\Big)C_A^2 +
\Big(-\frac{38}{9}+\frac{\pi^2}{3}\Big) C_A\, \bt_0 - 2 \bt_1
\,,\nn\\
\gamma_{J\,0}^q &= 6 C_F
\,,\nn\\
\gamma_{J\,1}^q
&= C_F \Bigl[
  \Bigl(\frac{146}{9} - 80 \zeta_3\Bigr) C_A
+ (3 - 4 \pi^2 + 48 \zeta_3) C_F
\nn \\ & \quad
+ \Bigl(\frac{121}{9} + \frac{2\pi^2}{3} \Bigr) \beta_0 \Bigr]
\,,\nn\\
\ga_{J\,0}^g &= 2 \bt_0
\,,\nn\\
\ga_{J\,1}^g
&= \Big(\frac{182}{9} - 32\zeta_3\Big)C_A^2 +
\Big(\frac{94}{9}-\frac{2\pi^2}{3}\Big) C_A\, \bt_0 + 2\bt_1
\,.\nn \end{align}
%%%

%===============================================================================
\section{Running Scales}
\label{app:scales}
%===============================================================================

We now present the remaining ingredients that enter in the running scales in \subsec{scales}. First of all, $\mu_\mathrm{run}$ is defined as
%%%
\begin{align}  \label{eq:murun}
& \mu_\mathrm{run}(\tau,\mu,r_t,t_i) =
\begin{cases}
\mu_0 &  0\le \tau \leq t_0 
\,,\\
\mu_0 + \frac{r_t \mu}{2(t_1-t_0)} (\tau-t_0)^2 & t_0 \leq \tau \leq t_1
\,,\\[5pt]
r_t \mu\, \tau  -  b & t_1 \leq \tau \leq t_2
\,,\\[3pt]
r_t \mu \tau + \mu + a(\tau) &  t_2 \leq \tau \leq t_3
\,,\\
\mu & t_3 \le \tau 
\,,\end{cases}
\end{align}
%%%
where the function 
%%%
\begin{align}\label{eq:atau}
  & a(\tau) =
  \begin{cases}
-b -\mu + (d-c) (\tau-t_2)^2  &\ \ t_2 \leq \tau \leq \frac{t_2+t_3}{2}
\,,\\[3pt]
-r_t \mu\tau -(d+c) (\tau-t_3)^2 &\ \  \frac{t_2+t_3}{2}\leq \tau \leq t_3
\,,\\
  \end{cases}
\end{align}
and the coefficients in \eqs{murun}{atau} are
\begin{align}
b &= \frac{r_t \mu (t_0+t_1) }{2} - \mu_0 
 \,,\qquad\quad
 c = \frac{r_t \mu}{2(t_3-t_2)}  
  \,, \nn\\
 d &= \frac{2(\mu-\mu_0)-r_t \mu (t_3+t_2-t_1-t_0)}{(t_3-t_2)^2} 
  \,.
\end{align}
The expressions for $a(\tau)$, $b$, $c$, and $d$ follow from demanding that $\mu_\mathrm{run}(\tau)$ is continuous and has a continuous derivative.  The independent parameters in $\mu_\mathrm{run}(\tau)$ are the scale $\mu_0$ at small $\tau$, the scale $\mu$ at large $\tau$, the dimensionless slope parameter $r_t$, and the implicit $t_i$ parameters that determine the location of the transition between the nonperturbative region $t\le t_1$ and resummation region $t_1\le \tau\le t_2$, and also the location of the transition to the fixed-order region $t\ge t_3$.

For  central parameter choices in \eq{runscales} we use
%%%
\begin{align}
&\mu = \pTJ
\,,\quad
e_i = e_{S_i} = 0
\,,\quad
\mu_0 = 2 \text{GeV}
\,,\quad
r=0.2
\,,\nn \\[3pt]
&t_0 = \frac{0.5 \text{GeV}}{\sqrt{Q_J\pTJ}} 
\,,\quad
  t_1 = \frac{2 \text{GeV}}{\sqrt{Q_J\pTJ}}
\,,\quad
  t_2 = 0.05
\,,\quad
   t_3 = 0.3 
\,,\nn \\[10pt]
\vspace{2cm}
&t_0' = \frac{2 \text{GeV}}{\pTJ} 
\,,\quad
  t_1' = \frac{8 \text{GeV}}{\pTJ}
\,,\quad
  t_2' = 0.3
\,,\quad
   t_3' = 0.6
\,.\end{align}
%%%
These $t_i$ parameters appear in the soft scale $\mu_{S_J}$ and jet scale $\mu_J$, while the $t_i'$ parameters appear in the beam related scales $\mu_{B_i}$ and $\mu_{S_B}$. The choice of $t_3=0.3$ ensures that we transition to the fixed-order region sufficiently before $m_J^2\simeq \pTJ R/\sqrt2$.

To estimate the perturbative uncertainty we vary the above parameters within reasonable ranges. The parameters $e_{S_J}$, $e_J$, $e_{S_B}$, and $e_B$ allow us to individually vary each of the scales $\mu_{S_J}$, $\mu_J$, $\mu_{B_{a,b}}$, and $\mu_B$. These variations are independent of varying the overall scale through changes in $\mu$. The parameter $r$  allows us to estimate uncertainty from the refactorization of the soft function. Since the cross section is most sensitive to $\mu$, $e_i$, $e_{S_i}$ and $r$, we restrict ourselves to the following separate variations,
%%%
\begin{align} \label{eq:scales}
\text{a)}& \
\mu = 2^{\pm 1} Q\,, \ \: e_J = e_B = e_{S_J} = e_{S_B} = 0\,, \ \: r=0.2
\,,\nn\\
\text{b)}& \
\mu = Q\,, \ \: e_J = \pm 0.5\,, \ \: e_B = e_{S_J} = e_{S_B} = 0\,, \ \: r=0.2
\,,\nn\\
\text{c)}& \
\mu = Q\,, \ \: e_B = \pm 0.5\,, \ \: e_J = e_{S_J} = e_{S_B} = 0\,, \ \: r=0.2
\,,\nn\\
\text{d)}& \
\mu = Q\,, \ \: e_{S_J} = \pm 0.5\,, \ \: e_J = e_B = e_{S_B} = 0\,, \ \: r=0.2
\,,\nn\\
\text{e)}& \
\mu = Q\,, \ \: e_{S_B} = \pm 0.5\,, \ \: e_J = e_B = e_{S_J} = 0\,, \ \: r=0.2
\,,\nn\\
\text{f)}& \
\mu = Q\,, \ \: e_J = e_B = e_{S_J} = e_{S_B} = 0\,, \ \: r=0.2\pm 0.2
\,.\end{align}
%%%
Following our discussion in Refs.~\cite{Stewart:2011cf, Dittmaier:2012vm}, we take the envelope of variations b) through f) and add this in quadrature with variation a).

\bibliographystyle{../physrev4}
\bibliography{../pp}

\providecommand{\href}[2]{#2}\begin{thebibliography}{10}

\bibitem{Altheimer:2012mn}
A.~Altheimer {\em et~al.},
\newblock J. Phys. G {\bf G39}, 063001 (2012),
  [\href{http://arXiv.org/abs/arXiv:1201.0008}{arXiv:1201.0008}].
%%CITATION = ARXIV:1201.0008;%%

\bibitem{Ellis:2009wj}
S.~D. Ellis, A.~Hornig, C.~Lee, C.~K. Vermilion, and J.~R. Walsh,
\newblock Phys. Lett. B {\bf 689}, 82 (2010),
  [\href{http://arXiv.org/abs/arXiv:0912.0262}{arXiv:0912.0262}].
%%CITATION = 0912.0262;%%

\bibitem{Bauer:2011uc}
C.~W. Bauer, F.~J. Tackmann, J.~R. Walsh, and S.~Zuberi,
\newblock Phys. Rev. D {\bf 85}, 074006 (2012),
  [\href{http://arXiv.org/abs/arXiv:1106.6047}{arXiv:1106.6047}].
%%CITATION = ARXIV:1106.6047;%%

\bibitem{Feige:2012vc}
I.~Feige, M.~D. Schwartz, I.~W. Stewart, and J.~Thaler,
\newblock Phys. Rev. Lett. {\bf 109}, 092001 (2012),
  [\href{http://arXiv.org/abs/arXiv:1204.3898}{arXiv:1204.3898}].
%%CITATION = ARXIV:1204.3898;%%

\bibitem{Larkoski:2012eh}
A.~J. Larkoski,
\newblock Phys. Rev. D {\bf 86}, 054004 (2012),
  [\href{http://arXiv.org/abs/arXiv:1207.1437}{arXiv:1207.1437}].
%%CITATION = ARXIV:1207.1437;%%

\bibitem{Krohn:2012fg}
D.~Krohn, M.~D. Schwartz, T.~Lin, and W.~J. Waalewijn,
\newblock Phys. Rev. Lett. {\bf 110}, 212001 (2013),
  [\href{http://arXiv.org/abs/1209.2421}{1209.2421}].
%%CITATION = ARXIV:1209.2421;%%

\bibitem{Waalewijn:2012sv}
W.~J. Waalewijn,
\newblock Phys. Rev. D {\bf 86}, 094030 (2012),
  [\href{http://arXiv.org/abs/arXiv:1209.3019}{arXiv:1209.3019}].
%%CITATION = ARXIV:1209.3019;%%

\bibitem{Ellis:2010rwa}
S.~D. Ellis, C.~K. Vermilion, J.~R. Walsh, A.~Hornig, and C.~Lee,
\newblock JHEP {\bf 11}, 101 (2010),
  [\href{http://arXiv.org/abs/arXiv:1001.0014}{arXiv:1001.0014}].
%%CITATION = 1001.0014;%%

\bibitem{Jouttenus:2009ns}
T.~T. Jouttenus,
\newblock Phys. Rev. D {\bf 81}, 094017 (2010),
  [\href{http://arXiv.org/abs/arXiv:0912.5509}{arXiv:0912.5509}].
%%CITATION = 0912.5509;%%

\bibitem{Banfi:2010pa}
A.~Banfi, M.~Dasgupta, K.~Khelifa-Kerfa, and S.~Marzani,
\newblock JHEP {\bf 08}, 064 (2010),
  [\href{http://arXiv.org/abs/arXiv:1004.3483}{arXiv:1004.3483}].
%%CITATION = 1004.3483;%%

\bibitem{Kelley:2011tj}
R.~Kelley, M.~D. Schwartz, and H.~X. Zhu,
\newblock \href{http://arXiv.org/abs/arXiv:1102.0561}{arXiv:1102.0561}.
%%CITATION = ARXIV:1102.0561;%%

\bibitem{Kelley:2011aa}
R.~Kelley, M.~D. Schwartz, R.~M. Schabinger, and H.~X. Zhu,
\newblock Phys. Rev. D {\bf 86}, 054017 (2012),
  [\href{http://arXiv.org/abs/arXiv:1112.3343}{arXiv:1112.3343}].
%%CITATION = ARXIV:1112.3343;%%

\bibitem{Li:2012bw}
H.-n. Li, Z.~Li, and C.-P. Yuan,
\newblock Phys. Rev. D {\bf 87}, 074025 (2012),
  [\href{http://arXiv.org/abs/1206.1344}{1206.1344}].
%%CITATION = ARXIV:1206.1344;%%

\bibitem{Dasgupta:2012hg}
M.~Dasgupta, K.~Khelifa-Kerfa, S.~Marzani, and M.~Spannowsky,
\newblock JHEP {\bf 1210}, 126 (2012),
  [\href{http://arXiv.org/abs/arXiv:1207.1640}{arXiv:1207.1640}].
%%CITATION = ARXIV:1207.1640;%%

\bibitem{Chien:2012ur}
Y.-T. Chien, R.~Kelley, M.~D. Schwartz, and H.~X. Zhu,
\newblock Phys. Rev. D {\bf 87}, 014010 (2013),
  [\href{http://arXiv.org/abs/arXiv:1208.0010}{arXiv:1208.0010}].
%%CITATION = ARXIV:1208.0010;%%

\bibitem{Krohn:2009th}
D.~Krohn, J.~Thaler, and L.-T. Wang,
\newblock JHEP {\bf 1002}, 084 (2010),
  [\href{http://arXiv.org/abs/arXiv:0912.1342}{arXiv:0912.1342}].
%%CITATION = ARXIV:0912.1342;%%

\bibitem{Butterworth:2008iy}
J.~M. Butterworth, A.~R. Davison, M.~Rubin, and G.~P. Salam,
\newblock Phys. Rev. Lett. {\bf 100}, 242001 (2008),
  [\href{http://arXiv.org/abs/arXiv:0802.2470}{arXiv:0802.2470}].
%%CITATION = ARXIV:0802.2470;%%

\bibitem{Ellis:2009su}
S.~D. Ellis, C.~K. Vermilion, and J.~R. Walsh,
\newblock Phys. Rev. D {\bf 80}, 051501 (2009),
  [\href{http://arXiv.org/abs/arXiv:0903.5081}{arXiv:0903.5081}].
%%CITATION = ARXIV:0903.5081;%%

\bibitem{Korchemsky:1999kt}
G.~P. Korchemsky and G.~Sterman,
\newblock Nucl. Phys. B {\bf 555}, 335 (1999),
  [\href{http://arXiv.org/abs/hep-ph/9902341}{hep-ph/9902341}].
%%CITATION = HEP-PH/9902341;%%

\bibitem{Bauer:2001yt}
C.~W. Bauer, D.~Pirjol, and I.~W. Stewart,
\newblock Phys. Rev. D {\bf 65}, 054022 (2002),
  [\href{http://arXiv.org/abs/hep-ph/0109045}{hep-ph/0109045}].
%%CITATION = HEP-PH 0109045;%%

\bibitem{Fleming:2007qr}
S.~Fleming, A.~H. Hoang, S.~Mantry, and I.~W. Stewart,
\newblock Phys. Rev. D {\bf 77}, 074010 (2008),
  [\href{http://arXiv.org/abs/hep-ph/0703207}{hep-ph/0703207}].
%%CITATION = HEP-PH/0703207;%%

\bibitem{Chien:2010kc}
Y.-T. Chien and M.~D. Schwartz,
\newblock JHEP {\bf 1008}, 058 (2010),
  [\href{http://arXiv.org/abs/arXiv:1005.1644}{arXiv:1005.1644}].
%%CITATION = ARXIV:1005.1644;%%

\bibitem{Abbate:2010xh}
R.~Abbate, M.~Fickinger, A.~H. Hoang, V.~Mateu, and I.~W. Stewart,
\newblock Phys. Rev. D {\bf 83}, 074021 (2011),
  [\href{http://arXiv.org/abs/arXiv:1006.3080}{arXiv:1006.3080}].
%%CITATION = ARXIV:1006.3080;%%

\bibitem{Gehrmann:2012sc}
T.~Gehrmann, G.~Luisoni, and P.~F. Monni,
\newblock Eur. Phys. J. C {\bf 73}, 2265 (2013),
  [\href{http://arXiv.org/abs/1210.6945}{1210.6945}].
%%CITATION = ARXIV:1210.6945;%%

\bibitem{Dixon:2008gr}
L.~J. Dixon, L.~Magnea, and G.~Sterman,
\newblock JHEP {\bf 08}, 022 (2008),
  [\href{http://arXiv.org/abs/arXiv:0805.3515}{arXiv:0805.3515}].
%%CITATION = 0805.3515;%%

\bibitem{ATLAS:2012am}
ATLAS Collaboration, G.~Aad {\em et~al.},
\newblock JHEP {\bf 1205}, 128 (2012),
  [\href{http://arXiv.org/abs/arXiv:1203.4606}{arXiv:1203.4606}].
%%CITATION = ARXIV:1203.4606;%%

\bibitem{Cacciari:2008gp}
M.~Cacciari, G.~P. Salam, and G.~Soyez,
\newblock JHEP {\bf 04}, 063 (2008),
  [\href{http://arXiv.org/abs/arXiv:0802.1189}{arXiv:0802.1189}].
%%CITATION = 0802.1189;%%

\bibitem{Banfi:2005gj}
A.~Banfi and M.~Dasgupta,
\newblock Phys. Lett. B {\bf 628}, 49 (2005),
  [\href{http://arXiv.org/abs/hep-ph/0508159}{hep-ph/0508159}].
%%CITATION = HEP-PH/0508159;%%

\bibitem{Delenda:2006nf}
Y.~Delenda, R.~Appleby, M.~Dasgupta, and A.~Banfi,
\newblock JHEP {\bf 0612}, 044 (2006),
  [\href{http://arXiv.org/abs/hep-ph/0610242}{hep-ph/0610242}].
%%CITATION = HEP-PH/0610242;%%

\bibitem{KhelifaKerfa:2011zu}
K.~Khelifa-Kerfa,
\newblock JHEP {\bf 1202}, 072 (2012),
  [\href{http://arXiv.org/abs/arXiv:1111.2016}{arXiv:1111.2016}].
%%CITATION = ARXIV:1111.2016;%%

\bibitem{Kelley:2012kj}
R.~Kelley, J.~R. Walsh, and S.~Zuberi,
\newblock JHEP {\bf 1209}, 117 (2012),
  [\href{http://arXiv.org/abs/arXiv:1202.2361}{arXiv:1202.2361}].
%%CITATION = ARXIV:1202.2361;%%

\bibitem{Dasgupta:2001sh}
M.~Dasgupta and G.~P. Salam,
\newblock Phys. Lett. B {\bf 512}, 323 (2001),
  [\href{http://arXiv.org/abs/hep-ph/0104277}{hep-ph/0104277}].
%%CITATION = HEP-PH/0104277;%%

\bibitem{Dasgupta:2002dc}
M.~Dasgupta and G.~P. Salam,
\newblock JHEP {\bf 0208}, 032 (2002),
  [\href{http://arXiv.org/abs/hep-ph/0208073}{hep-ph/0208073}].
%%CITATION = HEP-PH/0208073;%%

\bibitem{Clavelli:1979md}
L.~Clavelli,
\newblock Phys. Lett. B {\bf 85}, 111 (1979).
%\%CITATION = PHLTA,B85,111;\%\%

\bibitem{Chandramohan:1980ry}
T.~Chandramohan and L.~Clavelli,
\newblock Nucl. Phys. B {\bf 184}, 365 (1981).
%\%CITATION = NUPHA,B184,365;\%\%

\bibitem{Clavelli:1981yh}
L.~Clavelli and D.~Wyler,
\newblock Phys. Lett. B {\bf 103}, 383 (1981).
%\%CITATION = PHLTA,B103,383;\%\%

\bibitem{Berger:2003iw}
C.~F. Berger, T.~Kucs, and G.~Sterman,
\newblock Phys. Rev. D {\bf 68}, 014012 (2003),
  [\href{http://arXiv.org/abs/hep-ph/0303051}{hep-ph/0303051}].
%%CITATION = HEP-PH/0303051;%%

\bibitem{Fleming:2007xt}
S.~Fleming, A.~H. Hoang, S.~Mantry, and I.~W. Stewart,
\newblock Phys. Rev. D {\bf 77}, 114003 (2008),
  [\href{http://arXiv.org/abs/arXiv:0711.2079}{arXiv:0711.2079}].
%%CITATION = 0711.2079;%%

\bibitem{Appell:1988ie}
D.~Appell, G.~Sterman, and P.~B. Mackenzie,
\newblock Nucl. Phys. B {\bf 309}, 259 (1988).
%%CITATION = NUPHA,B309,259;%%

\bibitem{Catani:1998tm}
S.~Catani, M.~L. Mangano, and P.~Nason,
\newblock JHEP {\bf 07}, 024 (1998),
  [\href{http://arXiv.org/abs/hep-ph/9806484}{hep-ph/9806484}].
%%CITATION = HEP-PH/9806484;%%

\bibitem{Becher:2007ty}
T.~Becher, M.~Neubert, and G.~Xu,
\newblock JHEP {\bf 07}, 030 (2008),
  [\href{http://arXiv.org/abs/arXiv:0710.0680}{arXiv:0710.0680}].
%%CITATION = 0710.0680;%%

\bibitem{Kelley:2011ng}
R.~Kelley, M.~D. Schwartz, R.~M. Schabinger, and H.~X. Zhu,
\newblock Phys. Rev. D {\bf 84}, 045022 (2011),
  [\href{http://arXiv.org/abs/arXiv:1105.3676}{arXiv:1105.3676}].
%%CITATION = ARXIV:1105.3676;%%

\bibitem{Hornig:2011iu}
A.~Hornig, C.~Lee, I.~W. Stewart, J.~R. Walsh, and S.~Zuberi,
\newblock JHEP {\bf 1108}, 054 (2011),
  [\href{http://arXiv.org/abs/arXiv:1105.4628}{arXiv:1105.4628}].
%%CITATION = ARXIV:1105.4628;%%

\bibitem{Mellado:2004tj}
B.~Mellado, W.~Quayle, and S.~L. Wu,
\newblock Phys. Lett. B {\bf 611}, 60 (2005),
  [\href{http://arXiv.org/abs/hep-ph/0406095}{hep-ph/0406095}].
%%CITATION = HEP-PH/0406095;%%

\bibitem{Mellado:2007fb}
B.~Mellado, W.~Quayle, and S.~L. Wu,
\newblock Phys. Rev. D {\bf 76}, 093007 (2007),
  [\href{http://arXiv.org/abs/arXiv:0708.2507}{arXiv:0708.2507}].
%%CITATION = ARXIV:0708.2507;%%

\bibitem{Liu:2012sz}
X.~Liu and F.~Petriello,
\newblock Phys. Rev. D {\bf 87}, 014018 (2013),
  [\href{http://arXiv.org/abs/arXiv:1210.1906}{arXiv:1210.1906}].
%%CITATION = ARXIV:1210.1906;%%

\bibitem{Stewart:2010tn}
I.~W. Stewart, F.~J. Tackmann, and W.~J. Waalewijn,
\newblock Phys. Rev. Lett. {\bf 105}, 092002 (2010),
  [\href{http://arXiv.org/abs/arXiv:1004.2489}{arXiv:1004.2489}].
%%CITATION = 1004.2489;%%

\bibitem{Jouttenus:2011wh}
T.~T. Jouttenus, I.~W. Stewart, F.~J. Tackmann, and W.~J. Waalewijn,
\newblock Phys. Rev. D {\bf 83}, 114030 (2011),
  [\href{http://arXiv.org/abs/arXiv:1102.4344}{arXiv:1102.4344}].

\bibitem{Bauer:2000ew}
C.~W. Bauer, S.~Fleming, and M.~E. Luke,
\newblock Phys. Rev. D {\bf 63}, 014006 (2000),
  [\href{http://arXiv.org/abs/hep-ph/0005275}{hep-ph/0005275}].
%%CITATION = HEP-PH 0005275;%%

\bibitem{Bauer:2000yr}
C.~W. Bauer, S.~Fleming, D.~Pirjol, and I.~W. Stewart,
\newblock Phys. Rev. D {\bf 63}, 114020 (2001),
  [\href{http://arXiv.org/abs/hep-ph/0011336}{hep-ph/0011336}].
%%CITATION = HEP-PH 0011336;%%

\bibitem{Bauer:2001ct}
C.~W. Bauer and I.~W. Stewart,
\newblock Phys. Lett. B {\bf 516}, 134 (2001),
  [\href{http://arXiv.org/abs/hep-ph/0107001}{hep-ph/0107001}].
%%CITATION = HEP-PH 0107001;%%

\bibitem{Sjostrand:2006za}
T.~Sj{\"o}strand, S.~Mrenna, and P.~Skands,
\newblock JHEP {\bf 05}, 026 (2006),
  [\href{http://arXiv.org/abs/hep-ph/0603175}{hep-ph/0603175}].
%%CITATION = HEP-PH/0603175;%%

\bibitem{Sjostrand:2007gs}
T.~Sj{\"o}strand, S.~Mrenna, and P.~Skands,
\newblock Comput. Phys. Commun. {\bf 178}, 852 (2008),
  [\href{http://arXiv.org/abs/arXiv:0710.3820}{arXiv:0710.3820}].
%%CITATION = 0710.3820;%%

\bibitem{Cacciari:2007fd}
M.~Cacciari and G.~P. Salam,
\newblock Phys. Lett. B {\bf 659}, 119 (2008),
  [\href{http://arXiv.org/abs/arXiv:0707.1378}{arXiv:0707.1378}].
%%CITATION = ARXIV:0707.1378;%%

\bibitem{Soyez:2012hv}
G.~Soyez, G.~P. Salam, J.~Kim, S.~Dutta, and M.~Cacciari,
\newblock Phys. Rev. Lett. {\bf 110}, 162001 (2012),
  [\href{http://arXiv.org/abs/1211.2811}{1211.2811}].
%%CITATION = ARXIV:1211.2811;%%

\bibitem{Thaler:2011gf}
J.~Thaler and K.~Van~Tilburg,
\newblock JHEP {\bf 1202}, 093 (2012),
  [\href{http://arXiv.org/abs/arXiv:1108.2701}{arXiv:1108.2701}].
%%CITATION = ARXIV:1108.2701;%%

\bibitem{Schmidt:1997wr}
C.~R. Schmidt,
\newblock Phys. Lett. B {\bf 413}, 391 (1997),
  [\href{http://arXiv.org/abs/hep-ph/9707448}{hep-ph/9707448}].
%%CITATION = HEP-PH/9707448;%%

\bibitem{Stewart:2012yh}
I.~W. Stewart, F.~J. Tackmann, and W.~J. Waalewijn,
\newblock PoS {\bf LL2012}, 058 (2012),
  [\href{http://arXiv.org/abs/1211.2305}{1211.2305}].
%%CITATION = ARXIV:1211.2305;%%

\bibitem{Bauer:2003pi}
C.~W. Bauer and A.~V. Manohar,
\newblock Phys. Rev. D {\bf 70}, 034024 (2004),
  [\href{http://arXiv.org/abs/hep-ph/0312109}{hep-ph/0312109}].
%%CITATION = HEP-PH/0312109;%%

\bibitem{Fleming:2003gt}
S.~Fleming, A.~K. Leibovich, and T.~Mehen,
\newblock Phys. Rev. D {\bf 68}, 094011 (2003),
  [\href{http://arXiv.org/abs/hep-ph/0306139}{hep-ph/0306139}].
%%CITATION = HEP-PH/0306139;%%

\bibitem{Becher:2009th}
T.~Becher and M.~D. Schwartz,
\newblock JHEP {\bf 02}, 040 (2010),
  [\href{http://arXiv.org/abs/arXiv:0911.0681}{arXiv:0911.0681}].
%%CITATION = 0911.0681;%%

\bibitem{Stewart:2009yx}
I.~W. Stewart, F.~J. Tackmann, and W.~J. Waalewijn,
\newblock Phys. Rev. D {\bf 81}, 094035 (2010),
  [\href{http://arXiv.org/abs/arXiv:0910.0467}{arXiv:0910.0467}].
%%CITATION = 0910.0467;%%

\bibitem{Stewart:2010qs}
I.~W. Stewart, F.~J. Tackmann, and W.~J. Waalewijn,
\newblock JHEP {\bf 09}, 005 (2010),
  [\href{http://arXiv.org/abs/arXiv:1002.2213}{arXiv:1002.2213}].
%%CITATION = 1002.2213;%%

\bibitem{Mantry:2009qz}
S.~Mantry and F.~Petriello,
\newblock Phys. Rev. D {\bf 81}, 093007 (2010),
  [\href{http://arXiv.org/abs/arXiv:0911.4135}{arXiv:0911.4135}].
%%CITATION = 0911.4135;%%

\bibitem{Berger:2010xi}
C.~F. Berger, C.~Marcantonini, I.~W. Stewart, F.~J. Tackmann, and W.~J.
  Waalewijn,
\newblock JHEP {\bf 04}, 092 (2011),
  [\href{http://arXiv.org/abs/arXiv:1012.4480}{arXiv:1012.4480}].
%%CITATION = 1012.4480;%%

\bibitem{Korchemsky:1987wg}
G.~P. Korchemsky and A.~V. Radyushkin,
\newblock Nucl. Phys. B {\bf 283}, 342 (1987).
%%CITATION = NUPHA,B283,342;%%

\bibitem{Moch:2004pa}
S.~Moch, J.~A.~M. Vermaseren, and A.~Vogt,
\newblock Nucl. Phys. B {\bf 688}, 101 (2004),
  [\href{http://arXiv.org/abs/hep-ph/0403192}{hep-ph/0403192}].
%%CITATION = HEP-PH/0403192;%%

\bibitem{Kramer:1986sg}
G.~Kramer and B.~Lampe,
\newblock Z. Phys. C {\bf 34}, 497 (1987),
\newblock [Erratum-ibid.\ C {\bf 42}, 504 (1989)].
%%CITATION = ZEPYA,C34,497;%%

\bibitem{Harlander:2000mg}
R.~V. Harlander,
\newblock Phys. Lett. B {\bf 492}, 74 (2000),
  [\href{http://arXiv.org/abs/hep-ph/0007289}{hep-ph/0007289}].
%%CITATION = HEP-PH/0007289;%%

\bibitem{MertAybat:2006mz}
S.~{Mert Aybat}, L.~J. Dixon, and G.~Sterman,
\newblock Phys. Rev. D {\bf 74}, 074004 (2006),
  [\href{http://arXiv.org/abs/hep-ph/0607309}{hep-ph/0607309}].
%%CITATION = HEP-PH/0607309;%%

\bibitem{Becher:2006qw}
T.~Becher and M.~Neubert,
\newblock Phys. Lett. B {\bf 637}, 251 (2006),
  [\href{http://arXiv.org/abs/hep-ph/0603140}{hep-ph/0603140}].
%%CITATION = HEP-PH/0603140;%%

\bibitem{Ahrens:2008qu}
V.~Ahrens, T.~Becher, M.~Neubert, and L.~L. Yang,
\newblock Phys. Rev. D {\bf 79}, 033013 (2009),
  [\href{http://arXiv.org/abs/arXiv:0808.3008}{arXiv:0808.3008}].
%%CITATION = 0808.3008;%%

\bibitem{Chiu:2008vv}
J.-y. Chiu, R.~Kelley, and A.~V. Manohar,
\newblock Phys. Rev. D {\bf 78}, 073006 (2008),
  [\href{http://arXiv.org/abs/arXiv:0806.1240}{arXiv:0806.1240}].
%%CITATION = 0806.1240;%%

\bibitem{Becher:2009qa}
T.~Becher and M.~Neubert,
\newblock JHEP {\bf 06}, 081 (2009),
  [\href{http://arXiv.org/abs/arXiv:0903.1126}{arXiv:0903.1126}].
%%CITATION = 0903.1126;%%

\bibitem{Ligeti:2008ac}
Z.~Ligeti, I.~W. Stewart, and F.~J. Tackmann,
\newblock Phys. Rev. D {\bf 78}, 114014 (2008),
  [\href{http://arXiv.org/abs/arXiv:0807.1926}{arXiv:0807.1926}].
%%CITATION = 0807.1926;%%

\bibitem{:2012gk}
ATLAS Collaboration, G.~Aad {\em et~al.},
\newblock Phys. Lett. B {\bf 716}, 1 (2012),
  [\href{http://arXiv.org/abs/arXiv:1207.7214}{arXiv:1207.7214}].
%%CITATION = ARXIV:1207.7214;%%

\bibitem{:2012gu}
CMS Collaboration, S.~Chatrchyan {\em et~al.},
\newblock Phys. Lett. B {\bf 716}, 30 (2012),
  [\href{http://arXiv.org/abs/arXiv:1207.7235}{arXiv:1207.7235}].
%%CITATION = ARXIV:1207.7235;%%

\bibitem{Martin:2009bu}
A.~D. Martin, W.~J. Stirling, R.~S. Thorne, and G.~Watt,
\newblock Eur. Phys. J. C {\bf 64}, 653 (2009),
  [\href{http://arXiv.org/abs/arXiv:0905.3531}{arXiv:0905.3531}].
%%CITATION = 0905.3531;%%

\bibitem{Cacciari:2011ma}
M.~Cacciari, G.~P. Salam, and G.~Soyez,
\newblock Eur. Phys. J. C {\bf 72}, 1896 (2012),
  [\href{http://arXiv.org/abs/arXiv:1111.6097}{arXiv:1111.6097}].
%%CITATION = ARXIV:1111.6097;%%

\bibitem{Dokshitzer:1995qm}
Y.~L. Dokshitzer, G.~Marchesini, and B.~R. Webber,
\newblock Nucl. Phys. B {\bf 469}, 93 (1996),
  [\href{http://arXiv.org/abs/hep-ph/9512336}{hep-ph/9512336}].
%\%CITATION = HEP-PH/9512336;\%\%

\bibitem{Dokshitzer:1995zt}
Y.~L. Dokshitzer and B.~R. Webber,
\newblock Phys. Lett. B {\bf 352}, 451 (1995),
  [\href{http://arXiv.org/abs/hep-ph/9504219}{hep-ph/9504219}].
%\%CITATION = HEP-PH/9504219;\%\%

\bibitem{Dokshitzer:1997iz}
Y.~L. Dokshitzer, A.~Lucenti, G.~Marchesini, and G.~Salam,
\newblock Nucl. Phys. {\bf B511}, 396 (1998),
  [\href{http://arXiv.org/abs/hep-ph/9707532}{hep-ph/9707532}].
%%CITATION = HEP-PH/9707532;%%

\bibitem{Salam:2001bd}
G.~P. Salam and D.~Wicke,
\newblock JHEP {\bf 05}, 061 (2001),
  [\href{http://arXiv.org/abs/hep-ph/0102343}{hep-ph/0102343}].
%\%CITATION = HEP-PH/0102343;\%\%

\bibitem{Lee:2006fn}
C.~Lee and G.~F. Sterman,
\newblock eConf {\bf C0601121}, A001 (2006),
  [\href{http://arXiv.org/abs/hep-ph/0603066}{hep-ph/0603066}].
%%CITATION = HEP-PH/0603066;%%

\bibitem{Hoang:2007vb}
A.~H. Hoang and I.~W. Stewart,
\newblock Phys. Lett. B {\bf 660}, 483 (2008),
  [\href{http://arXiv.org/abs/arXiv:0709.3519}{arXiv:0709.3519}].
%%CITATION = 0709.3519;%%

\bibitem{Mateu:2012nk}
V.~Mateu, I.~W. Stewart, and J.~Thaler,
\newblock Phys. Rev. {\bf D87}, 014025 (2013),
  [\href{http://arXiv.org/abs/arXiv:1209.3781}{arXiv:1209.3781}].
%%CITATION = ARXIV:1209.3781;%%

\bibitem{Dasgupta:2007wa}
M.~Dasgupta, L.~Magnea, and G.~P. Salam,
\newblock JHEP {\bf 02}, 055 (2008),
  [\href{http://arXiv.org/abs/arXiv:0712.3014}{arXiv:0712.3014}].
%%CITATION = 0712.3014;%%

\bibitem{Fleming:2006cd}
S.~Fleming, A.~K. Leibovich, and T.~Mehen,
\newblock Phys. Rev. D {\bf 74}, 114004 (2006),
  [\href{http://arXiv.org/abs/hep-ph/0607121}{hep-ph/0607121}].
%%CITATION = HEP-PH/0607121;%%

\bibitem{Balzereit:1998yf}
C.~Balzereit, T.~Mannel, and W.~Kilian,
\newblock Phys. Rev. D {\bf 58}, 114029 (1998),
  [\href{http://arXiv.org/abs/hep-ph/9805297}{hep-ph/9805297}].
%%CITATION = HEP-PH/9805297;%%

\bibitem{Neubert:2004dd}
M.~Neubert,
\newblock Eur. Phys. J. C {\bf 40}, 165 (2005),
  [\href{http://arXiv.org/abs/hep-ph/0408179}{hep-ph/0408179}].
%%CITATION = HEP-PH/0408179;%%

\bibitem{Tarasov:1980au}
O.~V. Tarasov, A.~A. Vladimirov, and A.~Y. Zharkov,
\newblock Phys. Lett. B {\bf 93}, 429 (1980).
%%CITATION = PHLTA,B93,429;%%

\bibitem{Larin:1993tp}
S.~A. Larin and J.~A.~M. Vermaseren,
\newblock Phys. Lett. B {\bf 303}, 334 (1993),
  [\href{http://arXiv.org/abs/hep-ph/9302208}{hep-ph/9302208}].
%%CITATION = HEP-PH/9302208;%%

\bibitem{Idilbi:2006dg}
A.~Idilbi, X.~dong Ji, and F.~Yuan,
\newblock Nucl. Phys. B {\bf 753}, 42 (2006),
  [\href{http://arXiv.org/abs/hep-ph/0605068}{hep-ph/0605068}].
%%CITATION = HEP-PH/0605068;%%

\bibitem{Becher:2006mr}
T.~Becher, M.~Neubert, and B.~D. Pecjak,
\newblock JHEP {\bf 01}, 076 (2007),
  [\href{http://arXiv.org/abs/hep-ph/0607228}{hep-ph/0607228}].
%%CITATION = HEP-PH/0607228;%%

\bibitem{Stewart:2011cf}
I.~W. Stewart and F.~J. Tackmann,
\newblock Phys. Rev. D {\bf 85}, 034011 (2012),
  [\href{http://arXiv.org/abs/arXiv:1107.2117}{arXiv:1107.2117}].
%%CITATION = ARXIV:1107.2117;%%

\bibitem{Dittmaier:2012vm}
S.~Dittmaier {\em et~al.},
\newblock \href{http://arXiv.org/abs/arXiv:1201.3084}{arXiv:1201.3084}.
%%CITATION = ARXIV:1201.3084;%%

\end{thebibliography}

\end{document}